**ASSESSMENT OF SUSTAINABILITY VALUE AND DIGNIFIED WELL-BEING IN INCLUSIVE PRODUCT LIFECYCLES**

Naz Yaldız[1][https://orcid.org/0000-0001-8738-7137], Amaresh Chakrabarti[2][https://orcid.org/0000-0002-1809-1831]

[1,2]Department of Design and Manufacturing, Indian Institute of Science, Bengaluru, India
[1]Corresponding Author: nazyaldiz@iisc.ac.in
[2]ac123@iisc.ac.in

## ABSTRACT

Sustainability is a concept mainly assessed by the features of a product that are considered as specific metrics for sustainability in an engineering context. However, sustainability is a comprehensive process for improvement in sustainable development that can be structured with the influence of inclusivity of empowered people. We define sustainability value, which measures the merit of the process to contribute to sustainable development. Sustainability is interrelated with the well-being and dignity of people; we propose a formula for dignified well-being to evaluate the benefit of dignity and well-being in an inclusive product lifecycle. This study uses the framework for assessing inclusivity and empowerment to correlate the influence of inclusivity of empowered people on sustainability value and dignified well-being. Based on the strength of the correlation, we can support the hypothesis that the inclusivity of empowered people can influence the sustainability value of the process and the dignified well-being of stakeholders. Due to the integration of qualitative methods in our approaches, we tested it with an intercoder reliability test with three coders. Results show high coder agreement and the generalisability of the framework.

## 1. INTRODUCTION

The scope of sustainability can be varied based on the aim in a particular context. In engineering, sustainability is defined as "... the ability of a product to work continuously while ensuring the lowest environmental impacts and providing economic and social benefits to the stakeholders." (Ahmad et al., 2018). In design for sustainability, it is emphasised with the "combination of the dimensions of social, ecological and economic aspects. Complete and utilitarian sustainability exists when the maximum concordance of the benefits of all stakeholders is achieved." (Gräßler and Hesse, 2022). Mitake et al. (2020) noted, "Since sustainability is a dynamic system property and not predictive, it is guided by targets/visions, instead of traditional goal-based optimisation approaches." (Gaziulusoy and Brezet, 2015). Therefore, we interpret sustainability as a process that can be structured with the activities or functions of people involved to improve the environmental, social and economic aspects (ESEA). The sustainability functions are the inputs of the sustainability processes, which can contribute to ESEA. The outcome of the sustainability process can be the continuity in sustainable development (SD) (Ozili, 2022).

In product lifecycle processes, the functions can be specified with the use of materials (e.g., reducing cost and the number of parts of a product) (Ahmad et al., 2018; Gräßler and Hesse, 2022), designing lightweight products (Johnston-Lynch et al., 2024) or creating economic value via increasing the cost in reduction (Chen et al., 2020). However, these specified functions can limit the impact of sustainability functions only on ESEA. The sustainability functions can support the balanced development of ESEA in different contexts involving dignity, well-being, human rights and many more (May and Daly, 2020). Human dignity is contingent on having equal worth via social inclusion (May and Daly, 2019; Hojman and Miranda, 2018). Le Dantec and Edwards (2008) discuss using appropriate technology to increase the inclusion and dignity of homeless people in societies. Ivković et al. (2014) define the well-being of a society as "a benefit for all people in the society, implying accomplishment of adequate economic development (the objective dimension of well-being) and the resulting positive perception of people towards the proper stage in the society, i.e. the quality of life (the subjective dimension of well-being).". Human well-being is measured in societal contexts by evaluating income, longevity, education (Neumayer, 2004), housing, environmental quality, knowledge and skills, safety, social connections, etc. (OECD, 2024). We understand that the evaluation of dignity and well-being is mainly based on societal evaluation (objective approach by measuring housing, income, etc.) or subjective with well-framed contexts (e.g., the impact of emotions of workers on their designs). However, this understanding limits the assessment of well-being and dignity with specific measures for various contexts. There is a need to clarify the effects of sustainability functions in product lifecycle processes on dignity and well-being to extend the limitations in measures of well-being and dignity. Therefore, we propose a formula of dignified well-being (DWB) to evaluate it in different contexts in lifecycle processes.

Inclusivity of people in a product lifecycle process can empower them with their inclusion in functions (Yaldiz and Chakrabarti, 2024a). Yaldiz and Chakrabarti (2024a) discuss the empowering impact of inclusivity via the evaluation of the dependency level of stakeholders on each other. They highlight the exchange of resources to fulfil needs for the independence of people from a particular need. This reveals the nature of a product lifecycle contingent on the exchange networks among the stakeholders. However, it is necessary to understand the impact of inclusivity of people in exchange networks to accelerate sustainable development, dignity, and well-being. As mentioned in inclusive manufacturing (IM) is "... a new paradigm concept, where all parts of the lifecycle of a manufactured product is made accessible to people from all strata of the society, so as to, accelerate sustainable development and dignified well-being for all. Inclusive manufacturing aims at empowering people, especially those who are spatially, temporally, physically, economically and culturally disadvantaged, to actively participate in the conception, creation, distribution, transaction, use, and retirement of products and systems." (Roy et al., 2018). In our previous work (Yaldiz and Chakrabarti, 2024a), we propose metrics for inclusivity of people, and their empowerment in an inclusive product lifecycle, as espoused in the aim of IM. The work reported in this paper follows up on the above work, with the objective of understanding the influence of inclusivity and resulting empowerment on SD and DWB.

### 1.1 Objective and Research Questions

Building on our previous work, we aim to explore the impacts of inclusivity on lifecycle processes. We argue that the inclusion of people in the lifecycle processes based on the proposed inclusivity metrics will enhance their empowerment. The objective of the research reported in this paper is to understand the impact of enhanced empowerment (E) via the inclusivity of people on SD and DWB. Understanding the correlation between E&SD and E&DWB are crucial to fostering balanced societal development through inclusive product lifecycle processes.

We propose approaches to assess the impact of empowerment, which is enhanced by the inclusivity of people on SD and DWG. We define Sustainability Value (SV) as the potential contribution of a function to SD. We utilise the function analysis of ten case studies for the assessment of inclusivity and empowerment (Yaldiz and Chakrabarti, 2024a) to measure the SV of functions and DWB of empowered people. We apply the Pearson Coefficient Correlation test to assess the linear correlation between variables (E&SV and E&DWB). Our approach, tested for intercoder reliability among three coders, demonstrates trustworthiness in evaluating the influence of empowered people (or stakeholders) on SD and DWB.
In brief, this research replies to the following research questions (RQ):
RQ1 – How can the inclusivity of people in a product lifecycle process impact sustainable development and their dignified well-being?
RQ2 – How can we test our approach to assess its generalisability?
Section 2 reviews the literature; Section 3 outlines the methodology for addressing RQ1&Q2; Section 4 summarizes the framework application; Section 5 introduces metrics for SV and DWB; Section 6 discusses the correlation between variables; Section 7 validates the approach with results; Section 8 presents findings for RQ1&2, and the final section concludes with future work.

## 2. BACKGROUND

### 2.1 Sustainability

Sustainability, aspects of sustainability and sustainable development have various definitions, as in Supplementary_1. This can create an ambiguity in the operationalisation of sustainable development (Jabareen, 2008). In sustainable product development, the indicators for the competency of sustainability can be various when integrating sustainability into a product. Table 1 shows some of the characteristics which can be considered to increase the sustainability impact of products. However, the approach to identifying the characteristics of a product for sustainability can cause unequal improvement in all aspects of sustainability by not integrating factors of environmental, social, and economic into the product lifecycle process.

Table 1: Sustainability Characteristics for Product Development

| **Characteristics** | **References** (Supplementary_2) |
|---|---|
| Use of material (low weight, nontoxic, renewable) | Nishant et al. (2016), Waage (2007), Wlatz and Hallstedt (2018) |
| Enforce human rights | Waage (2007), Waage et al. (2005) |
| Integrating stakeholders in decision-making processes | Waage (2007), Waage et al. (2005) |
| Increasing awareness of stakeholders on rights to have services (education, health) | Waage (2007) |
| Innovation strategy, awareness of sustainability in the early phase of the design process | Ola et al. (2015), Mallalieu et al. (2024), Xu et al. (2023) |
| Business model and consideration of all lifecycle phases | Waage et al. (2005), Mallalieu et al. (2024), Nishant et al. (2016), Buchert et al (2017) |

Hallstedt (2017) proposes a framework to identify the sustainability criteria of a product by focusing on lifecycle phases and socio-ecological principles. Demssie et al. (2019) emphasise the context-specific nature of sustainability, requiring different competencies (communication and information acquiring, resource utilisation, social justice and inclusion, etc.) in various contexts. However, there is vagueness in how to measure the competencies to address the requirements of SD. Johnston-Lynch et al. (2024) highlight that "sustainability has roots in systems thinking (Osorio et al., 2009) and requires "*holism and system-wide approach[es] [...] in order to deal with complexity"* (Sala et al., 2013) of the almost unfathomable volume of simultaneous, interlinked activities on Earth affecting its behaviour.". The sustainability of a product is the merit of the system, which involves the lifecycle process with the interaction among stakeholders, societies, and their functions. Thus, it is necessary to restructure the assessment of sustainability by evaluating the impact of functions on ESEA.

**2.2 Inclusivity of People for Sustainability**

Inclusion and exclusion are interrelated notions with similarities in indicators to evaluate the boundaries to identify social inclusion in a particular context (Hansen, 2012; Rawal, 2008). Ozili (2020) points out the indicators of social inclusion as 'gender inequality, equity in the use of public resources, building human resources, environmental sustainability and social technology, etc.'. Silver (1994) and Rawal (2008) evaluate exclusion contingent on the nature of society, which shows the socio-economic structure, and cultural contexts. In the context of a product lifecycle, Yaldiz and Chakrabarti (2024a) discuss the interrelation of inclusion and exclusion by defining the inclusivity of people based on its metrics:

- The diversity of stakeholder groups.
- The number of lifecycle phases in which they are involved.
- The hierarchy among these groups.
- The number of interactions among the stakeholder groups.

In inclusive design, inclusivity is assed with the increase in the diversity of users with various kinds of abilities (Persad et al., 2007). Inclusion is considered as "... the opposite of exclusion" (Patrick and Hollenbeck, 2021).

Based on the list of definitions (Supplementary_3), the word clouds are created to visualise the conceptual evaluation of inclusion, exclusion, inclusive, inclusivity and inclusive design.

Figure 1 demonstrates the concepts to evaluate inclusion. Inclusion is highly interwoven with *social, individuals, practices, exclusion and financial*. Figure 2 shows the factors of exclusion with the interconnection of *inclusion in a societal context*. Figure 3 shows the factors for inclusive and inclusivity which are mainly considered in the domains of business and education. Figure 4 demonstrates the factors of inclusive design, and the focus is mostly on products, users and diversity. We interpret that the inclusion of people can contribute to social and economic aspects of sustainability. However, the impact of inclusivity on environmental sustainability is not reflected. It is required to assess equal enhancement on ESEA via social inclusion in different contexts to support SD and DWB of people.

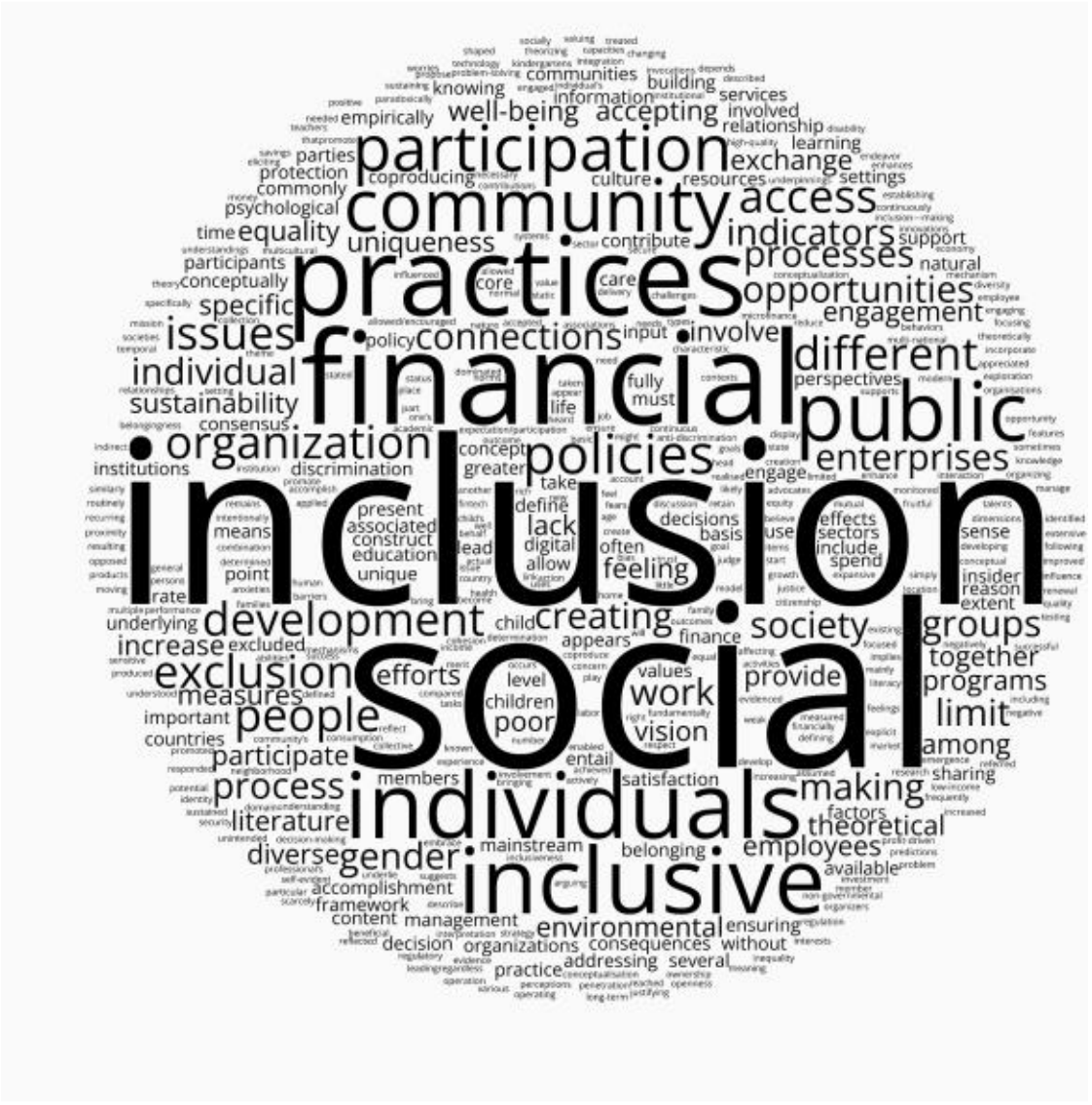

Figure 1: Inclusion

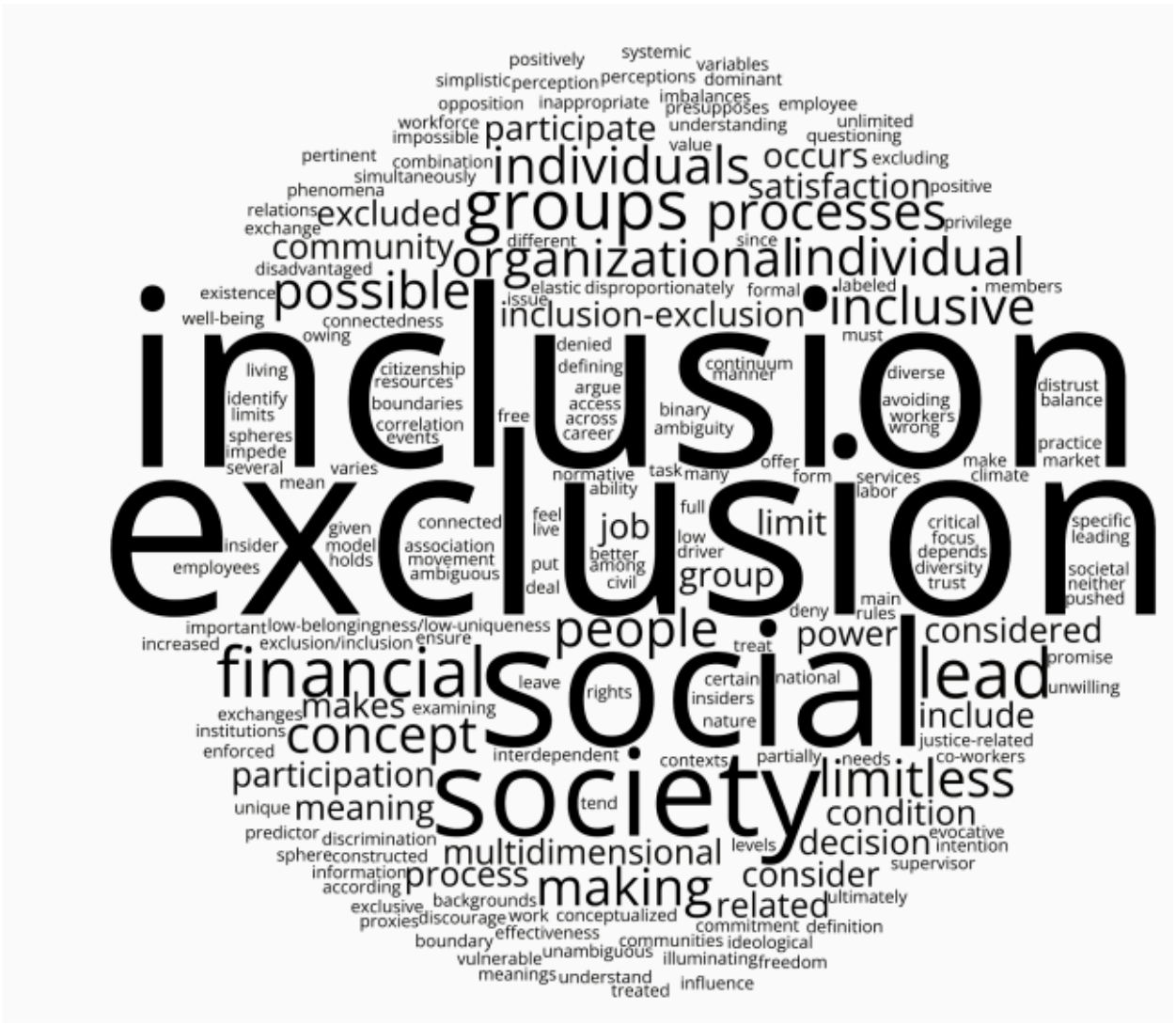

Figure 2: Exclusion

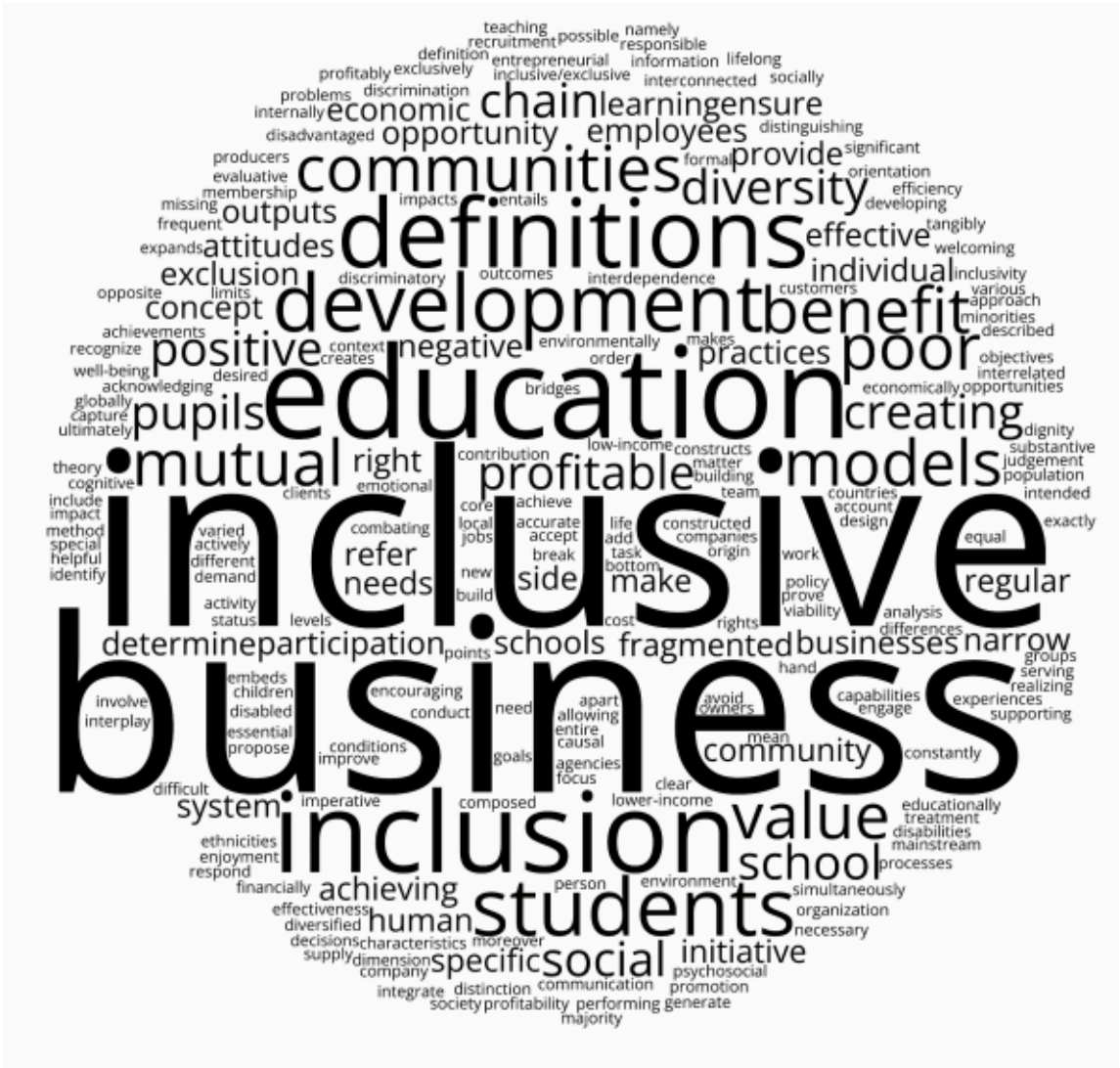

Figure 3: Inclusive and Inclusivity

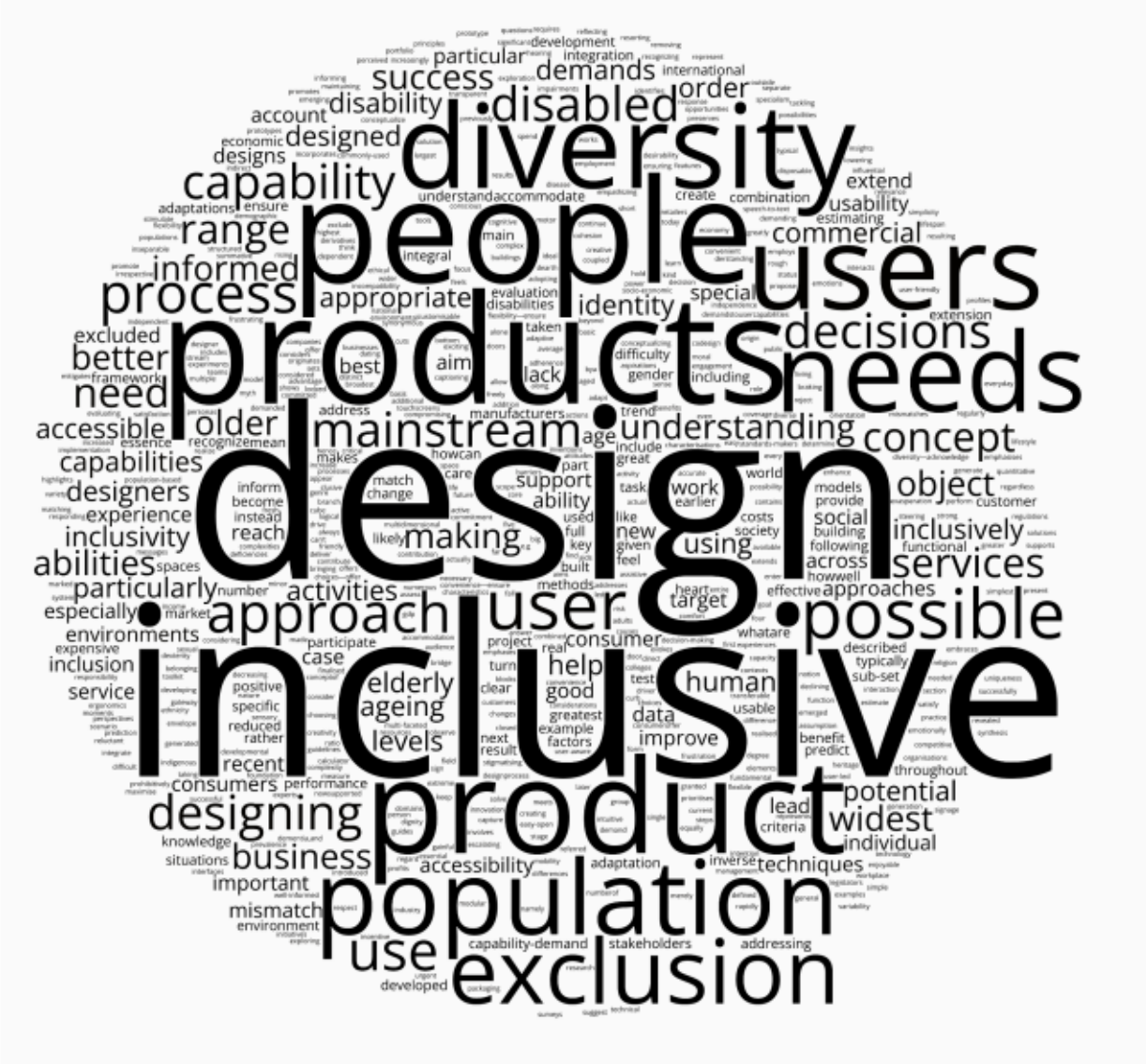

Figure 4: Inclusive design

**2.3 Empowerment in Exchange Networks for Sustainability**

Based on the definitions of empowerment (Supplementary_4), the word cloud is generated to understand the concepts that influence the empowerment of people. Considering the highlighted concepts in Figure 5, empowerment can be defined as a human development process that activates the power of people to take control of their decisions.

Empowerment is the process of sharing power with people with less power (Conger and Kanungo, 1988; Füller et al., 2009). Parson (1967) defines power as "... the capacity of a societal system to achieve collective goal" (Avelino, 2021). However, Zimmerman (1995) emphasised the context-dependent nature of empowering processes by pointing out that "Actual power or control is not necessary for empowerment because in some context and for some populations real control or power may not be the desired goal." In that case, it is necessary to understand the requirements for empowering people to obtain inclusive outcomes. These requirements can be fulfilled by generating social exchange networks, which are contingent on sharing resources (Cook et al., 2006), as in inclusive product lifecycle processes. However, the determination of the requirement of empowerment is not only connected with the mobilisation of resources. Yaldiz and Chakrabarti(2024a) define empowerment by identifying its metrics: impact of functions, the number of included stakeholders in a particular function and the number of means (tools) to complete a specific function in an inclusive product lifecycle.

Larsson and Jormfeldt (2017) emphasise that "Empowered persons experience themselves as capable persons; for example, they feel they have the power to do things in a good, effective, and constructive way (Kuokkanen & Leino-Kilpi, 2000), and this includes areas such as health and well-being." Sianipar et al. (2013) point out that "... empowerment will ensure the sustainability of given sustainable development.". However, the impact of the inclusivity of people in an empowering process (e.g. an inclusive product lifecycle) on SD and DWB needs to be elaborated for operationalisation.

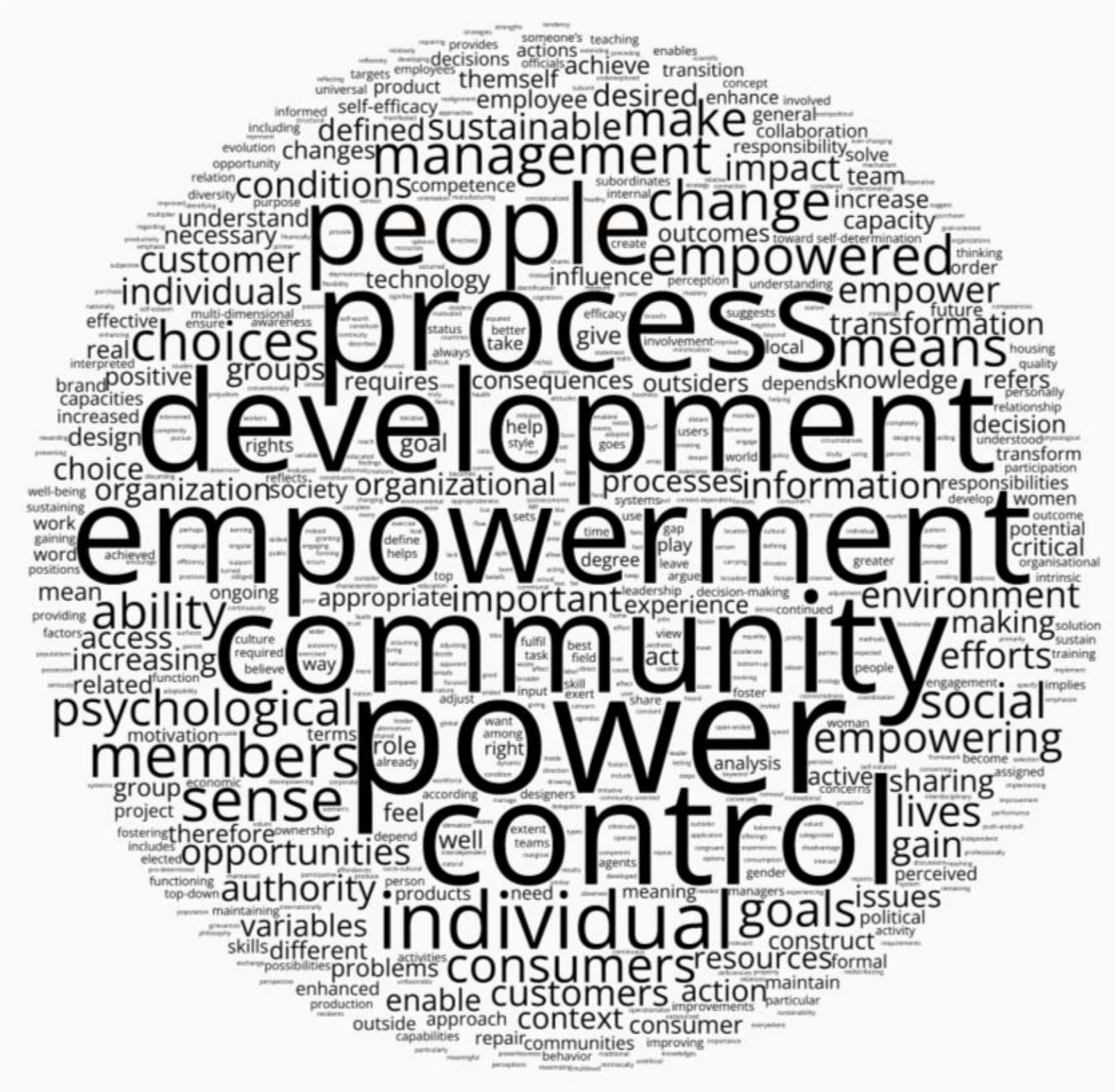

Figure 5: Empowerment

**2.4 Dignity and Well-Being**

Karlson (2004) defines dignity as "... taking active responsibility for one's own life project, may then look very differently to different individuals.". Lombe (2007) emphasises that "Inclusion is the realisation that everyone has essential dignity and everyone has something to contribute." Dignity is essential to contribute to environmental, social and economic aspects of sustainability (May and Daly, 2020). The relationship between environmental sustainability and dignity can be exemplified by providing services to societies to increase the accessibility to clean water or by designing environments by understanding nature and self-relation (May and Daly, 2020). The interrelation between social sustainability and dignity can be expressed as "All human beings are born free and equal in dignity and rights" in the Universal Declaration of Human Rights 1948 (UDHR, 1948). The relationship between the economic aspect of sustainability and dignity is explained by using income to fulfil personal needs and, at the end of the month, still having some amount for consuming interest (Goodland, 1995; Hicks, 1946; May and Daly, 2020). We interpret dignity as the power of self to take responsibility for decisions and contribute to SD via inclusion in societies.

Well-being is evaluated with objective (the well-being of a society is related to the conditions and results of the functions) and subjective (what people think and feel or perceive about the impacts of conditions) approaches (Ivković et al., 2014; Castellacci and Tveito, 2018). Goodland (1995) emphasises the priority of development as the enhancement of human well-being and the reduction in poverty, inequality, etc. Chiu (2003) related social sustainability with well-being by improving social inclusion and equitable division of resources. Kjell (2011) points out that "well-being research can clarify the aim of sustainability.". Sæteren and Nåden (2021) evaluate dignity as a resource of well-being. It can be interpreted that dignity and well-being are interconnected concepts which can contribute to SD.

## 3. METHODOLOGY

Based on the discussion in the literature (Sections 1&2), the interconnection among the notions (sustainability, well-being, dignity, inclusivity and empowerment) can be recognised. However, a context must be defined to clarify the relation among the notions. Otherwise, the definitions can be accepted as vague to operationalise, and the evaluation of sustainability, dignity, and well-being of people cannot be related to inclusivity and empowerment.

Our approach depends on operationalising the definition of IM (Section 1). Thus, we hypothesise that the inclusivity of people can empower them to enhance SD and DWB. We follow the methodology explained in subsections 3.1&3.2.

### 3.1 Methodology for addressing RQ1

RQ1 has two parts to be answered. First, it requires understanding how to assess sustainability in a product lifecycle. Second, it requires an evaluation of the impact of inclusivity on dignity and well-being. One of the aims of inclusion is empowering people, as presented in Yaldiz and Chakrabarti, 2024a. Thus, we focus on the correlation between empowerment and sustainability; empowerment and DWB.

For the first part, we examine the definitions of ESEA, sustainability and SD (Supplementary_1) to create a connection between functions and definitions. Based on the definitions of ESEA, we match the functions and aspects of sustainability. While matching aspects of sustainability and functions, we consider the conformity of the content of case studies and the definitions. This can also be defined as matching the context-content of the functions and the definition of the aspects of sustainability. With this approach, we determine the SV of each function. This analysis ensures the contribution of each stakeholder to sustainability based on the functions in which they are involved. We apply the Pearson Correlation Coefficient test to find the correlation between empowerment and SV. The result reveals that empowered people can contribute to SD by increasing the SV of functions.

For the second part, we define the formulas for dignity and well-being based on their definitions to evaluate the impact of the inclusivity of people on dignity and well-being. We calculated the ratio of dignity and well-being for each stakeholder based on the number of functions they were involved in. Then, we proposed a formula for DWB based on the interconnection between dignity and well-being. The result of the Pearson Correlation Coefficient test reveals the positive correlation between empowerment and DWB.

### 3.2 Methodology for addressing RQ2

We organised an intercoder reliability (ICR) test with three coders to validate our approach. The purposes of the ICR test are to check the agreement 1) in the match of functions and the definitions of the aspects of sustainability, 2) in the identification of functions from the same content, 3) in the identification of needs and resources of each stakeholder to be involved in a particular function. We provided a document to explain the steps of the application with the information as follows: 1) the content (highlighted paragraphs on the case study) to identify functions and stakeholders for each function, 2) the Matrix of Needs&Satisfiers (Max-Neef, 1992) to the determination of needs of stakeholders to involve in a function, and information of possible resources(experience, capability and position/duty) 3) a list of definitions of environmental, social and economic sustainability. We compared the answers of coders based on the determined criteria (Section 7) and used percentage agreement to calculate intercoder reliability. Based on the results, we present the high reliability of our approach.

## 4. FRAMEWORK OF EMPOWERMENT AND INCLUSIVITY ASSESSMENT

The framework (Yaldiz and Chakrabarti 2024a&b) has three steps to evaluate the correlation between inclusivity and empowerment in an inclusive product lifecycle process. The fourth and fifth steps of the framework are proposed in this study to measure SV and DWB. The first step is identifying stakeholders to determine in which phase of the lifecycle they are involved. This is followed by identifying functions and the included stakeholders in the function. This creates the exchange relation among the stakeholders due to the reason (fulfilment of needs by sharing resources) for involvement in the functions. The second step is to measure the empowering impact of each function on stakeholders based on their interactions with other stakeholders. The last step is to measure the amount of power of stakeholders in connection with their dependencies on each other and the potential of their involvement to extend the boundaries of the context. The consideration of extension (the tool used to influence people to be included in lifecycle processes and the frequency or the number of functions completed per stakeholder) can increase the inclusivity of people to influence their power.

With the aim of this research, we reveal the impact of functions on the sustainability, dignity, and well-being of people. Therefore, it is important to clarify acceptance in identifying types of functions and limitations to identify them from a case study or a particular content. Figure 6 demonstrates the type of functions in an exchange network among stakeholders A, B, and C. F1 and F2 are the group functions or function pairs that create a dependency of stakeholders on each other based on their needs and resources. The other group functions are F4-5, F8-10, F13-15, F18-20, F23-25, F28-30. Due to the mobilisation of the resources of stakeholders by involving in these function groups, they can increase their power by empowering themselves. The individual functions are also the way to empowerment; however, they do not impact the increase in power because individual functions are not to create an exchange network with other stakeholders. Thus, individual functions cannot influence the inclusion of new stakeholders in a product lifecycle process. The Activity Theory influences the acceptance of limiting the number of functions. Engeström (1987) emphasises that "activity theory demands analysing at least two interacting activity systems." We adopt this to exchange networks, which are dependent on group functions as well as individual functions. In exchange networks, we detail the interactions a maximum of two times to complete a goal. As in Figure 6, F1 is the goal of included people and requires a coalition among A-B-C to achieve it. F30 is the achievement of the goal of coalition A-B-C. The link between B-C and A-C shows the interactions that occurred two times. The number of individual functions is connected with different group functions, and the detailing of the interactions is limited to a maximum of two functions.

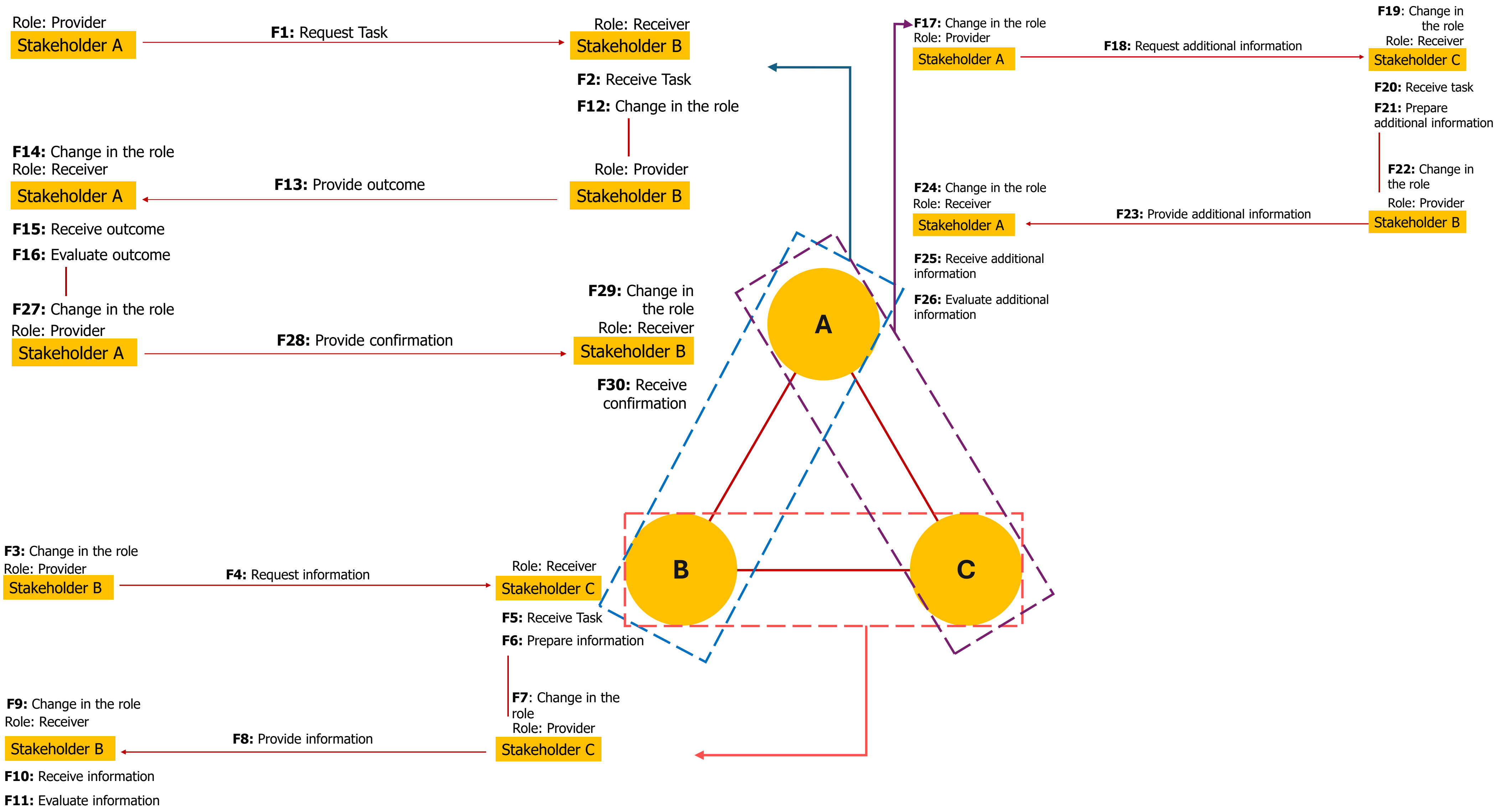

Figure 6: An example of the exchange network with the types of functions

## 5. METRICS OF SV AND DWB

### 5.1 Sustainability Value

As discussed in Sections 1 and 2.1, sustainability is a context-based process with the influence of various factors based on ESEA. Intending to operationalise the IM definition, we aim to measure the SV of each function in an inclusive product lifecycle process. Thus, we can evaluate the overall value of the process to contribute to SD. Figure 7 demonstrates the calculation of the SV of functions. Based on the function explanation in a specific context (e.g., an activity in a particular location during the planning phase of the product lifecycle process) with clarity in content (e.g., the content of a case study), the influence of the function on sustainability aspects can be quantified. If a function contributes to environmental, social and economic sustainability, the quantity of impact is 1 for each aspect. The calculation of the SV is the multiplication of each aspect. If the content can contribute to all aspects of sustainability, the SV of the function is accepted as 1. Otherwise, it is 0, even if it impacts one or two aspects of sustainability.

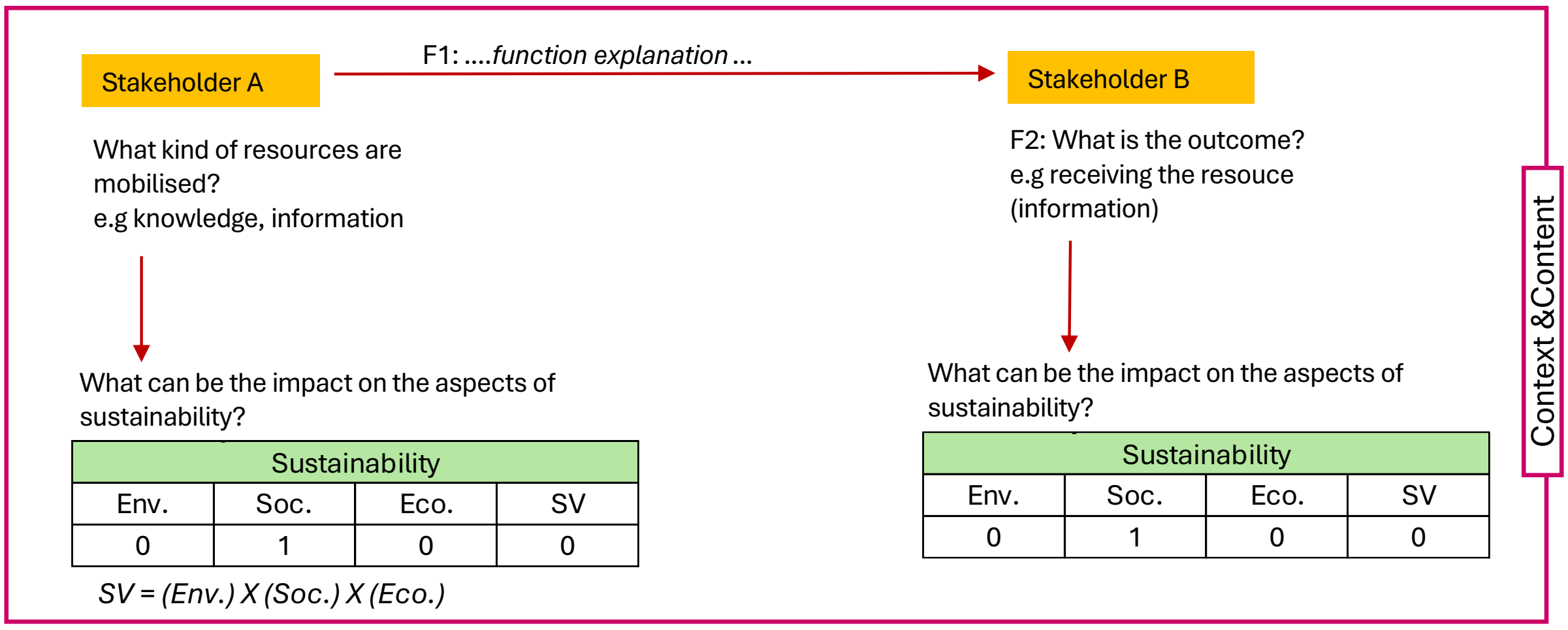

Figure 7: Measuring SV of functions

### 5.2 Dignified Well-Being

Based on the discussion in Section 2.4, we interpret influencing factors of dignity as social inclusion and being responsible for inclusion and decision-making. In the lifecycle context, it is necessary to consider the factors that influence people to be included in functions. The factors can be their roles as providers or receivers. The motivation of having these roles is to fulfilment of needs by sharing resources. Thus, we accept that having a role to be included in a function and becoming independent via fulfilling needs can impact the dignity of stakeholders. We measure the increase in the dignity of each stakeholder by counting the number of group functions in which a stakeholder is included. Dignity is a resource of well-being. Thus, we formulated well-being connected with the approach that we followed to measure dignity. We consider group functions to be within the scope and accept well-being as a multiplication of provided resources and the number of fulfilled needs by the provider. We limit this formulation with the acceptance of high satisfaction from fulfilling the needs due to the progress in the lifecycle process. Still, based on the approaches of subjective well-being, the circumstances might be insufficient for personal satisfaction. However, our acceptance considers stakeholders as a group, not individuals (Yaldiz and Chakrabarti, 2024a). Therefore, our approach is to interrelate well-being and dignity by supporting the independence of stakeholder groups to increase their power. As in the IM definition, we concede DWB and formulate it with the multiplication of metrics of dignity and well-being, as in Equation (1).

$$\text{Dignified Well} - \text{Being } (\text{DWB})_{\text{stakeholder}} = (\text{number of function groups}) \times (\text{number of provided resources}) \times (\text{number fulfilled needs by provider}) \quad \textit{Eq. (1)}$$

## 6. CASE STUDY

The case studies (Supplementary_5) were analysed based on the framework for inclusivity and empowerment assessment (Yaldiz and Chakrabarti, 2024a).

To evaluate the impact of empowering people (via their inclusion based on inclusivity metrics) on inclusive outcomes, we measured the SV of each function and DWB for each stakeholder. The results of the correlation (based on the Pearson Correlation Coefficient Test with a significance level of $p<0.05$) between empowerment, aspects of sustainability, SV and DWB are presented in Table 2. The number of stakeholders (#S) and the total number of functions (#F) are different in case studies (CS) based on the explanations in related references. For the variables, environmental (ENV.), social (SOC.), economic (ECO.) and SV, the sample sizes are the total number of functions (#F). #F is the summation of the functions completed in all lifecycle phases in a particular CS. #LC is the number of lifecycle phases based on the explanation in case studies. The correlation between empowerment(E) and inclusivity(I) is presented in Table 2 and formulation of them is discussed in our previous work (Yaldiz and Chakrabarti, 2024a). The sample size is varied for DWB due to the different number of times stakeholders are involved in various functions. Table 3 shows the total number of 'provide' functions (D: the metric of Dignity), the total number of fulfilled needs(N) and mobilised resources (R) and the total amount of DWG for all stakeholders in each case study. The ENV, SOC, ECO and SV show the number of functions that can impact related variables.

Except for CS2-7, the correlation of ENV.&E and SV.&E have the same ratio. Even though the ratio of SOC.&E and ECO.&SV are higher than ENV.&E, the correlation of SV.&E is the same with the correlation of ENV.&E. This highlights the importance of the influence of empowered people on environmental sustainability and reveals the strong connection between environmental sustainability and SV. In CS2-7, the correlation of ENV.&E, SV.&E are closer to each other. The strong correlation between SOC&E can be explained by the influence of function on social sustainability due to the empowering aim of social sustainability (Supplementary_1). Economic sustainability and empowerment have a strong correlation.

The results of CS3-6-7 are closer to the ideal situation by having higher correlations. The overall understanding from the analysis of the inclusivity of people (considering the relationship between the metrics of empowerment and inclusivity) can impact SD based on the strong correlation between SV&E and the DWG of people. In addition, the results provide information about what needs to be sustained. For example, in Table 3, CS_1 impacts SD with the amount of SV (32). The amounts of SOC (179) and ECO (56) listed in Table 3 CS_1 can be elaborated based on the related functions to address what needs to be improved to contribute to another aspect of sustainability. In essence, if one function has an impact on only the social aspect, the function can be analysed in detail to address what needs to be improved to contribute to both economic and environmental sustainability. Thus, the SV of the lifecycle process can be increased, which can influence SD.

Table 2: Comparison of CS

| CS | #S | #F | #LC | ENV.&E | | SOC.&E | | ECO.&E | | SV&E | | DWB&E | | E&I | |
|---|---|---|---|---|---|---|---|---|---|---|---|---|---|---|---|
| | | | | r | p<0.05 | r | p<0.05 | r | p<0.05 | r | p<0.05 | r | p<0.05 | r | p <0.1 |
| 1 | 11 | 175 | 3 | 0.74 | 0.01 | 0.98 | 0.00 | 0.94 | 0.00 | 0.74 | 0.01 | 0.99 | 0.00 | 0.89 | 0.00 |
| 2 | 9 | 383 | 4 | 0.68 | 0.04 | 0.94 | 0.00 | 0.87 | 0.00 | 0.69 | 0.04 | 0.76 | 0.02 | 0.66 | 0.05 |
| 3 | 8 | 205 | 4 | 0.96 | 0.00 | 1.00 | 0.00 | 0.96 | 0.00 | 0.96 | 0.00 | 0.95 | 0.00 | 0.96 | 0.01 |
| 4 | 8 | 300 | 4 | 0.92 | 0.00 | 0.98 | 0.00 | 0.95 | 0.00 | 0.92 | 0.00 | 0.98 | 0.00 | 0.66 | 0.02 |
| 5 | 7 | 168 | 2 | 0.87 | 0.01 | 0.99 | 0.00 | 0.82 | 0.02 | 0.87 | 0.01 | 0.93 | 0.00 | 0.77 | 0.04 |
| 6 | 4 | 177 | 3 | 0.98 | 0.01 | 0.98 | 0.02 | 0.98 | 0.01 | 0.98 | 0.01 | 1.00 | 0.00 | 0.91 | 0.06 |
| 7 | 8 | 309 | 4 | 0.93 | 0.00 | 0.99 | 0.00 | 0.96 | 0.00 | 0.90 | 0.00 | 0.94 | 0.00 | 0.93 | 0.00 |
| 8 | 11 | 304 | 4 | 0.87 | 0.00 | 0.98 | 0.00 | 0.91 | 0.00 | 0.87 | 0.00 | 0.79 | 0.00 | 0.78 | 0.00 |
| 9 | 11 | 315 | 4 | 0.87 | 0.00 | 0.95 | 0.00 | 0.87 | 0.00 | 0.87 | 0.00 | 0.95 | 0.00 | 0.74 | 0.01 |
| 10 | 15 | 340 | 3 | 0.88 | 0.00 | 0.98 | 0.00 | 0.89 | 0.00 | 0.88 | 0.00 | 0.91 | 0.00 | 0.50 | 0.06 |

Table 3: The number of metrics for variables

| CS | D | N | R | DWB | ENV | SOC | ECO | SV |
|---|---|---|---|---|---|---|---|---|
| 1 | 84 | 46 | 23 | 1050 | 32 | 179 | 56 | 32 |
| 2 | 194 | 130 | 69 | 26346 | 71 | 385 | 92 | 69 |
| 3 | 101 | 60 | 34 | 1646 | 42 | 205 | 65 | 42 |
| 4 | 152 | 86 | 43 | 9198 | 74 | 300 | 116 | 74 |
| 5 | 78 | 48 | 24 | 2246 | 29 | 168 | 44 | 28 |
| 6 | 78 | 48 | 24 | 2008 | 21 | 177 | 21 | 21 |
| 7 | 151 | 88 | 44 | 5500 | 80 | 309 | 73 | 61 |
| 8 | 152 | 100 | 50 | 8916 | 72 | 304 | 161 | 72 |
| 9 | 145 | 96 | 48 | 3618 | 84 | 316 | 104 | 84 |
| 10 | 172 | 102 | 55 | 8526 | 35 | 347 | 58 | 35 |

## 7. VALIDATION

We followed the steps below to validate our approach.

1. We chose the sections of the framework that use subjective approaches. These are: assessment of SV due to the qualitative method as matching the function and the definition of sustainability aspects; function identification with the determination of needs and resources. We identified criteria for agreement analysis for each subjective approach.
   The criteria to compare function identification are as follows:
   - Use of Model: This evaluates the use of provided function generalisation methods. Request Task/information& Receive Task/information, and so on (Section 4). This evaluation is critical to understanding the required improvement in the identification of functions for the generalisation of the framework.
   - Detailing Interactions: This evaluates the detailing interactions a max of two times (Section 4).
   - Roles: This evaluates the consideration of the change functions, e.g. change in the roles for the following of the roles of stakeholders (Provider to Receiver and vice versa). Understanding providers and receivers is required to identify the mobilised resources and the fulfilment of the needs.
   - Identification of Stakeholders: This evaluates the identification of stakeholders as groups between coders based on the content or statements from the case study.
   - Individual and Group Functions: This evaluates the identification of group and individual functions. It is essential to find function pairs for the following steps of the framework (e.g. power relations). Understanding the nature of exchange networks by identifying function pairs in lifecycle processes is critical to creating new exchange networks.

   We defined criteria as the group of needs (e.g. participation and doing) and a group of resources (position, experience, capability) for each stakeholder group to compare the analysis of needs and resources. We followed the method mentioned in Section 5.1 to compare the SV.
2. We selected a case study from Supplementary_5. Due to the time limitation of coders, we highlighted some paragraphs related only to the Planning Phase.
3. Coders were chosen based on their current positions; they are PhD students in engineering design with architecture and mechanical engineering backgrounds.
4. We prepared information (Supplementary_6) for coders to follow while applying the chosen sections of the framework.
5. We requested coders to read the notes and follow the steps accordingly to find the SV of the identified functions, needs and resources of stakeholders.
6. We did not have any interaction with coders during the experiment. At the end of the coding process, coders submitted their analysis as a hard copy.
7. We separately compared our results (Coder 1) with all coders (2, 3 and 4).
8. Based on the criteria and approaches presented in previous sections, we compare the results of Coder_1 with those of all coders.

### 7.1 Results

The results are compared (Tables 4, 5 and 6) based on the analysis of percentage agreement due to the lower amount of data. The small dataset and imbalanced distribution are inadequate for analysing the result based on Cohen's Kappa. Boughorbel et al. (2017) proposed the Matthews Correlation Coefficient metric for the imbalanced data; however, the sample is not as big as suggested (N should be around 100). As listed in Tables 4,5 and 6, the results (listed under the Before) are not as high as expected. Therefore, we had a discussion with coders to understand the different perspectives for improving our approach. Based on the elaboration of the coders' analysis, we found a need for improvement in our explanation (Supplementary_6) provided for the ICR test. Thus, we detailed our explanation (Supplementary_7) and requested coders to redo the analysis based on the improved notes.

Table 4 lists the comparison of percentage agreement among coders 1, 2, 3 and 4. The variety in results of the comparison (as listed in the Before columns in Table 4) shows the inconsistency in understanding the application of the approach. We detailed and clarified the approach for evaluating SV with more examples. As listed in Table 4, comparing the results of the ICR test based on the improved explanation (demonstrated under the After columns) supports the consistency in the application. It increases the reliability of the SV evaluation.

Table 4: Comparison of ICR results for variables ENV, SOC, ECO and SV

| Coders | % Agreement Between Coders | | | | | | | |
|---|---|---|---|---|---|---|---|---|
| | SUSTAINABILITY | | | | | | | |
| | ENV | | SOC | | ECO | | SV | |
| | Before | After | Before | After | Before | After | Before | After |
| C_1&2 | 0.81 | 0.98 | 1 | 1 | 0.67 | 0.93 | 0.81 | 0.98 |
| C_1&3 | 0.79 | 0.93 | 0.7 | 1 | 0.63 | 0.84 | 0.79 | 0.93 |
| C_1&4 | 0.42 | 0.91 | 1 | 1 | 0.51 | 0.88 | 0.77 | 0.95 |
| C_2&3 | 0.79 | 0.91 | 0.7 | 1 | 0.72 | 0.81 | 0.95 | 0.98 |
| C_2&4 | 0.37 | 0.88 | 1 | 1 | 0.42 | 0.81 | 0.86 | 0.93 |
| C_3&4 | 0.53 | 0.88 | 0.7 | 1 | 0.51 | 0.81 | 0.84 | 0.88 |

Based on the results listed in the columns of Before in Table 5, improvements were required in clarification in identifying the types of functions (provide, receive, etc.), application of the detailing interactions, clarification on the importance of determining the roles of stakeholders, and identification of individual or group functions. With the improvements in the content, the results have higher reliability of the approach, as listed in the After columns in Table 5.

Table 5: Comparison of ICR results for the function identification

| Coders | % Agreement Between Coders | | | | | | | | | |
|---|---|---|---|---|---|---|---|---|---|---|
| | FUNCTION IDENTIFICATION | | | | | | | | | |
| | Use of Model | | Detailing Interactions | | Roles | | Stakeholders | | Individual & Group Functions | |
| | Before | After | Before | After | Before | After | Before | After | Before | After |
| C_1&2 | 0.55 | 1 | 0 | 0.65 | 0.55 | 1 | 0.55 | 1 | 0.55 | 1 |
| C_1&3 | 0.74 | 0.88 | 0.37 | 0.76 | 0.56 | 0.8 | 0.81 | 0.9 | 0.81 | 1 |
| C_1&4 | 0.56 | 0.82 | 0.11 | 0.63 | 0.3 | 0.84 | 0.52 | 0.89 | 0.7 | 0.89 |

The percentage of GoN and GoR was lower because of unclarity in matching function explanation and the need based on the 'Matrix of Needs and Satisfiers' (proposed by Max-Neef, 1992). By providing more examples about the determination of the group of needs, the results of the ICR test are improved, as listed in Table 6.

Table 6: Comparison of ICR results for the analysis of N&R

| Coders | % Agreement Between Coders N&R Analysis | | | | | | | |
|---|---|---|---|---|---|---|---|---|
| | Group of Needs (GoN) | | Number of Need Groups (#NG) | | Group of Resources (GoR) | | Number of Resources (#R) | |
| | Before | After | Before | After | Before | After | Before | After |
| C_1&2 | 0.52 | 0.84 | 1 | 1 | 0.74 | 0.84 | 0.97 | 0.97 |
| C_1&3 | 0.37 | 0.85 | 0.59 | 0.85 | 0.39 | 0.81 | 0.44 | 0.81 |
| C_1&4 | 0.28 | 0.85 | 0.85 | 0.85 | 0.39 | 0.85 | 0.63 | 0.85 |

The overall understanding is that the high percentage agreement between the coders shows the generalisable potential of the framework and the approach to evaluate the SV of functions.

## 8. DISCUSSION

Sustainability is a process of achieving balanced improvements in environmental, social, and economic sustainability for SD. Evaluating sustainability as a feature of a product can cause imbalanced improvements in the ESEA and cannot be sufficient for improvements in SD. Therefore, we support balanced development in ESEA of sustainability with functions by defining SV in an inclusive product lifecycle to contribute to SD. This approach can refocus the scope of sustainability from product to humanity. Thus, the inclusivity of people is necessary for balanced improvements in all aspects of sustainability. It is essential to understand who can be included in which phase of a lifecycle process for the sustainability of the process and the opportunity to empower themselves. This reveals the interrelation between inclusivity and sustainability. In literature, inclusivity is evaluated with increasing diversity of people in a particular activity (e.g. inclusive design; increasing diversity of users); however, inclusivity of people can also be evaluated as the attribute of sustainability to empower people and improve their well-being and dignity. Well-being and dignity are connected concepts; improvement in one of them can benefit the other. In different contexts and circumstances, the well-being and dignity of people can be affected by various factors. In an inclusive product lifecycle context, the well-being and dignity of people can be improved with the fulfilment of their needs based on the function type (group functions) and with which roles (provider and receiver) in which they are involved.

At this point, to reply to RQ1&2, we determine the limitations. This research is limited to the application of the framework for the assessment of empowerment and inclusivity; evaluation of the SV of functions is limited to the definitions of sustainability and their match with function explanations; DWB of people is limited to objective assessment in a product lifecycle context and only for stakeholder groups.

Based on the above analysis, the following is found as the answer to RQ1: The inclusivity of people with the metrics of it as in the definition (Section 2.2) can empower people based on the definition of empowerment (Section 2.3) by creating a positive correlation with the acceptance of $r \geq 0.5$ and $p<0.1$ (as in Table 2), and that can influence improvement on:

- the SD, based on the range of the r-value of the Pearson Correlation Coefficient test of SV&E, which is between 0.69 and 0.98 with a significance level of $p<0.05$ and
- the DWB of people, with the range of r-value of the correlation of DWB&E, which is between 0.76 and 1.00, with a significance level of $p<0.05$.

This is interpreted as follows.

Inclusivity of people should relate to a purpose. Diverse stakeholder groups can be involved in various function groups to achieve the collective goal based on their needs and resources. In this way, stakeholder groups can empower themselves. Empowering people influences sustainability for the continuity of the process in dynamic conditions. An increase in the inclusivity of people in empowering functions can influence SD due to the increase in SV of the process. In this manner, the well-being and dignity of people can be improved.
Based on the above analysis, the following is found as the answer to RQ2: The approach can be generalisable in:

- the assessment of sustainability value as in the minimum agreement percentage 0.93 (Table 4) and
- the function identification with the agreement percentage above 0.65 (Table 5) and
- the need and resource analysis with the agreement percentage above 0.81 (Table 6).

Generalising the approach is important to make it beneficial to address improvements in empowerment, sustainability, well-being and dignity via the inclusivity of people in lifecycle processes. Thus, we applied the ICR test with three coders and compared the results based on the percentage agreement due to the lower amount of data. Based on the coders' feedback, we detailed our explanation of the application of the framework. With this, we assess the high reliability among four coders, including the authors. This reinforces our confidence in the qualitative analysis and supports the potential for the generalisability of the framework and the measurement of SV and DWB.

## 9. CONCLUSIONS AND FUTURE WORK

This research addresses the influence of empowerment on SD and DWB by operationalising the definitions of inclusive manufacturing. First, we analysed the definitions of sustainability and SD. We found that, in engineering, the dimensions of sustainability are more related to product attributes. Thus, we accept the dimensions as the aspects of sustainability: environmental, social, and economic. With this, the impacts of functions on dimensions of sustainability can be measured by matching the function and its definitions. Second, we measured the DWB of stakeholders by operationalising the definitions of well-being and dignity. Third, we measure the SV of functions and DWB of stakeholders included in product lifecycle processes based on ten case studies. Fourth, we correlated the relation among the concepts, and we found that empowered people based on inclusivity metrics can contribute to SD with the amount of SV, which can also improve their DWB. Evaluating the SV is fundamental to address what is required to be sustained for the improvements in SD. We organised an intercoder reliability test with three coders to test our qualitative approach. The test results ensure a high agreement between the coders greater than what would occur by random chance. The limitation of the research is the application of the framework for the assessment of inclusivity and empowerment. The influence of the increase in empowerment via the inclusivity of stakeholders (e.g. increase in interactions and generation of diverse exchange networks among stakeholders) on SV and DWB will be discussed in our following work.

**Supplementary_1**

| No | Definitions | | References |
|---|---|---|---|
| **1** | Environmental Sustainability | Goodland (1995) defined environmental sustainability as “a set of constraints on the four major activities regulating the scale of the human economic subsystems: the use of renewable and nonrenewable resources on the source side, and pollution and waste assimilation on the sink side” | Chen, X., Despeisse, M., & Johansson, B. (2020). Environmental sustainability of digitalization in manufacturing: A review. Sustainability, 12(24), 10298. |
| **2** | Environmental Sustainability | environmental sustainability can be defined as the development of “meeting the resource and services needs of current and future generations without compromising the health of the ecosystems that provide them” [18,21]. | Chen, X., Despeisse, M., & Johansson, B. (2020). Environmental sustainability of digitalization in manufacturing: A review. Sustainability, 12(24), 10298. |
| **3** | Environmental Sustainability | Environmental sustainability in manufacturing involves stabilizing the balance between manufacturing activities and their impact on the natural environment. | Chen, X., Despeisse, M., & Johansson, B. (2020). Environmental sustainability of digitalization in manufacturing: A review. Sustainability, 12(24), 10298. |
| **4** | Environmental Sustainability | According to indicator categorization from the National Institute of Standards and Technology (NIST) [25], environmental indicators are categorized by the impact of emissions, resource consumption, pollutions, and the natural habitat conservation [25]. | Chen, X., Despeisse, M., & Johansson, B. (2020). Environmental sustainability of digitalization in manufacturing: A review. Sustainability, 12(24), 10298. |
| **5** | Environmental Sustainability | Environmental sustainability refers to systemic conditions where neither on a planetary nor on a regional level do human activities disturb the natural cycles more than planetary resilience allows, and at the same time do not impoverish the natural capital that has to be shared with future generations. These two limitations, based on a prevalently physical character, will be aligned with a third limitation, based on ethics: the principle of equity states that in a sustainable framework, every person, including those from future generations, has the right to the same environmental space, that is, the right to access the same amount of natural resources. | Vezzoli, C., & Manzini, E. (2008). Design for environmental sustainability (p. 4). London: Springer. |
| **6** | Environmental Sustainability | enviromental sustainability as the maintenance of important environmental functions, and hence the maintenance of the capacity of the capital stock to provide those functions. | Ekins, P. (2011). Environmental sustainability: From environmental valuation to the sustainability gap. Progress in Physical Geography, 35(5), 629-651. |
| **7** | Environmental Sustainability | If the key consideration for environmental sustainability is the maintenance of the functions that are important for human welfare, then in the first instance, at least, it is on the ‘functions for people’ on which attention should be focused. ‘functions for people’ were fundamentally dependent on the Life-Support ‘functions of nature’. This suggests that principles of environmental sustainability will need to maintain important environmental functions. | Ekins, P. (2011). Environmental sustainability: From environmental valuation to the sustainability gap. Progress in Physical Geography, 35(5), 629-651. |
| **8** | Environmental Sustainability | as Chiu (2003, p. 26) has described it, identify ‘the social conditions necessary to support ecological sustainability’. | Vallance, S., Perkins, H. C., & Dixon, J. E. (2011). What is social sustainability? A clarification of concepts. Geoforum, 42(3), 342-348. |
| **9** | Environmental Sustainability | Inequality exerts adverse impact on environmental outcomes through several channels, including the household, community, national, and international channels. These channels however overlap with one another and can thus reinforce the impact of inequality. Other dimensions of inequality, in particular gender inequality, also impact environmental quality negatively. | Islam, S. N. (2015). Inequality and environmental sustainability. |
| **10** | Environmental Sustainability | income and wealth inequality can be harmful for environmental sustainability. | Islam, S. N. (2015). Inequality and environmental sustainability. |
| **11** | Environmental Sustainability | gender inequalities tend to be lower in more economically equitable countries, and this may reinforce the environment-favoring impact of the reduction of inequality). | Islam, S. N. (2015). Inequality and environmental sustainability. |
| **12** | Environmental Sustainability | Although Environmental Sustainability is needed by humans and originated because of social concerns, environmental sustainability itself seeks to improve human welfare by protecting the sources of raw materials used for human needs and ensuring that the sinks for human wastes are not exceeded, in order to prevent harm to humans. | Goodland, R. (1995). The concept of environmental sustainability. Annual review of ecology and systematics, 1-24. |
| **13** | Environmental Sustainability | Environmental sustainability or maintenance of life-support systems is a prerequisite for social sustainability. | Goodland, R. (1995). The concept of environmental sustainability. Annual review of ecology and systematics, 1-24. |
| **14** | Environmental Sustainability | Environmental sustainability adds consideration of the physical inputs into production, emphasizing environmental life-support systems without which neither production nor humanity could exist. These life support systems include atmosphere, water, and soil-all of these need to be healthy, meaning that their environmental service capacity must be maintained | Goodland, R. (1995). The concept of environmental sustainability. Annual review of ecology and systematics, 1-24. |
| **15** | Environmental Sustainability | Environmentally sustainable development implies sustainable levels of both production (sources), and consumption (sinks), rather than sustained economic growth. | Goodland, R. (1995). The concept of environmental sustainability. Annual review of ecology and systematics, 1-24. |
| **16** | Environmental Sustainability | Ecodesign focuses on reducing the environmental footprint of a product throughout its entire life cycle, whilst ensuring that essential product criteria such as performance and cost are not compromised (Johansson, 2002). | Villers, M., Pigosso, D. C., Howard, T. J., & McAloone, T. C. (2024). Towards a unified absolute environmental sustainability decoupling indicator. Proceedings of the Design Society, 4, 1507-1516. |
| **17** | Social Sustainability | The social aspect relates to the organizational vision to generate value in order to perform fair business practices to benefit the employees, the community, and society [28]. | Chen, X., Despeisse, M., & Johansson, B. (2020). Environmental sustainability of digitalization in manufacturing: A review. Sustainability, 12(24), 10298. |
| **18** | Social Sustainability | Traditional 'hard' social sustainability themes such as employment and poverty alleviation are increasingly been complemented or replaced by 'soft' and less measurable concepts such as happiness, social mixing, and a sense of place in the social sustainability debate. This is adding complexity to the analysis of social sustainability, especially from an assessment point of view. | Colantonio, A. (2009). Social sustainability: a review and critique of traditional versus emerging themes and assessment methods. |
| **19** | Social Sustainability | the development of new sustainability indicators is increasingly focused on measuring emerging themes rather than on improving the assessment of more traditional concepts such as equity and fairness. | Colantonio, A. (2009). Social sustainability: a review and critique of traditional versus emerging themes and assessment methods. |
| **20** | Social Sustainability | the OECD (2001) points out that social sustainability is currently dealt with in connection with the social implications of environmental politics rather than as an equally constitutive component of sustainable development. | Colantonio, A. (2009). Social sustainability: a review and critique of traditional versus emerging themes and assessment methods. |
| **21** | Social Sustainability | Hardoy et al (1992) dispute interpretations according to which social sustainability is defined purely as the social conditions necessary to support environmental sustainability. | Colantonio, A. (2009). Social sustainability: a review and critique of traditional versus emerging themes and assessment methods. |
| **22** | Social Sustainability | Assefa and Frostell, 2007 contend that social sustainability is the finality of development whilst economic and environmental sustainabilities are both the goals of sustainable development and instruments to its achievement. | Colantonio, A. (2009). Social sustainability: a review and critique of traditional versus emerging themes and assessment methods. |
| **23** | Social Sustainability | social sustainability can be interpreted as a socio-historical process rather than a state. In this perspective, the understanding of social sustainability cannot be reduced to a static zero-one situation where zero suggests an unsustainable situation and one indicates presence of sustainability. | Colantonio, A. (2009). Social sustainability: a review and critique of traditional versus emerging themes and assessment methods. |
| **24** | Social Sustainability | Biart (2002: 6) highlights the importance of social requirements for the sustainable development of societies. Despite the confusion over the meaning of social capital, his approach emphasises the importance of 'time–frames' and 'social conditions' for the long-term functioning of societal systems. | Colantonio, A. (2009). Social sustainability: a review and critique of traditional versus emerging themes and assessment methods. |
| **25** | Social Sustainability | A more comprehensive definition of social sustainability with a special focus on urban environments is provided by Polese and Stren (2000: 15-16). They emphasise the economic (development) and social (civil society, cultural diversity and social integration) dimensions of sustainability, highlighting the tensions and trade-offs between development and social disintegration intrinsic to the concept of sustainable development. | Colantonio, A. (2009). Social sustainability: a review and critique of traditional versus emerging themes and assessment methods. |
| **26** | Social Sustainability | social sustainability interpretations emphasising social equity and justice may assist cities in evolving to become 'good' places by facilitating a fairer distribution of resources and a long term vision (Ancell and Thomposon-Fawcett, 2008). | Colantonio, A. (2009). Social sustainability: a review and critique of traditional versus emerging themes and assessment methods. |
| **27** | Social Sustainability | Chiu (2003) identifies three main approaches to the interpretation of social sustainability.<br>The first interpretation equates social sustainability to environmental sustainability. As a result, the social sustainability of an activity depends upon specific social relations, customs, structure and value, representing the social limits and constraints of development.<br>The second interpretation, which she labels 'environment-oriented', refers to the social preconditions required to achieve environmental sustainability. According to this interpretation, social structure, values and norms can be changed in order to carry out human activities within the physical limits of the planet.<br>Lastly, the third 'people-oriented', interpretation refers to improving the well-being of people and the equitable distribution of resources whilst reducing social exclusions and destructive conflict. | Colantonio, A. (2009). Social sustainability: a review and critique of traditional versus emerging themes and assessment methods. |
| **28** | Social Sustainability | equity is considered a crucial component of social sustainability because of the increasing evidence that societies with lower levels of disparity have longer life expectancies, less homicides and crime, stronger patterns of civic engagement and more robust economic vitality (GVRD, 2004) | Colantonio, A. (2009). Social sustainability: a review and critique of traditional versus emerging themes and assessment methods. |
| **29** | Social Sustainability | Key themes for the operationalisation of social sustainability<br>• Livelihood<br>• Equity<br>• Capability to withstand external pressures<br>• Safety nets • security (crime)<br>• Inclusion • pride and sense of place • community participation<br>• Poverty • basic needs<br>• Livelihood<br>• Democracy, Human rights<br>• Social homogeneity • cultural and community diversity<br>• Equitable income distribution<br>• Equitable access to resources and social services<br>• paid and voluntary work<br>• social security<br>• equal opportunities to participate in a democratic society<br>• enabling of social innovation<br>• social justice<br>• solidarity • community stability<br>• participation • interactions in the community/social networks<br>• security<br>• education<br>• skills<br>• experience • employment<br>• consumption • social capital<br>• personal disability<br>• needs of future generations<br>• empowerment and participation | Colantonio, A. (2009). Social sustainability: a review and critique of traditional versus emerging themes and assessment methods. |
| **30** | Social Sustainability | Traditional Social Sustainability Key Themes<br>• Basic needs, including housing and environmental health<br>• Education and skills<br>• Employment<br>• Equity<br>• Human rights and gender<br>• Poverty<br>• Social justice | Colantonio, A. (2009). Social sustainability: a review and critique of traditional versus emerging themes and assessment methods. |
| **31** | Social Sustainability | Emerging Social Sustainability Key Themes<br>• Demographic change (aging, migration and mobility)<br>• Social mixing and cohesion<br>• Identity, sense of place and culture<br>• Empowerment, participation and access<br>• Health and Safety<br>• Social capital<br>• Well being, Happiness and Quality of Life | Colantonio, A. (2009). Social sustainability: a review and critique of traditional versus emerging themes and assessment methods. |
| **32** | Social Sustainability | social sustainability blends traditional social policy areas and principles such as equity and health, with issues concerning participation, needs, social capital, the economy, the environment, and more recently, with the notions of happiness, well-being and quality of life. | Colantonio, A. (2009). Social sustainability: a review and critique of traditional versus emerging themes and assessment methods. |
| **33** | Social Sustainability | For a community to function and be sustainable, the basic needs of its residents must be met. A socially sustainable community must have the ability to maintain and build on its own resources and have the resiliency to prevent and/or address problems in the future (City of Vancouver, 2005 : 12). | Colantonio, A. (2009). Social sustainability: a review and critique of traditional versus emerging themes and assessment methods. |
| **34** | Social Sustainability | A strong definition of social sustainability must rest on the basic values of equity and democracy, the latter meant as the effective appropriation of all human rights – political, civil, economic, social and cultural – by all people Sachs (1999: 27) | Colantonio, A. (2009). Social sustainability: a review and critique of traditional versus emerging themes and assessment methods. |
| **35** | Social Sustainability | Social sustainability is given if work within a society and the related institutional arrangements satisfy an extended set of human needs [and] are shaped in a way that nature and its reproductive capabilities are preserved over a long period of time and the normative claims of social justice, human dignity and participation are fulfilled. Littig and Grießler (2005: 72). | Colantonio, A. (2009). Social sustainability: a review and critique of traditional versus emerging themes and assessment methods. |
| **36** | Social Sustainability | The relationship between urban form and social sustainability is explored and two main dimensions of social sustainability are identifi ed and discussed in detail: equitable access and the sustainability of the community itself. | Dempsey, N., Bramley, G., Power, S., & Brown, C. (2011). The social dimension of sustainable development: Defining urban social sustainability. Sustainable development, 19(5), 289-300. |
| **37** | Social Sustainability | Social sustainability has to be considered as a dynamic concept, which will change over time (from year to year/decade to decade) in a place.<br>This may come about through external influences: for example, social cohesion and interaction may increase, prompted by changes in local authority service delivery or the threat of airport expansion. Economic, environmental and political crises at a local or broader scale may also influence social activity at the local scale. | Dempsey, N., Bramley, G., Power, S., & Brown, C. (2011). The social dimension of sustainable development: Defining urban social sustainability. Sustainable development, 19(5), 289-300. |
| **38** | Social Sustainability | While few would argue for dirty, unsafe spaces without vegetation over clean, safe and green public spaces (Dempsey, 2008b), to presume that urban social sustainability can only occur in neighbourhoods with 'high' environmental quality is short-sighted. | Dempsey, N., Bramley, G., Power, S., & Brown, C. (2011). The social dimension of sustainable development: Defining urban social sustainability. Sustainable development, 19(5), 289-300. |
| **39** | Social Sustainability | The perceived safety of a neighbourhood is said to be a fundamental part of social sustainability (Barton, 2000a). In its defi nition of social cohesion, the UK House of Commons Committee positioned perceived safety within Maslows hierarchy of needs, with the fulfi lment of basic needs required before social cohesion can be achieved (House of Commons, 2004; Maslow, 1954). | Dempsey, N., Bramley, G., Power, S., & Brown, C. (2011). The social dimension of sustainable development: Defining urban social sustainability. Sustainable development, 19(5), 289-300. |
| **40** | Social Sustainability | a dimension of social sustainability related to social coherence and social network integration (Littig and Griessler, 2005). | Dempsey, N., Bramley, G., Power, S., & Brown, C. (2011). The social dimension of sustainable development: Defining urban social sustainability. Sustainable development, 19(5), 289-300. |
| **41** | Social Sustainability | The social sustainability indicators and associated framework established in this work contribute to the collective understanding of what it means to be sustainable. | Hutchins, M. J., Richter, J. S., Henry, M. L., & Sutherland, J. W. (2019). Development of indicators for the social dimension of sustainability in a US business context. Journal of Cleaner Production, 212, 687-697. |
| **42** | Social Sustainability | "[social sustainability] is a concept in chaos, and we argue that this severely compromises its importance and utility." (Vallance et al. 2011). | Woodcraft, S. (2012). Social sustainability and new communities: Moving from concept to practice in the UK. Procedia-Social and Behavioral Sciences, 68, 29-42. |
| **43** | Social Sustainability | Urban social sustainability: contributory factors as identified in the review of literature (in no particular<br>order) by Dempsey et al., 2009 (As quoted Dempsey et al. 2011) | Woodcraft, S. (2012). Social sustainability and new communities: Moving from concept to practice in the UK. Procedia-Social and Behavioral Sciences, 68, 29-42. |
| **44** | Social Sustainability | Urban social sustainability: contributory factors as identified in the review of literature (in no particular<br>order) by Dempsey et al., 2009 (As quoted Dempsey et al. 2011) | Woodcraft, S. (2012). Social sustainability and new communities: Moving from concept to practice in the UK. Procedia-Social and Behavioral Sciences, 68, 29-42. |
| **45** | Social Sustainability | abe to be maintained at a certain rate or level (oxford dictionaries 2012)<br>social as 'relating to society or its organization' | Woodcraft, S. (2012). Social sustainability and new communities: Moving from concept to practice in the UK. Procedia-Social and Behavioral Sciences, 68, 29-42. |
| **46** | Social Sustainability | (Sachs 1999; Agyeman 2008) argue social sustainability must be grounded in equality, democracy and social justice. | Woodcraft, S. (2012). Social sustainability and new communities: Moving from concept to practice in the UK. Procedia-Social and Behavioral Sciences, 68, 29-42. |
| **47** | Social Sustainability | The Berkeley Group: "Social sustainability is about people's quality of life, now and in the future. It describes the extent to which a neighbourhood supports individual and collective wellbeing. Social sustainability combines design of the physical environment with a focus on how the people who live in and use a space relate to each other and function as a community. It is enhanced by development which provides the right infrastructure to support a strong social and cultural life, opportunities for people to get involved, and scope for the place and the community to evolve." (Bacon et al. 2012, p.9) | Woodcraft, S. (2012). Social sustainability and new communities: Moving from concept to practice in the UK. Procedia-Social and Behavioral Sciences, 68, 29-42. |
| **48** | Social Sustainability | Vallance et al (2011) identify work addressing underdevelopment, basic needs, and the promotion of stronger environmental ethics. Other authors emphasize the preservation of social values, cultural traditions and ways of life (Barbier 1987; Koning 2002; Vallance et al. 2011). | Woodcraft, S. (2012). Social sustainability and new communities: Moving from concept to practice in the UK. Procedia-Social and Behavioral Sciences, 68, 29-42. |
| **49** | Social Sustainability | social sustainability incorporates a set of underlying themes that could be described as social capital, human capital, and well-being (Colantonio & T. Dixon 2010; Dempsey et al. 2011; Weingaertner & Moberg 2011; Murphy 2012; Magee et al. 2012). | Woodcraft, S. (2012). Social sustainability and new communities: Moving from concept to practice in the UK. Procedia-Social and Behavioral Sciences, 68, 29-42. |
| **50** | Social Sustainability | social sustainability has become shorthand in policy discourse for creating places 'that work' - where people want to live now and in the future (Bacon et al. 2012; Woodcraft 2011). the usefulness of social sustainability as a planning tool depends on how it is enacted in practice. | Woodcraft, S. (2012). Social sustainability and new communities: Moving from concept to practice in the UK. Procedia-Social and Behavioral Sciences, 68, 29-42. |
| **51** | Social Sustainability | Bramley and Power (2009), for example, have argued that social sustainability in this context is often equated with social capital, social cohesion and social exclusion. | Vallance, S., Perkins, H. C., & Dixon, J. E. (2011). What is social sustainability? A clarification of concepts. Geoforum, 42(3), 342-348. |
| **52** | Social Sustainability | three types of social sustainability: 'development sustainability' which addresses poverty and inequity; 'bridge sustainability' with its concerns about changes in behavior so as to achieve bio-physical environmental goals; and 'maintenance sustainability' which refers to the preservation of socio-cultural patterns and practices in the context of social and economic change. | Vallance, S., Perkins, H. C., & Dixon, J. E. (2011). What is social sustainability? A clarification of concepts. Geoforum, 42(3), 342-348. |
| **53** | Social Sustainability | Achieved only by systematic community participation and strong civil society. Cohesion of community, cultural identity, diversity, sodality, comity, tolerance, humility, compassion, patience, forbearance, fellowship, fraternity, institution, love, pluralism, commonly accepted standards of honesty, laws, discipline, etc, constitute part of social sustainability. This 'moral capital' as some call it, requires maintenance and replenishment by shared values and equal rights, and by community, religious, and cultural interactions. The creation of social capital as needed for social sustainability is not yet adequately recognized. | Goodland, R. (1995). The concept of environmental sustainability. Annual review of ecology and systematics, 1-24. |
| **54** | Social Sustainability | Criteria across the social, environmental and economical dimensions | Gräßler, I., & Hesse, P. (2022). Approach to Sustainability-Based Assessment of Solution Alternatives in Early Stages of Product Engineering. Proceedings of the Design Society, 2, 1001-1010. |
| **55** | Economic Sustainability | Being economically sustainable involves the organizational vision, to create economic value either through increased added value or through cost reduction in production, with the purpose of assuring the possibility of delivering products and services to the market while having a profit between revenues and costs [27]. | Chen, X., Despeisse, M., & Johansson, B. (2020). Environmental sustainability of digitalization in manufacturing: A review. Sustainability, 12(24), 10298. |
| **56** | Economic Sustainability | In the economic aspect, it is necessary to implement "resource efficiency"—managing losses and surpluses to maximize economic efficiency, in addition to focusing on the market to obtain results and maintain business strategies [14–16]. | Maynard, D. D. C., Vidigal, M. D., Farage, P., Zandonadi, R. P., Nakano, E. Y., & Botelho, R. B. A. (2020). Environmental, social and economic sustainability indicators applied to food services: A systematic review. Sustainability, 12(5), 1804. |
| **57** | Economic Sustainability | sustainable economic development tend to focus on increasing the stock of man-made capital and the degree to which other capital stocks may be reduced on this account (OECD, 2001). | Spangenberg, J. H. (2005). Economic sustainability of the economy: concepts and indicators. International journal of sustainable development, 8(1-2), 47-64. |
| **58** | Economic Sustainability | Describing the sustainability-relevant aspects of a simple economic process illustrates the limits to understanding the impacts in all four dimensions imposed by this approach: If a new machine (man-made capital) replaces skilled workers, this may be an effective substitution regarding production and value creation, but in terms of resource consumption (environmental capital), income generation (social capital) and skills training (human capital), the outcome is definitively different, a fact which is not captured as long as all impacts are reduced to their function in the production process, and measured according to the assumption of strong comparability. | Spangenberg, J. H. (2005). Economic sustainability of the economy: concepts and indicators. International journal of sustainable development, 8(1-2), 47-64. |
| **59** | Economic Sustainability | Economic sustainability focuses on that portion of the natural resource base that provides physical inputs, both renewable (e.g. forests) and exhaustible (e.g. minerals), into the production process. | Goodland, R. (1995). The concept of environmental sustainability. Annual review of ecology and systematics, 1-24. |
| **60** | Economic Sustainability | Hicks' (48) definition of income - "the amount one can consume during a period and still be as well off at the end of the period" -can define economic sustainability, as it devolves on consuming interest, rather than capital. | Goodland, R. (1995). The concept of environmental sustainability. Annual review of ecology and systematics, 1-24. |
| **61** | Sustainability | the Brundtland Report (1987): "development that meets the needs of the present without compromising the ability of future generations to meet their own needs" | Chen, X., Despeisse, M., & Johansson, B. (2020). Environmental sustainability of digitalization in manufacturing: A review. Sustainability, 12(24), 10298. |
| **62** | Sustainability | Sustainability as a policy concept has its origin in the Brundtland Report of 1987. | Kuhlman, T., & Farrington, J. (2010). What is sustainability?. Sustainability, 2(11), 3436-3448. |
| **63** | Sustainability | If 'sustainability' is anything more than a slogan or expression of emotion, it must amount to an injunction to preserve productive capacity for the indefinite future. | Kuhlman, T., & Farrington, J. (2010). What is sustainability?. Sustainability, 2(11), 3436-3448. |
| **64** | Sustainability | Sustainability may then be defined as maintaining well-being over a long, perhaps even an indefinite period.<br>We propose to replace the social and economic dimensions of sustainability as conventionally used with a single dimension called well-being, which is a policy goal that must be balanced with another one called sustainability. | Kuhlman, T., & Farrington, J. (2010). What is sustainability?. Sustainability, 2(11), 3436-3448. |
| **65** | Sustainability | Sustainability can now loosely be defined as a state of affairs where the sum of natural and man-made resources remains at least constant for the foreseeable future, in order that the well-being of future generations does not decline. | Kuhlman, T., & Farrington, J. (2010). What is sustainability?. Sustainability, 2(11), 3436-3448. |
| **66** | Sustainability | Sustainability, then, is a matter of what resources—natural resources, quality of the environment, and capital—we bequeath to coming generations. | Kuhlman, T., & Farrington, J. (2010). What is sustainability?. Sustainability, 2(11), 3436-3448. |
| **67** | Sustainability | [Sustainability] aims to determine the minimal social requirements for long-term development (sometimes called critical social capital) and to identify the challenges to the very functioning of society, in the long run, Biart (2002:6) | Colantonio, A. (2009). Social sustainability: a review and critique of traditional versus emerging themes and assessment methods. |
| **68** | Sustainability | Debates about sustainability no longer consider sustainability solely as an environmental concern, but also incorporate economic and social dimensions. | Dempsey, N., Bramley, G., Power, S., & Brown, C. (2011). The social dimension of sustainable development: Defining urban social sustainability. Sustainable development, 19(5), 289-300. |
| **69** | Sustainability | The sustainability of community is about the ability of society itself, or its manifestation as local community, to sustain and reproduce itself at an acceptable level of functioning. This is associated with social capital and social cohesion as concepts that encompass social networks, norms of reciprocity and features of social organization (Coleman, 1988), and the integration of resulting social behaviour (Dempsey, 2008a). | Dempsey, N., Bramley, G., Power, S., & Brown, C. (2011). The social dimension of sustainable development: Defining urban social sustainability. Sustainable development, 19(5), 289-300. |
| **70** | Sustainability | Inherent in definitions of sustainability is the concept of inter-generational equity. | Dempsey, N., Bramley, G., Power, S., & Brown, C. (2011). The social dimension of sustainable development: Defining urban social sustainability. Sustainable development, 19(5), 289-300. |
| **71** | Sustainability | Sustainability requires that corporations maintain the integrity of social and environmental systems while undertaking this reconfiguration (the economic health of the business will almost always be a priority). | Hutchins, M. J., Richter, J. S., Henry, M. L., & Sutherland, J. W. (2019). Development of indicators for the social dimension of sustainability in a US business context. Journal of Cleaner Production, 212, 687-697. |
| **72** | Sustainability | the Workshop on Urban Sustainability of the US National Science Foundation (2000, p. 1) pointed out, sustainability is 'laden with so many definitions that it risks plunging into meaninglessness, at best, and becoming a catchphrase for demagogy, at worst. [It] is used to justify and legitimate a myriad of policies and practices ranging from communal agrarian utopianism to large-scale capital-intensive market development'. | Hopwood, B., Mellor, M., & O'Brien, G. (2005). Sustainable development: mapping different approaches. Sustainable development, 13(1), 38-52. |
| **73** | Sustainability | sustainability metrics can be structured into three categories: renewable, nonrenewable, and pollution (Daly 1990). | Melville, N. P. (2010). Information systems innovation for environmental sustainability. MIS quarterly, 1-21. |
| **74** | Sustainability | 'To think that their present societal arrangements might be sustained – that is an unsustainable thought for the majority of the world's people' (1998, p. 103; see also Gunder, 2006). | Vallance, S., Perkins, H. C., & Dixon, J. E. (2011). What is social sustainability? A clarification of concepts. Geoforum, 42(3), 342-348. |
| **75** | Sustainability | sustainability is a dynamic system property and not predictive, it is guided by targets/visions, instead of traditional goal-based optimization approaches (Gaziulusoy and Brezet, 2015). | Mitake, Y., Hiramitsu, K., Nagayama, A., Muraoka, N., Sholihah, M., & Shimomura, Y. (2020, May). A Conceptual Framework of Product-Service Systems Design for Sustainability Transitions. In Proceedings of the Design Society: DESIGN Conference (Vol. 1, pp. 2069-2078). Cambridge University Press. |
| **76** | Sustainability | sustainability has roots in systems thinking (Osorio et al., 2009), and requires "holism and system-wide approach[es] [...] in order to deal with complexity" (Sala et al., 2013) of the almost unfathomable volume of simultaneous, interlinked activities on Earth affecting its behaviour. | Johnston-Lynch, K., Whitfield, R. I., & Evans, D. (2024). Systems thinking towards holistic, sustainability-oriented assessment and decision-making for lightweighting. Proceedings of the Design Society, 4, 1319-1328. |
| **77** | Sustainability | Kupfer et al. (2022), recently, acknowledged that "pressure to provide sustainable products is forcing the lightweight industry to rethink [...] to integrate sustainability criteria in all decisions". | Johnston-Lynch, K., Whitfield, R. I., & Evans, D. (2024). Systems thinking towards holistic, sustainability-oriented assessment and decision-making for lightweighting. Proceedings of the Design Society, 4, 1319-1328. |
| **78** | Sustainability | increasingly sustainability is interpreted as principally anthropogenically relevant (Lövbrand et al., 2009), and important to species on Earth to ensure their habitat is protected in the form in which they know how to survive and live well. | Johnston-Lynch, K., Whitfield, R. I., & Evans, D. (2024). Systems thinking towards holistic, sustainability-oriented assessment and decision-making for lightweighting. Proceedings of the Design Society, 4, 1319-1328. |
| **79** | Sustainability | lightweighting to being an "effective enabler for sustainability" (Herrmann et al., 2018). | Johnston-Lynch, K., Whitfield, R. I., & Evans, D. (2024). Systems thinking towards holistic, sustainability-oriented assessment and decision-making for lightweighting. Proceedings of the Design Society, 4, 1319-1328. |
| **80** | Sustainability | Sustainability remains a persistent, common goal across all stakeholders (Kupfer et al., 2022) despite perceptively there being multiple, dynamic goals. | Johnston-Lynch, K., Whitfield, R. I., & Evans, D. (2024). Systems thinking towards holistic, sustainability-oriented assessment and decision-making for lightweighting. Proceedings of the Design Society, 4, 1319-1328. |
| **81** | Sustainability | Manufactured products impact all three facets of sustainability; economy, environment and society throughout their entire life cycle; material extraction, manufacturing, transportation, use and disposal (Tarne et al., 2017). It was found that about 80% of sustainability impacts are decided at the product design stage. | Ahmad, S., Wong, K. Y., Tseng, M. L., & Wong, W. P. (2018). Sustainable product design and development: A review of tools, applications and research prospects. Resources, Conservation and Recycling, 132, 49-61. |
| **82** | Sustainability | sustainability can be defined as the ability of a product to work continuously while ensuring lowest environmental impacts and providing economic and social benefits to the stakeholders. | Ahmad, S., Wong, K. Y., Tseng, M. L., & Wong, W. P. (2018). Sustainable product design and development: A review of tools, applications and research prospects. Resources, Conservation and Recycling, 132, 49-61. |
| **83** | Sustainability | all three aspects of sustainability must be considered as an integral part of a sustainable<br>design (Gennari, 2000). | Ahmad, S., Wong, K. Y., Tseng, M. L., & Wong, W. P. (2018). Sustainable product design and development: A review of tools, applications and research prospects. Resources, Conservation and Recycling, 132, 49-61. |
| **84** | Sustainability | Sustainability is understood as the combination of the dimensions of social, ecological and economical aspects. Complete and utilitarian sustainability exists when the maximum concordance of the benefits of all stakeholders is achieved. | Gräßler, I., & Hesse, P. (2022). Approach to Sustainability-Based Assessment of Solution Alternatives in Early Stages of Product Engineering. Proceedings of the Design Society, 2, 1001-1010. |
| **85** | Sustainability | "Economic sustainability could be defined as realization of growth, efficiency and “equitable” distribution of wealth. Social sustainability implies participation in making decisions, mobility and cohesion, fulfilment of social identity, development of institutions and alike. The third aspect of sustainability is environmental sustainability. It respects the wholeness of different eco-systems, the carrying (receiving) capacity and protection of<br>natural resources including biological diversity as well" [10]. | Ivković, A. F., Ham, M., & Mijoč, J. (2014). Measuring objective well-being and sustainable development management. Journal of Knowledge Management, Economics and Information Technology, 4(2), 1-29. |
| **86** | Sustainability | Sustainability advances environmental, social and economic equity in a variety of contexts, including dignity,38 human rights,39 climate change, access to and availability of fresh water,40 shale gas development,41 corporate practices, and higher education. | May, J. R., & Daly, E. (2020). The indivisibility of human dignity and sustainability. The Cambridge Handbook on Environmental Justice and Sustainable Development (Sumudu Atapattu, Carmen G. Gonzalez and Sara Seck, eds) Cambridge University Press (2020). |
| **87** | Sustainable Development | Glavic and Lukman (2007) reviewed sustainable development as a process that "emphasizes the evolution of human society from the responsible economic point of view, in accordance with environmental and natural processes". | Chen, X., Despeisse, M., & Johansson, B. (2020). Environmental sustainability of digitalization in manufacturing: A review. Sustainability, 12(24), 10298. |
| **88** | Sustainable Development | the definition of sustainability adopted by the United Nations in its Agenda for Development: Development is a multidimensional undertaking to achieve a higher quality of life for all people. Economic development, social development, and environmental protection are interdependent and mutually reinforcing components of sustainable development [12]. | Kuhlman, T., & Farrington, J. (2010). What is sustainability?. Sustainability, 2(11), 3436-3448. |
| **89** | Sustainable Development | Development (and/or growth) that is compatible with harmonious evolution of civil society, fostering an environment conducive to the compatible cohabitation of culturally and socially diverse groups while at the same time encouraging social integration, with improvements in the quality of life for all segments of the population Polese and Stren (2000: 15-16) | Colantonio, A. (2009). Social sustainability: a review and critique of traditional versus emerging themes and assessment methods. |
| **90** | Sustainable Development | Sustainable communities are here defined as places where people want to live and work, now and in the future. They meet the diverse needs of existing and future residents, are sensitive to their environment, and contribute to a high quality of life. They are safe and inclusive, well planned, built and run, and offer equality of opportunity and good services for all (ODPM, 2006, p. 12). | Dempsey, N., Bramley, G., Power, S., & Brown, C. (2011). The social dimension of sustainable development: Defining urban social sustainability. Sustainable development, 19(5), 289-300. |
| **91** | Sustainable Development | the concept of sustainable development is an attempt to combine growing concerns about a range of environmental issues with socio-economic issues. Sustainable development has the potential to address fundamental challenges for humanity, now and into the future. *(Strategy/policy generation for sustainable development)* | Hopwood, B., Mellor, M., & O'Brien, G. (2005). Sustainable development: mapping different approaches. Sustainable development, 13(1), 38-52. |
| **92** | Sustainable Development | The concept of sustainable development is potentially an important shift in understanding relationships of humanity with nature and between people. | Hopwood, B., Mellor, M., & O'Brien, G. (2005). Sustainable development: mapping different approaches. Sustainable development, 13(1), 38-52. |
| **93** | Sustainable Development | The concept of sustainable development is the result of the growing awareness of the global links between mounting environmental problems, socio-economic issues to do with poverty and inequality and concerns about a healthy future for humanity. It strongly links environmental and socio-economic issues. | Hopwood, B., Mellor, M., & O'Brien, G. (2005). Sustainable development: mapping different approaches. Sustainable development, 13(1), 38-52. |
| **94** | Sustainable Development | Social justice today and in the future is a crucial component of the concept of sustainable development. | Hopwood, B., Mellor, M., & O'Brien, G. (2005). Sustainable development: mapping different approaches. Sustainable development, 13(1), 38-52. |
| **95** | Sustainable Development | Daly (1993) criticized the notion of 'sustainable growth' as 'thought-stopping' and oxymoronic in a world in which ecosystems are finite. At some point, economic growth with ever more use of resources and production of waste is unsustainable. Instead Daly argued for the term 'sustainable development' by which he, much more clearly than Brundtland, meant qualitative, rather than quantitative, improvements. Development<br>is open to confusion, with some seeing it as an end in itself, so it has been suggested that greater clarity would be to speak of 'sustainable livelihoods', which is the aim that Brundtland outlined (Workshop on Urban Sustainability, 2000). | Hopwood, B., Mellor, M., & O'Brien, G. (2005). Sustainable development: mapping different approaches. Sustainable development, 13(1), 38-52. |
| **96** | Sustainable Development | The concept of sustainable development represents a shift in the understanding of humanity's place on the planet, but it is open to interpretation of being anything from almost meaningless to of extreme importance to humanity. | Hopwood, B., Mellor, M., & O'Brien, G. (2005). Sustainable development: mapping different approaches. Sustainable development, 13(1), 38-52. |
| **97** | Sustainable Development | Haughton (1999) has usefully summarized the ideas of sustainable development in five principles based on equity: futurity – inter-generational equity; social justice – intra-generational equity; trans frontier responsibility – geographical equity; procedural equity – people treated openly and fairly; inter species equity – importance of biodiversity. | Hopwood, B., Mellor, M., & O'Brien, G. (2005). Sustainable development: mapping different approaches. Sustainable development, 13(1), 38-52. |
| **98** | Sustainable Development | the nature of the changes necessary in society's political and economic structures and human–environment relationships to achieve sustainable development: that it can be achieved within the present structures – status quo; that fundamental reform is necessary but without a full rupture with the existing arrangements – reform; and that as the roots of the problems are the very economic and power structures of society a radical transformation is needed – transformation (Rees, 1995). | Hopwood, B., Mellor, M., & O'Brien, G. (2005). Sustainable development: mapping different approaches. Sustainable development, 13(1), 38-52. |
| **99** | Sustainable Development | O'Riordan states that 'wealth creation based on renewability and replenishment rather than exploitation . . . is a contradiction in terms for modern capitalism', so that real sustainable development requires a 'massive redistribution of wealth and power' (1989, p. 93). | Hopwood, B., Mellor, M., & O'Brien, G. (2005). Sustainable development: mapping different approaches. Sustainable development, 13(1), 38-52. |
| **100** | Sustainable Development | sustainable development needs to be based on the appreciation of the close links between the environment and society with feedback loops both ways, and that social and environmental equity are fundamental ideas. Given the need for fundamental change, a deep connection between human life and the environment and a common linkage of power structures that exploit both people and planet, we would argue that transformation is essential. | Hopwood, B., Mellor, M., & O'Brien, G. (2005). Sustainable development: mapping different approaches. Sustainable development, 13(1), 38-52. |
| **101** | Sustainable Development | De Groot (1992: 130) has identified nine different types of values of environmental functions, grouped under the three dimensions of sustainable development: | Ekins, P. (2011). Environmental sustainability: From environmental valuation to the sustainability gap. Progress in Physical Geography, 35(5), 629-651. |
| **102** | Sustainable Development | Daly (1991), working with a 'strong' to 'very strong' sustainability framework, has suggested four principles of sustainable development:<br>(1) limit the human scale (throughput) to that which is within the earth's carrying capacity;<br>(2) ensure that technological progress is efficiency-increasing rather than through-put-increasing;<br>(3) for renewable resources harvesting rates should not exceed regeneration rates (sustained yield); waste emissions should not exceed the assimilative capacities of the<br>receiving environment;<br>(4) non-renewable resources should be exploited no faster than the rate of creation of renewable substitutes. | Ekins, P. (2011). Environmental sustainability: From environmental valuation to the sustainability gap. Progress in Physical Geography, 35(5), 629-651. |
| **103** | Sustainable Development | inter and intra-generational equity, the distribution of power and resources, employment, education, the provision of basic infra- structure and services, freedom, justice, access to influential decision-making fora, and general 'capacity-building' have all been identified as important aspects of the development paradigm | Vallance, S., Perkins, H. C., & Dixon, J. E. (2011). What is social sustainability? A clarification of concepts. Geoforum, 42(3), 342-348. |
| **104** | Sustainable Development | 'Worth-Living Development' that seeks to ensure that "each generation will hand over to the next one a better place to live in" [3]. | Howes, M., Wortley, L., Potts, R., Dedekorkut-Howes, A., Serrao-Neumann, S., Davidson, J., ... & Nunn, P. (2017). Environmental sustainability: a case of policy implementation failure?. Sustainability, 9(2), 165. |
| **105** | Sustainable Development | there are no standardised measures for sustainable development because of low levels of operationalisation of the concept. | Howes, M., Wortley, L., Potts, R., Dedekorkut-Howes, A., Serrao-Neumann, S., Davidson, J., ... & Nunn, P. (2017). Environmental sustainability: a case of policy implementation failure?. Sustainability, 9(2), 165. |
| **106** | Sustainable Development | In the economic debate, sustainable development is most often described as the need to<br>maintain a permanent income for humankind, generated from non-declining capital stocks (Hicksian income). | Spangenberg, J. H. (2005). Economic sustainability of the economy: concepts and indicators. International journal of sustainable development, 8(1-2), 47-64. |
| **107** | Sustainable Development | Sustainable development (SD) should integrate social, environmental, and economic sustainability and use these three to start to make development sustainable. | Goodland, R. (1995). The concept of environmental sustainability. Annual review of ecology and systematics, 1-24. |
| **108** | Sustainable Development | The priority for development should be improvement in human well-being-the reduction of poverty, illiteracy, hunger, disease, and inequity. | Goodland, R. (1995). The concept of environmental sustainability. Annual review of ecology and systematics, 1-24. |
| **109** | Sustainable Development | World Wildlife Fund's (107) definition of sustainable development is similar: "Improvement in the quality of human life within the carrying capacity of supporting ecosystems." | Goodland, R. (1995). The concept of environmental sustainability. Annual review of ecology and systematics, 1-24. |
| **110** | Sustainable Development | As Jacobs (1991) identified, sustainable development is a 'contestable concept', that like<br>'democracy' or 'equality' has a basic meaning that almost everyone is in favour of, but there are deep conflicts around how they should be understood and fostered. | Elliott, J. (2012). An introduction to sustainable development. Routledge. |
| **111** | Sustainable Development | ‘In principle, such an optimal (sustainable growth) policy would seek to maintain an “acceptable” rate of growth in per-capita real incomes without depleting the national capital asset stock or the natural environmental asset stock.’ (Turner, 1988: 12) | Elliott, J. (2012). An introduction to sustainable development. Routledge. |
| **112** | Sustainable Development | ‘A sustainable society is one in which peoples’ ability to do what they have good reason to value is continually enhanced.’ (Sen, 1999) | Elliott, J. (2012). An introduction to sustainable development. Routledge. |
| **113** | Sustainable Development | ‘Sustainable development appears to be an over-used, misunderstood phrase.’ (Mawhinney, 2002: 5) | Elliott, J. (2012). An introduction to sustainable development. Routledge. |
| **114** | Sustainable Development | sustainable development as a moral concept that seeks to define a ‘fair and just’ development (Starkey and Walford, 2001). The notion of ‘environmental justice’ is now a prominent part of contemporary discussions of the meaning and practice of sustainable development (see Walker, 2012). | Elliott, J. (2012). An introduction to sustainable development. Routledge. |
| **115** | Sustainable Development | Strong sustainability demands the protection of critical natural capital because once lost, these assets are lost forever, and they cannot be recreated. | Elliott, J. (2012). An introduction to sustainable development. Routledge. |
| **116** | Sustainable Development | Sustainable development is not an identifiable ‘end point’ or ‘state’ but requires ongoing critical consideration of the processes (the ‘means’) of development and decision making across all spheres of life. Hence, there is no blueprint for how to achieve sustainable development; rather the nature of sustainable development will be specific to particular places and points in time. | Elliott, J. (2012). An introduction to sustainable development. Routledge. |
| **117** | Sustainable Development | development needed to be sustainable; it must encompass not only economic and social activities, but also those related to population, the use of natural resources and the resulting impacts on the environment. | Elliott, J. (2012). An introduction to sustainable development. Routledge. |
| **118** | Sustainable Development | Sustainable development as a concept is thus closely associated with the desire to develop a harmonious society oriented towards greater economic prosperity, social cohesion and environmental protection. | Ivković, A. F., Ham, M., & Mijoč, J. (2014). Measuring objective well-being and sustainable development management. Journal of Knowledge Management, Economics and Information Technology, 4(2), 1-29. |
| **119** | Sustainable Product Design | In short, traditional product design focuses on product functionalities, quality and costs for meeting customer requirements, whereas sus tainable product design (SPD) systematically views the entire product life cycle for functional, environmental, and economic performances(Lu et al., 2011). | Ahmad, S., Wong, K. Y., Tseng, M. L., & Wong, W. P. (2018). Sustainable product design and development: A review of tools, applications and research prospects. Resources, Conservation and Recycling, 132, 49-61. |
| **120** | Sustainable Product Design | Boyko (2009) improved the conventional design process by in corporating ‘sustainability tasks’ along all design stages. | Ahmad, S., Wong, K. Y., Tseng, M. L., & Wong, W. P. (2018). Sustainable product design and development: A review of tools, applications and research prospects. Resources, Conservation and Recycling, 132, 49-61. |
| **121** | Sustainable Product Development | SPD is defined as the "systematic incorporation of environmental, economic and social considerations into the Product Development Process (PDP) to fulfil the elementary needs of society while improving the environmental and economic performance". Successful SPD has the potential to help companies in minimising the impact throughout the products' lifecycle, as well as to gain and maintain competitive advantage (Clark et al., 2009; Hwang et al., 2013). | Vilochani, S., McAloone, T. C., & Pigosso, D. C. (2023). Management practices for Sustainable Product Development: Insights from a systematic literature review. Proceedings of the Design Society, 3, 2505-2514. |
| **122** | Sustainable Product Deevelopment | Integrating the breadth of sustainability into product development is labelled sustainable design or sustainable product development (SPD) (Gagnon et al., 2012), which means that a strategic sustainability perspective is integrated and implemented into the early phases of product innovation, including life-cycle thinking (Hallstedt and Isaksson, 2017). | Kwok, S. Y., Schulte, J., & Hallstedt, S. I. (2020, May). Approach for sustainability criteria and product life-cycle data simulation in concept selection. In Proceedings of the Design Society: DESIGN Conference (Vol. 1, pp. 1979-1988). Cambridge University Press. |
| **123** | Sustainable Product Deevelopment | Hallstedt (2017) presents an approach for how a Sustainability Design Space (SDS) can be defined using an approach that combined forecasting and backcasting to identify leading sustainability criteria and indicators. The SDS has several purposes: i) to represent the most important sustainability aspects that can be accomplished within the time constrained early development situation; ii) to be used as input for simulation and modelling for sustainability (Jaghbeer et al., 2017); iii) to inform sustainability models to be used in an automated decision support environment (Bertoni et al., 2018a); iv) to enable performance evaluation against the identified sustainability criteria (Watz and Hallstedt, 2018). | Kwok, S. Y., Schulte, J., & Hallstedt, S. I. (2020, May). Approach for sustainability criteria and product life-cycle data simulation in concept selection. In Proceedings of the Design Society: DESIGN Conference (Vol. 1, pp. 1979-1988). Cambridge University Press. |
| **124** | Sustainable product development | Sustainable product development, can be defined as the integration of sustainability into the early phases of the product innovation process and includes a life-cycle thinking, and is therefore an important measure in the development of a more sustainable product design and production (Watz and Hallstedt, 2022). | Leonard, P. L., Hallstedt, S. I., Nylander, J. W., & Isaksson, O. (2024). Design Aspects of Importance in a Sustainability Transition. DS 130: Proceedings of NordDesign 2024, Reykjavik, Iceland, 12th-14th August 2024, 106-114. |
| **125** | Sustainable product development | Sustainable product development capabilities can be defined as: "Skills and knowledge in the field of sustainable product development, exercised through support methods and tools applied in routines and organizational processes, which enable firms to coordinate activities on strategic, tactical and operational organizational levels to accelerate towards a sustainability transformation and make use of their assets." (Hallstedt et al., 2023b). | Leonard, P. L., Hallstedt, S. I., Nylander, J. W., & Isaksson, O. (2024). Design Aspects of Importance in a Sustainability Transition. DS 130: Proceedings of NordDesign 2024, Reykjavik, Iceland, 12th-14th August 2024, 106-114. |

**Supplementary_2**

| No | References |
|---|---|
| 1 | Nishant, R., Goh, M., & Kitchen, P. J. (2016). Sustainability and differentiation: Understanding materiality from the context of Indian firms. Journal of Business Research, 69(5), 1892-1897. |
| 2 | Waage, S. A., Geiser, K., Irwin, F., Weissman, A. B., Bertolucci, M. D., Fisk, P., ... & McPherson, A. (2005). Fitting together the building blocks for sustainability: a revised model for integrating ecological, social, and financial factors into business decision-making. Journal of Cleaner Production, 13(12), 1145-1163. |
| 3 | Isaksson, O., Bertoni, M., Hallstedt, S., & Lavesson, N. (2015). Model based decision support for value and sustainability in product development. In 20th International Conference on Engineering Design (ICED), Milan. The Design Society. |
| 4 | Watz, M., & Hallstedt, S. I. (2018). Integrating sustainability in product requirements. In DS 92: Proceedings of the DESIGN 2018 15th International Design Conference (pp. 1405-1416). |
| 5 | Waage, S. A. (2007). Re-considering product design: a practical “road-map” for integration of sustainability issues. Journal of Cleaner production, 15(7), 638-649. |
| 6 | Xu, T., Liu, W., & Zhu, Z. (2023). Multi-criteria Decision-making Methods Applied to Achieve Sustainable Design: A Systematic Review. Proceedings of the Design Society, 3, 2725-2734. |
| 7 | Mallalieu, A. M., Jonasson, A., Petersson, S., Rosendal, M., Hallstedt, S. I., Almefelt, L., & Isaksson, O. (2024). Sustainability criteria for introducing new technologies in low-income contexts. Proceedings of the Design Society, 4, 1359-1368. |

**Supplementary_3**

| No | Domain | Definition | | References |
|---|---|---|---|---|
| 1 | Education | Inclusion | Advocates of inclusion have responded by arguing that if such negative social outcomes are produced then the process involved cannot be inclusion | Frederickson,N.,Simmonds, E., Evans, L., & Soulsby, C. (2007). Assessing the social and affective outcomes of inclusion. British Journal of Special Education, 34(2), 105-115. |
| 2 | Education | Inclusion | inclusion, as opposed to ‘poor integration’ | Frederickson,N.,Simmonds, E., Evans, L., & Soulsby, C. (2007). Assessing the social and affective outcomes of inclusion. British Journal of Special Education, 34(2), 105-115. |
| 3 | Education | Inclusion | ‘the concept of inclusion must embrace the feeling of belonging, since such a feeling appears to be necessary both for successful learning and for more general well-being’. (Warnock, 2005, p. 15) | Frederickson,N.,Simmonds, E., Evans, L., & Soulsby, C. (2007). Assessing the social and affective outcomes of inclusion. British Journal of Special Education, 34(2), 105-115. |
| 4 | Management | Inclusion | inclusion remains a new concept without consensus on the nature of this construct or its theoretical underpinnings. | Shore, L. M., Randel, A. E., Chung, B. G., Dean, M. A., Holcombe Ehrhart, K., & Singh, G. (2011). Inclusion and diversity in work groups: A review and model for future research. Journal of management, 37(4), 1262-1289. |
| 5 | Management | Inclusion | we present a 2 × 2 framework of inclusion in which we propose that uniqueness and belongingness work together to create feelings of inclusion. | Shore, L. M., Randel, A. E., Chung, B. G., Dean, M. A., Holcombe Ehrhart, K., & Singh, G. (2011). Inclusion and diversity in work groups: A review and model for future research. Journal of management, 37(4), 1262-1289. |
| 6 | Management | Inclusion | uniqueness will provide opportunities for improved group performance when a unique individual is an accepted member of the group and the group values the particular unique characteristic (“Inclusion” cell in Figure 1). | Shore, L. M., Randel, A. E., Chung, B. G., Dean, M. A., Holcombe Ehrhart, K., & Singh, G. (2011). Inclusion and diversity in work groups: A review and model for future research. Journal of management, 37(4), 1262-1289. |
| 7 | Management | Inclusion | Individual is treated as an insider and also allowed/encouraged to retain uniqueness within the work group. | Shore, L. M., Randel, A. E., Chung, B. G., Dean, M. A., Holcombe Ehrhart, K., & Singh, G. (2011). Inclusion and diversity in work groups: A review and model for future research. Journal of management, 37(4), 1262-1289. |
| 8 | Management | Inclusion | Pelled, Ledford, and Mohrman (1999: 1014) defined inclusion as “the degree to which an employee is accepted and treated as an insider by others in a work system.” | Shore, L. M., Randel, A. E., Chung, B. G., Dean, M. A., Holcombe Ehrhart, K., & Singh, G. (2011). Inclusion and diversity in work groups: A review and model for future research. Journal of management, 37(4), 1262-1289. |
| 9 | Management | Inclusion | Roberson (2006: 217) argued that inclusion refers to “the removal of obstacles to the full participation and contribution of employees in organizations,” | Shore, L. M., Randel, A. E., Chung, B. G., Dean, M. A., Holcombe Ehrhart, K., & Singh, G. (2011). Inclusion and diversity in work groups: A review and model for future research. Journal of management, 37(4), 1262-1289. |
| 10 | Management | Inclusion | Miller (1998: 151) similarly described inclusion as the extent to which diverse individuals “are allowed to participate and are enabled to contribute fully.” | Shore, L. M., Randel, A. E., Chung, B. G., Dean, M. A., Holcombe Ehrhart, K., & Singh, G. (2011). Inclusion and diversity in work groups: A review and model for future research. Journal of management, 37(4), 1262-1289. |
| 11 | Management | Inclusion | Lirio, Lee, Williams, Haugen, and Kossek (2008: 443) referred to inclusion as "when individuals feel a sense of belonging, and inclusive behaviors such as eliciting and valuing contributions from all employees are part of the daily life in the organization." | Shore, L. M., Randel, A. E., Chung, B. G., Dean, M. A., Holcombe Ehrhart, K., & Singh, G. (2011). Inclusion and diversity in work groups: A review and model for future research. Journal of management, 37(4), 1262-1289. |
| 12 | Management | Inclusion | Avery, McKay, Wilson, and Volpone (2008: 6) stated that inclusion is "the extent to which employees believe their organizations engage in efforts to involve all employees in the mission and operation of the organization with respect to their individual talents." | Shore, L. M., Randel, A. E., Chung, B. G., Dean, M. A., Holcombe Ehrhart, K., & Singh, G. (2011). Inclusion and diversity in work groups: A review and model for future research. Journal of management, 37(4), 1262-1289. |
| 13 | Management | Inclusion | Wasserman, Gallegos, and Ferdman (2008: 176) define a culture of inclusion as existing when "people of all social<br>identity groups [have] the opportunity to be present, to have their voices heard and appreciated, and to engage in core activities on behalf of the collective." | Shore, L. M., Randel, A. E., Chung, B. G., Dean, M. A., Holcombe Ehrhart, K., & Singh, G. (2011). Inclusion and diversity in work groups: A review and model for future research. Journal of management, 37(4), 1262-1289. |
| 14 | Management | Inclusion | Holvino, Ferdman, and Merrill-Sands (2004: 249) define a multicultural, inclusive organization as "one in which the diversity of knowledge and perspectives that members of different groups bring to the organization has shaped its strategy, its work, its management and operating systems, and its core values and norms for success." | Shore, L. M., Randel, A. E., Chung, B. G., Dean, M. A., Holcombe Ehrhart, K., & Singh, G. (2011). Inclusion and diversity in work groups: A review and model for future research. Journal of management, 37(4), 1262-1289. |
| 15 | Management | Inclusion | Pelled and colleagues (1999) focused on three practices as indicators of inclusion: decision-making influence, access to sensitive work information, and job security. | Shore, L. M., Randel, A. E., Chung, B. G., Dean, M. A., Holcombe Ehrhart, K., & Singh, G. (2011). Inclusion and diversity in work groups: A review and model for future research. Journal of management, 37(4), 1262-1289. |
| 16 | Management | Inclusion | First, practices that are associated with insider status, including sharing information, participation in decision making, and having voice, are reflected in measures of inclusion. While these practices are assumed to enhance employee perceptions of inclusion, more explicit theorizing about the psychological mechanisms that underlie this link is needed. Second, there is a clear theme that inclusion has positive consequences for individuals and organizations, but as yet, little is known about how or why this occurs. This last point suggests that, fundamentally, the inclusion construct and its underlying theoretical basis need greater development. | Shore, L. M., Randel, A. E., Chung, B. G., Dean, M. A., Holcombe Ehrhart, K., & Singh, G. (2011). Inclusion and diversity in work groups: A review and model for future research. Journal of management, 37(4), 1262-1289. |
| 17 | Management | Inclusion | social exchange theory (Blau, 1964), provide a basis for making predictions about the effects of inclusion. Justice is associated with high-quality social exchange relationships (Masterson, Lewis, Goldman, & Taylor, 2000; Moorman,<br>Blakely, & Niehoff, 1998; Wayne, Shore, Bommer, & Tetrick, 2002) that involve mutual investment by both parties and concern for the interests of the other party in the relationship (Shore, Tetrick, Lynch, & Barksdale, 2006).<br>trust is an important underlying mechanism of social exchange (Konovsky & Pugh, 1994; Shore et al., 2006), | Shore, L. M., Randel, A. E., Chung, B. G., Dean, M. A., Holcombe Ehrhart, K., & Singh, G. (2011). Inclusion and diversity in work groups: A review and model for future research. Journal of management, 37(4), 1262-1289. |
| 18 | Management | Inclusion | more extensive research testing the effects of inclusion on an individual's well-being (e.g., stress, health) is likely to be a fruitful endeavor. There is some evidence to support the relationship between one's feeling of inclusion by others in the organization and psychological well-being (Mor Barak et al., 1998; Shaufeli, van Dierdendonck, & van Gorp, 1996) | Shore, L. M., Randel, A. E., Chung, B. G., Dean, M. A., Holcombe Ehrhart, K., & Singh, G. (2011). Inclusion and diversity in work groups: A review and model for future research. Journal of management, 37(4), 1262-1289. |
| 19 | Management | Inclusion | Another important potential outcome of inclusion is career opportunities for diverse people. | Shore, L. M., Randel, A. E., Chung, B. G., Dean, M. A., Holcombe Ehrhart, K., & Singh, G. (2011). Inclusion and diversity in work groups: A review and model for future research. Journal of management, 37(4), 1262-1289. |
| 20 | Finance | Inclusion | Social inclusion is the process of ensuring that all members of society have equal access to the same opportunities (Oxoby, 2009; Martin and Cobigo, 2011; Silver, 2010). | Ozili, P. K. (2020). Social inclusion and financial inclusion: international evidence. International Journal of Development Issues, 19(2), 169-186. |
| 21 | Finance | Inclusion | Social inclusion policies and institutions are interventions that promote full participation in society by all members of society (Collins, 2003), by eliminating the barriers that prevent individuals from fully participating in society in a meaningful way (Percy-Smith, 2000; Marston and Dee, 2015). | Ozili, P. K. (2020). Social inclusion and financial inclusion: international evidence. International Journal of Development Issues, 19(2), 169-186. |
| 22 | Finance | Inclusion | Some barriers or factors affecting the rate of social inclusion include: the different interpretation of social inclusion (Littlewood et al., 2017), lack of funds to finance social enterprises development (Biancone and Radwan, 2018), lack of community enterprises (Barraket and Archer, 2010) and a weak social inclusion model in several countries (Daly, 2008). | Ozili, P. K. (2020). Social inclusion and financial inclusion: international evidence. International Journal of Development Issues, 19(2), 169-186. |
| 23 | Finance | Inclusion | Several indicators of social inclusion have been identified in the policy and academic literature such as gender equality, equity in the use of public resources, building human resources, social protection, discrimination, environmental sustainability and social technology, etc. (World bank, 2014; Warschauer, 2004; Griessler and Littig, 2005) and there is currently no consensus in the literature on which social inclusion indicators reflect the actual social inclusion level in a country (Atkinson et al.,2004). | Ozili, P. K. (2020). Social inclusion and financial inclusion: international evidence. International Journal of Development Issues, 19(2), 169-186. |
| 24 | Finance | Inclusion | the reason for focusing on the three measures in this study (gender equality, environmental sustainability and social protection) is mainly because they are a recurring issue in modern public life and because data is available for these social inclusion indicators, compared to the broader measures of social inclusion for which data is scarcely available. | Ozili, P. K. (2020). Social inclusion and financial inclusion: international evidence. International Journal of Development Issues, 19(2), 169-186. |
| 25 | Finance | Inclusion | Financial inclusion is the process of ensuring that all individuals have access to basic financial services through their participation in the formal financial sector (Ozili, 2018). | Ozili, P. K. (2020). Social inclusion and financial inclusion: international evidence. International Journal of Development Issues, 19(2), 169-186. |
| 26 | Finance | Inclusion | One merit of financial inclusion is that it can increase the number of account ownership and increase access to credit for individuals so that people can have money to spend on consumption, savings, education and health care for their families (Allen et al.,2016). | Ozili, P. K. (2020). Social inclusion and financial inclusion: international evidence. International Journal of Development Issues, 19(2), 169-186. |
| 27 | Finance | Inclusion | some factors can promote financial inclusion such as: proximity to a microfinance institution (Brown et al., 2015), the level of education, income and age (Tuesta et al.,2015), financial literacy (Grohmann et al., 2018; Ozili, 2020a, 2020b), financial innovations (Yawe and Prabhu, 2015; Shen et al., 2019), institutional regulation (Chen and Divanbeigi, 2019) and regulatory support for the development and growth of social enterprises (Wilson, 2012). | Ozili, P. K. (2020). Social inclusion and financial inclusion: international evidence. International Journal of Development Issues, 19(2), 169-186. |
| 28 | Finance | Inclusion | social inclusion policies can contribute to financial inclusion by establishing social enterprises or institutions that<br>promote gender equality, anti-discrimination and environmental sustainability so that access to, and the delivery of, financial products and services to the poor and low-income individuals is not influenced by social discrimination, gender inequality and other bias in society. | Ozili, P. K. (2020). Social inclusion and financial inclusion: international evidence. International Journal of Development Issues, 19(2), 169-186. |
| 29 | Finance | Inclusion | financial inclusion policies are monitored to ensure that financial inclusion policies not lead to unintended social consequences, that lead to social exclusion. | Ozili, P. K. (2020). Social inclusion and financial inclusion: international evidence. International Journal of Development Issues, 19(2), 169-186. |
| 30 | Finance | Inclusion | in developing countries, social inclusion is often achieved through the means of financial inclusion policies or digital financial inclusion such as mobile phone penetration, increased use of digital finance apps and bank apps and the emergence of fintech, etc, which is not only promoted by development non-governmental organisations and the state but also by profit-driven multi-national financial institutions. | Ozili, P. K. (2020). Social inclusion and financial inclusion: international evidence. International Journal of Development Issues, 19(2), 169-186. |
| 31 | Finance | Inclusion | in financially inclusive societies, individuals both poor and rich individuals may have fears, worries and challenges and these anxieties can make individuals reduce their rate of participation in society, leading to social exclusion (Fraioli, 2012). | Ozili, P. K. (2020). Social inclusion and financial inclusion: international evidence. International Journal of Development Issues, 19(2), 169-186. |
| 32 | Social Psychology | Inclusion | inclusion is seen as the satisfaction of individual needs within a group. | Jansen, W. S., Otten, S., Van der Zee, K. I., & Jans, L. (2014). Inclusion: Conceptualization and measurement. European journal of social psychology, 44(4), 370-385. |
| 33 | Social Psychology | Inclusion | inclusion consists of two components: belonging<br>ness and uniqueness. | Jansen, W. S., Otten, S., Van der Zee, K. I., & Jans, L. (2014). Inclusion: Conceptualization and measurement. European journal of social psychology, 44(4), 370-385. |
| 34 | Social Psychology | Inclusion | the definition by Shore et al. (2011), inclusion is established when individuals have a sense of belonging to the group and, at the same time, perceive themselves to be a distinct and unique individual. | Jansen, W. S., Otten, S., Van der Zee, K. I., & Jans, L. (2014). Inclusion: Conceptualization and measurement. European journal of social psychology, 44(4), 370-385. |
| 35 | Social Psychology | Inclusion | inclusion should be measured with items in which the group is defined as the source and the individual as the target of inclusion | Jansen, W. S., Otten, S., Van der Zee, K. I., & Jans, L. (2014). Inclusion: Conceptualization and measurement. European journal of social psychology, 44(4), 370-385. |
| 36 | Education | Inclusion | Inclusion expresses a vision of the direction in which the political level wants to develop the institutions of the welfare state, such as public schools. | Hansen, J. H. (2012). Limits to inclusion. International Journal of Inclusive Education, 16(1), 89-98. |
| 37 | Education | Inclusion | inclusion needs to be based on a specific pedagogical understanding: that participation in communities is a prerequisite for development and learning. | Hansen, J. H. (2012). Limits to inclusion. International Journal of Inclusive Education, 16(1), 89-98. |
| 38 | Education | Inclusion | the determination of inclusion and inclusive processes do not provide a theoretical or conceptual understanding of this limit. In other words, the empirically evidenced limit to inclusion is not theoretically determined in the conceptualisation of inclusion and the exploration of inclusive processes. | Hansen, J. H. (2012). Limits to inclusion. International Journal of Inclusive Education, 16(1), 89-98. |
| 39 | Education | Inclusion | a vision, and the main goal is to secure the participation of all children while accepting that the vision never can be fully realised, because it is a vision. | Hansen, J. H. (2012). Limits to inclusion. International Journal of Inclusive Education, 16(1), 89-98. |
| 40 | Education | Inclusion | it is argued that inclusion is not a condition, but a process that never ends. By understanding inclusion as a process, it is possible to focus on intervention forms, which are increasingly helping to realise the vision, knowing that it can never be fully realised. | Hansen, J. H. (2012). Limits to inclusion. International Journal of Inclusive Education, 16(1), 89-98. |
| 41 | Education | Inclusion | simply accepting that inclusion has a limit in practice and then accepting without question that teachers sometimes judge that it is not beneficial to a specific child's learning and development to participate in the classroom. | Hansen, J. H. (2012). Limits to inclusion. International Journal of Inclusive Education, 16(1), 89-98. |
| 42 | Education | Inclusion | justifying the limit to inclusion in practice as the professional's own lack of inclusiveness (Galloway et al 1979, Sakellariadis 2008). | Hansen, J. H. (2012). Limits to inclusion. International Journal of Inclusive Education, 16(1), 89-98. |
| 43 | Sociology and Anthropology | Inclusion | Where a conceptualization of inclusion does appear in the social exclusion literature, it is often only indirect.<br>Frequently, for example, it appears in invocations of 'normal' social expectation/participation or, more commonly,' mainstream' applied to various things that people are understood to be excluded from: labor market, economy, society, culture, citizenship, etc. The meaning and location of the mainstream is routinely taken to be self-evident. As this implies, social inclusion is most commonly defined only negatively – as whatever is not socially excluded. For this reason, much of the discussion of social inclusion is conceptually dominated by exclusion – social exclusion is the datum point against which social inclusion is both empirically measured and conceptually defined (Cameron, 2006:397) | Rawal, N. (2008). Social inclusion and exclusion: A review. Dhaulagiri Journal of Sociology and Anthropology, 2, 161-180. |
| 44 | Education | Inclusion | Participation practices entail efforts to increase public input oriented primarily to the content of programs and policies. Inclusion practices entail continuously creating a community involved in coproducing processes, policies, and programs for defining and addressing public issues. | Quick, K. S., & Feldman, M. S. (2011). Distinguishing participation and inclusion. Journal of planning education and research, 31(3), 272-290. |
| 45 | Education | Inclusion | inclusion and participation are two different dimensions of public engagement and that organizing public management to incorporate both enhances the quality of the decisions reached and the community's long-term capacities. Specifically, participation is oriented to increasing input for decisions. | Quick, K. S., & Feldman, M. S. (2011). Distinguishing participation and inclusion. Journal of planning education and research, 31(3), 272-290. |
| 46 | Education | Inclusion | Inclusion is oriented to making connections among people, across issues, and over time. It is an expansive and ongoing framework for interaction that uses the opportunities to take action on specific items in the public domain as a means of intentionally creating a community engaged in an ongoing stream of issues. | Quick, K. S., & Feldman, M. S. (2011). Distinguishing participation and inclusion. Journal of planning education and research, 31(3), 272-290. |
| 47 | Education | Inclusion | inclusion to describe connections made not only among individuals' and groups' points of view but connections across issues, sectors, and engagement efforts. | Quick, K. S., & Feldman, M. S. (2011). Distinguishing participation and inclusion. Journal of planning education and research, 31(3), 272-290. |
| 48 | Education | Inclusion | Inclusive practices display the following features: engaging multiple ways of knowing, coproducing the process and content of decision making, and sustaining temporal openness. | Quick, K. S., & Feldman, M. S. (2011). Distinguishing participation and inclusion. Journal of planning education and research, 31(3), 272-290. |
| 49 | Education | Inclusion | Inclusive practices allow participants to experience the creation of a problem-solving community as well as the accomplishment of specific tasks or goals, resulting in a greater sense of satisfaction. | Quick, K. S., & Feldman, M. S. (2011). Distinguishing participation and inclusion. Journal of planning education and research, 31(3), 272-290. |
| 50 | Education | Inclusion | inclusion—making connections and building communities— and other understandings of inclusion in engagement. | Quick, K. S., & Feldman, M. S. (2011). Distinguishing participation and inclusion. Journal of planning education and research, 31(3), 272-290. |
| 51 | Education | Inclusion | An inclusive “way of knowing” a public policy problem is an ongoing accomplishment that must be sustained through the “continuous renewal of associations” among parties and perspectives (Feldman et al. 2006). | Quick, K. S., & Feldman, M. S. (2011). Distinguishing participation and inclusion. Journal of planning education and research, 31(3), 272-290. |
| 52 | Education | Inclusion | inclusive practices involve creating community through sharing practices, bringing together what in other contexts might be different “cores”—such as different sectors or types of expertise—and creating together a moving, changing combination of them. To accomplish this, organizers of public involvement manage these processes in ways that allow participants to coproduce the practices through which they develop ways of addressing issues and become a community in which these practices take place. | Quick, K. S., & Feldman, M. S. (2011). Distinguishing participation and inclusion. Journal of planning education and research, 31(3), 272-290. |
| 53 | Education | Inclusion | “inclusion builds community” | Quick, K. S., & Feldman, M. S. (2011). Distinguishing participation and inclusion. Journal of planning education and research, 31(3), 272-290. |
| 54 | Education | Inclusion | Paradoxically, the cohesion of an inclusive community depends on its not being a static collection of persons or practices. | Quick, K. S., & Feldman, M. S. (2011). Distinguishing participation and inclusion. Journal of planning education and research, 31(3), 272-290. |
| 55 | Education | Inclusion | Inclusion, as a value, supports the right of all children, regardless of their diverse abilities, to participate actively in natural settings within their communities. A natural setting is one in which the child would spend time had he or she not had a disability. Such settings include but are not limited to home and family, play groups, child care, nursery schools, Head Start programs, kindergartens, and neighborhood school classrooms. | Bricker, D. (1995). The challenge of inclusion. Journal of early intervention, 19(3), 179-194. |
| 56 | Management | Exclusion | low-belongingness/low-uniqueness combination that we have labeled exclusion. | Shore, L. M., Randel, A. E., Chung, B. G., Dean, M. A., Holcombe Ehrhart, K., & Singh, G. (2011). Inclusion and diversity in work groups: A review and model for future research. Journal of management, 37(4), 1262-1289. |
| 57 | Management | Exclusion | Individual is not treated as an organizational insider with unique value in the work group but there are other employees or groups who are insiders. | Shore, L. M., Randel, A. E., Chung, B. G., Dean, M. A., Holcombe Ehrhart, K., & Singh, G. (2011). Inclusion and diversity in work groups: A review and model for future research. Journal of management, 37(4), 1262-1289. |
| 58 | Management | Exclusion | Mor Barak (2000: 52) stated that “employee perception of inclusion-exclusion is conceptualized as a continuum of the degree to which individuals feel a part of critical organizational processes. These processes include access to information and resources, connectedness to supervisor and co-workers, and ability to participate in and influence the decision making process.” | Shore, L. M., Randel, A. E., Chung, B. G., Dean, M. A., Holcombe Ehrhart, K., & Singh, G. (2011). Inclusion and diversity in work groups: A review and model for future research. Journal of management, 37(4), 1262-1289. |
| 59 | Management | Exclusion | perceptions of inclusion-exclusion, which would then lead to job satisfaction, organizational commitment, individual well-being, and task effectiveness. | Shore, L. M., Randel, A. E., Chung, B. G., Dean, M. A., Holcombe Ehrhart, K., & Singh, G. (2011). Inclusion and diversity in work groups: A review and model for future research. Journal of management, 37(4), 1262-1289. |
| 60 | Management | Exclusion | “Diversity climate is related to the inclusion or exclusion of people from diverse backgrounds (Mor Barak et al., 1998), and . . . to the justice-related events pertinent to the balance of power and relations across social groups (Kossek & Zonia, 1993).” | Shore, L. M., Randel, A. E., Chung, B. G., Dean, M. A., Holcombe Ehrhart, K., & Singh, G. (2011). Inclusion and diversity in work groups: A review and model for future research. Journal of management, 37(4), 1262-1289. |
| 61 | Management | Exclusion | inclusion is positively related to job satisfaction (Acquavita et al., 2009) and that exclusion from decision making is a predictor of intention to leave (Mor Barak et al., 2006) | Shore, L. M., Randel, A. E., Chung, B. G., Dean, M. A., Holcombe Ehrhart, K., & Singh, G. (2011). Inclusion and diversity in work groups: A review and model for future research. Journal of management, 37(4), 1262-1289. |
| 67 | Finance | Exclusion | low levels of social inclusion in a society can lead to increased distrust among members of the society, which can make people unwilling to deal with financial institutions or the financial services they offer, thereby leading to financial exclusion. Societal trust is the main driver of the positive association or correlation between social inclusion and financial inclusion. | Ozili, P. K. (2020). Social inclusion and financial inclusion: international evidence. International Journal of Development Issues, 19(2), 169-186. |
| 68 | Finance | Exclusion | in financially inclusive societies, individuals both poor and rich individuals may have fears, worries and challenges and these anxieties can make individuals reduce their rate of participation in society, leading to social exclusion (Fraioli, 2012). | Ozili, P. K. (2020). Social inclusion and financial inclusion: international evidence. International Journal of Development Issues, 19(2), 169-186. |
| 69 | Education | Exclusion | we cannot investigate inclusion without investigating exclusion. | Hansen, J. H. (2012). Limits to inclusion. International Journal of Inclusive Education, 16(1), 89-98. |
| 70 | Education | Exclusion | it is not possible to consider inclusion as limitless. If inclusion was considered as limitless, inclusion would ultimately lead to exclusion because an unlimited inclusive practice cannot deny exclusion by which the inclusive community may lead to an exclusive community. | Hansen, J. H. (2012). Limits to inclusion. International Journal of Inclusive Education, 16(1), 89-98. |
| 71 | Education | Exclusion | Inclusion and exclusion are two connected and interdependent processes. Exclusion makes inclusion possible and simultaneously makes inclusion impossible as a limitless condition. As a concept, inclusion therefore presupposes exclusion, because an inclusive society needs to include a certain degree of exclusion to ensure its own existence. | Hansen, J. H. (2012). Limits to inclusion. International Journal of Inclusive Education, 16(1), 89-98. |
| 72 | Education | Exclusion | we must focus on examining the boundary between inclusion and exclusion in the specific communities by questioning how this limit is constructed. | Hansen, J. H. (2012). Limits to inclusion. International Journal of Inclusive Education, 16(1), 89-98. |
| 73 | Education | Exclusion | we cannot consider inclusion in itself by excluding its other: exclusion. We can identify inclusion neither by defining a normative limit between inclusion and exclusion nor by avoiding limits and making inclusion unambiguous. Thus, it is not possible to put meaning into the concept of inclusion without its otherness, exclusion. | Hansen, J. H. (2012). Limits to inclusion. International Journal of Inclusive Education, 16(1), 89-98. |
| 74 | Sociology and Anthropology | Exclusion | social exclusion is more illuminating and holds the promise of understanding disadvantaged groups better, others argue that this concept is so evocative, ambiguous, multidimensional and elastic that it can be defined in many different ways and owing to its ambiguity in definition it may mean all things to all people. | Rawal, N. (2008). Social inclusion and exclusion: A review. Dhaulagiri Journal of Sociology and Anthropology, 2, 161-180. |
| 75 | Sociology and Anthropology | Exclusion | the meaning of social exclusion depends on the nature of the society, or the dominant model of the society from which exclusion occurs and it varies in meanings according to national and ideological contexts (Silver, 1994:539). | Rawal, N. (2008). Social inclusion and exclusion: A review. Dhaulagiri Journal of Sociology and Anthropology, 2, 161-180. |
| 76 | Sociology and Anthropology | Exclusion | Social exclusion has been defined as 'the process through which individuals or groups are wholly or partially excluded from full participation in the society within which they live' (European foundation, 1195, p.4, quoted in de Haan, 1998, cited in Francis, 2002). | Rawal, N. (2008). Social inclusion and exclusion: A review. Dhaulagiri Journal of Sociology and Anthropology, 2, 161-180. |
| 77 | Sociology and Anthropology | Exclusion | social exclusion as multidimensional phenomena and have considered several important living condition variables as proxies for social exclusion. They are: 1) Exclusion from formal citizenship rights: 2) Exclusion from labor market; 3) Exclusion from participation in civil society and 4) Exclusion from social arenas | Rawal, N. (2008). Social inclusion and exclusion: A review. Dhaulagiri Journal of Sociology and Anthropology, 2, 161-180. |
| 78 | Sociology and Anthropology | Exclusion | exclusion is a form of discrimination, which occurs when individuals are denied free movement and exchange between spheres, when rules inappropriate to a given sphere are enforced or when group boundaries impede individual freedom to participate in social exchanges. | Rawal, N. (2008). Social inclusion and exclusion: A review. Dhaulagiri Journal of Sociology and Anthropology, 2, 161-180. |
| 79 | Sociology and Anthropology | Exclusion | it would be wrong to treat the issue of exclusion in a simplistic manner or understand it through the binary opposition of exclusion/inclusion. | Rawal, N. (2008). Social inclusion and exclusion: A review. Dhaulagiri Journal of Sociology and Anthropology, 2, 161-180. |
| 80 | Psychology | Exclusion | vulnerable workers tend to be disproportionately excluded (Burgess and Connell, 2015) or even pushed out of the workforce (Guichard, 2009) since systemic imbalances in power and privilege discourage them from career paths (Blustein et al., 2014). | Di Fabio, A., & Svicher, A. (2021). The Psychology of Sustainability and Sustainable Development: Advancing decent work, inclusivity, and positive strength-based primary preventive interventions for vulnerable workers. Frontiers in Psychology, 12, 718354. |
| 81 | Management | Inclusive | Janssens and Zanoni (2007) concluded that inclusive work contexts tend to involve practices encouraging the same treatment of employees while simultaneously acknowledging individual differences, for example, recruitment of ethnic minorities based on individual capabilities rather than on ethnic membership; teams composed of different ethnicities performing jobs of the same status; and high task interdependence allowing for frequent, substantive communication among team members. | Shore, L. M., Randel, A. E., Chung, B. G., Dean, M. A., Holcombe Ehrhart, K., & Singh, G. (2011). Inclusion and diversity in work groups: A review and model for future research. Journal of management, 37(4), 1262-1289. |
| 82 | Business - Social Responsibility | Inclusive | The inclusive business model embeds its origin in the bottom of the pyramid theory (BOP) proposed by Prahalad (2005), which is based on the concept of ''serving the poor profitably'' (Prahalad and Hart, 2002; Prahalad and Hammond, 2002). | Michelini, L., & Fiorentino, D. (2012). New business models for creating shared value. Social Responsibility Journal, 8(4), 561-577. |
| 83 | Business - Social Responsibility | Inclusive | the inclusive business model is based (Marquez et al., 2010): the initiative must actively seek business opportunity with LIC (Lower-income communities); it must prove significant for the organization; and it has to generate both economic and social value. | Michelini, L., & Fiorentino, D. (2012). New business models for creating shared value. Social Responsibility Journal, 8(4), 561-577. |
| 84 | Business - Management | Inclusive | Inclusive Business entails creating a net positive development impact through an financially profitable business model. | Wach, E. (2012). Measuring the 'inclusivity'of inclusive business. IDS Practice Papers, 2012(9), 01-30. |
| 85 | Business - Management | Inclusive | Inclusive Business as 'profitable core business activity that also tangibly expands opportunities for the poor and disadvantaged in developing countries' (BIF 2011). | Wach, E. (2012). Measuring the 'inclusivity'of inclusive business. IDS Practice Papers, 2012(9), 01-30. |
| 86 | Business - Management | Inclusive | Inclusive Businesses make 'A positive contribution to the development of companies, the local population and<br>the environment' (Gradl and Knobloch 2010: 10). | Wach, E. (2012). Measuring the 'inclusivity'of inclusive business. IDS Practice Papers, 2012(9), 01-30. |
| 87 | Business - Management | Inclusive | Inclusive Business as 'An economically profitable, environmentally and socially responsible entrepreneurial initiative' (2011: 13). | Wach, E. (2012). Measuring the 'inclusivity'of inclusive business. IDS Practice Papers, 2012(9), 01-30. |
| 88 | Business - Management | Inclusive | 'Inclusive business integrates people living in poverty into the value chain as consumers or producers' (Gradl and Knobloch 2010: 10). | Wach, E. (2012). Measuring the 'inclusivity'of inclusive business. IDS Practice Papers, 2012(9), 01-30. |
| 89 | Business - Management | Inclusive | 'Engage the poor as employees, suppliers, distributors, or consumers and expand their economic opportunities in a wide variety of ways' (BIF 2011). | Wach, E. (2012). Measuring the 'inclusivity'of inclusive business. IDS Practice Papers, 2012(9), 01-30. |
| 90 | Business - Management | Inclusive | Inclusive Business models include the poor on the demand side as clients and customers, and on the supply side as employees, producers and business owners at various points in the value chain' (UNDP 2008: 2). | Wach, E. (2012). Measuring the 'inclusivity'of inclusive business. IDS Practice Papers, 2012(9), 01-30. |
| 91 | Business - Management | Inclusive | The concept of mutual benefit also recurs in definitions of Inclusive Business: UNDP states that Inclusive Business models 'Build bridges between business and the poor for mutual benefit' (UNDP 2008: 2), and WBCSD states that Inclusive Business models 'Integrate low-income communities in its value chain for the mutual benefit of both the company and the community' (WBCSD and SNV 2011:13). | Wach, E. (2012). Measuring the 'inclusivity'of inclusive business. IDS Practice Papers, 2012(9), 01-30. |
| 92 | Business - Management | Inclusive | one of the key characteristics of an 'Inclusive Business' model is profitability or mutual benefit (UNDP 2008; WBCSD 2011), businesses – and the development agencies supporting them – need to accept that, in order to ensure economic viability, there are limits to the development objectives they can achieve, and the negative impacts they can avoid. | Wach, E. (2012). Measuring the 'inclusivity'of inclusive business. IDS Practice Papers, 2012(9), 01-30. |
| 93 | Business - Management | Inclusive | 'Inclusive Business' models, it is necessary to (a) conduct a causal chain analysis to determine how the outputs of a business model are intended to link to desired outcomes, (b) capture all outputs (or as many as possible) of the proposed business model, both positive and negative, and (c) determine and make clear who makes the judgement as to what is 'inclusive'. | Wach, E. (2012). Measuring the 'inclusivity'of inclusive business. IDS Practice Papers, 2012(9), 01-30. |
| 94 | Business - Management | Inclusive | 'Inclusive Business' as a method of achieving development goals (Nelson and Prescott 2008), it is imperative that they have accurate information on which to base these decisions. | Wach, E. (2012). Measuring the 'inclusivity'of inclusive business. IDS Practice Papers, 2012(9), 01-30. |
| 95 | Education | Inclusive | Feldman and Khademian (2007) described inclusive practices as creating "communities of participation." | Quick, K. S., & Feldman, M. S. (2011). Distinguishing participation and inclusion. Journal of planning education and research, 31(3), 272-290. |
| 96 | Education | Inclusive | inclusive education should be understood in the context of an approach to the 'problems' of social diversity in societies that are highly diversified internally and yet globally interconnected. | Armstrong, D., Armstrong, A. C., & Spandagou, I. (2011). Inclusion: By choice or by chance?. International journal of inclusive education, 15(1), 29-39. |
| 97 | Education | Inclusive | Narrow definitions of inclusion refer to the promotion of the inclusion of specific group of students, mainly, but not exclusively, disabled students and/or students with special education needs in ‘mainstream’ or ‘regular’ education. ‘Broad’ definitions of inclusion, on the other hand, do not focus on specific groups of students, but rather on diversity and how schools respond to the diversity of all students (and even every other member of the school community). We can also add another dimension to this distinction, namely ‘fragmented’ definitions. Both narrow and broad definitions can be fragmented when they break down the group that they refer to. This way of distinguishing definitions of inclusion can be helpful when looking at what exactly policy documents propose. For example, the Ofsted Guidance for Evaluating Educational Inclusion (2000) states that:<br><br>An educationally inclusive school is one in which the teaching and learning, achievements, attitudes and well-being of every young person matter [...] This does not mean treating all pupils in the same way. Rather it involves taking account of pupils’ varied life experiences and needs. [...] They identify any pupils who may be missing out, difficult to engage, or feeling in some way to be apart from what the school seeks to provide. (Ofsted 2000, 7) | Armstrong, D., Armstrong, A. C., & Spandagou, I. (2011). Inclusion: By choice or by chance?. International journal of inclusive education, 15(1), 29-39. |
| 98 | Education | Inclusive | States Parties recognize the right of persons with disabilities to education.<br>With a view to realizing this right without discrimination and on the basis of equal opportunity, States Parties shall ensure an inclusive education system at all levels and lifelong learning. (UN 2006, 16, emphasis added) | Armstrong, D., Armstrong, A. C., & Spandagou, I. (2011). Inclusion: By choice or by chance?. International journal of inclusive education, 15(1), 29-39. |
| 99 | Education | Inclusive | ‘[i]nclusion and participation are essential to human dignity and to the exercise and enjoyment of human rights’ (11). | Armstrong, D., Armstrong, A. C., & Spandagou, I. (2011). Inclusion: By choice or by chance?. International journal of inclusive education, 15(1), 29-39. |
| 100 | Education | Inclusive | Regular schools with this inclusive orientation are the most effective means of combating discriminatory attitudes, creating welcoming communities, building an inclusive society and achieving education for all; moreover, they provide an effective education to the majority of children and improve the efficiency and ultimately the cost effectiveness of the entire education system. (UNESCO 1994, 3) | Armstrong, D., Armstrong, A. C., & Spandagou, I. (2011). Inclusion: By choice or by chance?. International journal of inclusive education, 15(1), 29-39. |
| 101 | Education | Inclusive | According to UNESCO (2001) achieving education for all will require sustained, intensive and co-ordinated action on several fronts. Transforming resource inputs into learning outcomes requires not just financial investment but also effective education systems, the right mix of resources (e.g. teachers and learning materials) and an overall national context of sound economic and social policies. (6) | Armstrong, D., Armstrong, A. C., & Spandagou, I. (2011). Inclusion: By choice or by chance?. International journal of inclusive education, 15(1), 29-39. |
| 102 | Education | Inclusive | ‘Inclusion’ cannot simply be constructed as the opposite of ‘exclusion’. Inclusion and exclusion are interrelated processes and their interplay constantly creates new inclusive/exclusive conditions and possibilities. | Armstrong, D., Armstrong, A. C., & Spandagou, I. (2011). Inclusion: By choice or by chance?. International journal of inclusive education, 15(1), 29-39. |
| 103 | Psychology | Inclusivity | A formal definition and four evaluative constructs of psychosocial inclusivity in design are described: Cognitive, Emotional, Social and Value. | Lim, Y., Giacomin, J., & Nickpour, F. (2021). What is psychosocially inclusive design? A definition with constructs. The Design Journal, 24(1), 5-28. |
| 104 | Design/ Engineering | Inclusive Design | Inclusive design meets the needs of people of all abilities. It extends and preserves independence but it need not be expensive. For products to be inclusively designed, all stakeholders need to be informed and active: demanding consumers, influential consumer organisations, creative designers and design colleges, even reluctant manufacturers, legislators and standards-makers. | Etchell, L., & Yelding, D. (2004). Inclusive design: products for all consumers. Consumer Policy Review, 14(6), 186-193. |
| 105 | Design/ Engineering | Inclusive Design | Inclusive design at its simplest means designing for as many people as possible, taking into account the diversity of their abilities. | Etchell, L., & Yelding, D. (2004). Inclusive design: products for all consumers. Consumer Policy Review, 14(6), 186-193. |
| 106 | Design/ Engineering | Inclusive Design | Inclusive design should mean that the need for assistive products or adaptations is reduced. | Etchell, L., & Yelding, D. (2004). Inclusive design: products for all consumers. Consumer Policy Review, 14(6), 186-193. |
| 107 | Design/ Engineering | Inclusive Design | Inclusive design is a design philosophy that aims to consider the needs of people with reduced functional capability during the design process. The goal is to design consumer products that are accessible to as many people as possible, without being stigmatising or resorting to special aids and adaptations [25]. | Persad, U., Langdon, P., & Clarkson, J. (2007). Characterising user capabilities to support inclusive design evaluation. Universal Access in the Information Society, 6, 119-135. |
| 108 | Design/ Engineering | Inclusive Design | Inclusive design rejects the notion of the 'average user' as a myth, and embraces diversity in users' sensory, cognitive and motor characteristics. Though ageing, disease and trauma account for this diversity, the characterisations of users based on levels of capability to perform actions in the world appear to be the most appropriate for product design and evaluation [9, 21, 52]. | Persad, U., Langdon, P., & Clarkson, J. (2007). Characterising user capabilities to support inclusive design evaluation. Universal Access in the Information Society, 6, 119-135. |
| 109 | Design/ Engineering | Inclusive Design | In order to predict exclusion, it is necessary to determine if a task demands fall outside the user's performance envelope as previously described. | Persad, U., Langdon, P., & Clarkson, J. (2007). Characterising user capabilities to support inclusive design evaluation. Universal Access in the Information Society, 6, 119-135. |
| 110 | Design/ Engineering | Inclusive Design | inclusive design considers the needs and capabilities of the whole population and is based on the assumption that considering the full diversity of users will result in a better product (Johnson, Clarkson, & Huppert, 2010). | Patrick, V. M., & Hollenbeck, C. R. (2021). Designing for all: Consumer response to inclusive design. Journal of consumer psychology, 31(2), 360-381. |
| 111 | Design/ Engineering | Inclusive Design | Inclusive design does not aim to change, but rather to improve the design of products and environments by making them more usable, safe, and appealing to people with a wide range of abilities (Carroll & Kincade, 2007). | Patrick, V. M., & Hollenbeck, C. R. (2021). Designing for all: Consumer response to inclusive design. Journal of consumer psychology, 31(2), 360-381. |
| 112 | Design/ Engineering | Inclusive Design | Inclusive design, according to design expert Kat Holmes, is simply the opposite of exclusion. | Patrick, V. M., & Hollenbeck, C. R. (2021). Designing for all: Consumer response to inclusive design. Journal of consumer psychology, 31(2), 360-381. |
| 113 | Design/ Engineering | Inclusive Design | The implementation of inclusive design needs adherence to the following five basic principles (adapted from CABE 2008) in order to provide all people with dignity, comfort, and convenience, as well as the ability to be independent and participate equally in activities:<br>• People—Place people at the heart of the design process.<br>• Diversity—Acknowledge diversity and individual difference.<br>• Choices—Offer choices where a single design solution cannot accommodate all users.<br>• Flexibility—Ensure flexibility in use.<br>• Convenience—Ensure that design products are convenient and enjoyable to use for everyone. | Patrick, V. M., & Hollenbeck, C. R. (2021). Designing for all: Consumer response to inclusive design. Journal of consumer psychology, 31(2), 360-381. |
| 114 | Design/ Engineering | Inclusive Design | we use the term inclusive design and conceptualize it as the actual or perceived match between the user and the design object. We propose that when the identity of the user actually does not—or feels as if it does not—map onto the form and function of the design object, the user is excluded. As such, inclusive design makes products, services, spaces, and experiences not only effective and easy to use, but also emotionally positive, accessible to, and usable by people with the widest range of abilities within the widest range of situations without the need for special adaptation or accommodation. In sum, inclusive design does not reject users because of who they are, but instead works with excluded users to create solutions that work well and benefit everyone. | Patrick, V. M., & Hollenbeck, C. R. (2021). Designing for all: Consumer response to inclusive design. Journal of consumer psychology, 31(2), 360-381. |
| 115 | Design/ Engineering | Inclusive Design | identity is multidimensional (Oyserman & Schwarz, 2020) and includes a combination of age, developmental and acquired disabilities, religion, ethnicity, socioeconomic status, sexual orientation, indigenous heritage, national origin, and gender, we use the ADDRESSING framework, first defined by Hays (2008), to capture the full range of identity considerations that can inform inclusive design (see Table 2). The ADDRESSING framework can help designers understand the scope of possible users and the complexities of human experience and identity (Hays, 1996, 2008). | Patrick, V. M., & Hollenbeck, C. R. (2021). Designing for all: Consumer response to inclusive design. Journal of consumer psychology, 31(2), 360-381. |
| 116 | Design/ Engineering | Inclusive Design | inclusive design does not focus solely on a process, the designer, the user, or adapt ability for the sake of profits alone; rather, inclusivity involves a moral responsibility and an ethical commitment to decreasing the mismatch between a user and the design object. | Patrick, V. M., & Hollenbeck, C. R. (2021). Designing for all: Consumer response to inclusive design. Journal of consumer psychology, 31(2), 360-381. |
| 117 | Design/ Engineering | Inclusive Design | a three-dimensional framework for inclusive design: (a) Recognize, respect, and design for human uniqueness and variability; (b) Use inclusive, open, and transparent processes, and codesign with people who hold a diversity of perspectives, including those who can't use or have difficulty using the current designs; and (c) Realize that you are designing in a complex adaptive system (Treviranus, 2018). | Patrick, V. M., & Hollenbeck, C. R. (2021). Designing for all: Consumer response to inclusive design. Journal of consumer psychology, 31(2), 360-381. |
| 118 | Design/ Engineering | Inclusive Design | Inclusive design demands diverse teams that are committed to empathizing with consumers and exploring all the elements and possibilities that make up an activity (Etchell & Yelding, 2004). | Patrick, V. M., & Hollenbeck, C. R. (2021). Designing for all: Consumer response to inclusive design. Journal of consumer psychology, 31(2), 360-381. |
| 119 | Design/ Engineering | Inclusive Design | Mismatches between products and their users represent the "building blocks of exclusion. They can feel like  little moments of exasperation when a technology product doesn't work the way we think it should. Or they can feel like running into a locked door marked with a big sign that says Keep Out" (Holmes, p. 2). | Patrick, V. M., & Hollenbeck, C. R. (2021). Designing for all: Consumer response to inclusive design. Journal of consumer psychology, 31(2), 360-381. |
| 120 | Design/ Engineering | Inclusive Design | inclusivity is about creating the best possible match between the environment and all potential users, recognizing that inclusion and beautification improve the overall experience for everyone. Indeed, numerous inventions that we all use today are the direct or indirect result of inclusive design: touchscreens, flexible straws, closed captioning, automatic doors, curb cuts, audiobooks,  and speech-to-text software to name a few. | Patrick, V. M., & Hollenbeck, C. R. (2021). Designing for all: Consumer response to inclusive design. Journal of consumer psychology, 31(2), 360-381. |
| 121 | Design/ Engineering | Inclusive Design | the best inclusive designs are user-led and aim for usability by the greatest number of people, including those with disabilities. | Patrick, V. M., & Hollenbeck, C. R. (2021). Designing for all: Consumer response to inclusive design. Journal of consumer psychology, 31(2), 360-381. |
| 122 | Design/ Engineering | Inclusive Design | conceptualizing inclusivity using three levels (see Figure 2) based on the degree of mismatch between the individual and the designed object (e.g., a product, service, or space). Level 1 Inclusive Design merely addresses accessibility, typically reflecting industry regulations. Level 2 Inclusive Design incorporates greater engagement and evokes positive emotions from consumers, while Level 3 Inclusive Design represents the ideal wherein there is little or no mismatch between any consumer and the designed object. | Patrick, V. M., & Hollenbeck, C. R. (2021). Designing for all: Consumer response to inclusive design. Journal of consumer psychology, 31(2), 360-381. |
| 123 | Design/ Engineering | Inclusive Design | In particular, where some users are excluded from using a product or service, many more are likely to find it difficult or frustrating to use (Clarkson et al., 2003). Hence, inclusive design has become synonymous with good design, where accessibility, as an extension of usability, can lead to increased commercial success. | Langdon, P., Johnson, D., Huppert, F., & Clarkson, P. J. (2015). A framework for collecting inclusive design data for the UK population. Applied ergonomics, 46, 318-324. |
| 124 | Design/ Engineering | Inclusive Design | barriers remain to industry uptake of the ethos of inclusive design (Dong et al., 2004) in the form of: the lack of a perceived justifiable business case to support inclusive design; the lack of knowledge and tools to practice inclusive design; a lack of understanding of the difficulties experienced by users of new technology products; poor access to appropriate user sets. | Langdon, P., Johnson, D., Huppert, F., & Clarkson, P. J. (2015). A framework for collecting inclusive design data for the UK population. Applied ergonomics, 46, 318-324. |
| 125 | Design/ Engineering | Inclusive Design | an inclusive design approach for bringing disabled people into mainstream life by improving product design. In order to predict product exclusion and difficulty, recent accurate data on capability in the population, from post-census, disability surveys, has been combined with models of human product interaction. The research to do this has generated new models and methods for assessing levels of design inclusion. | Langdon, P., Johnson, D., Huppert, F., & Clarkson, P. J. (2015). A framework for collecting inclusive design data for the UK population. Applied ergonomics, 46, 318-324. |
| 126 | Design/ Engineering | Inclusive Design | There is an increasingly urgent need to design products to accommodate the rising prevalence of functional impairments, both at more extreme levels,such as dementia,and at the extension to the mainstream, for example, in elderly friendly ICT. Analytical inclusive design is nowsupported bya plethora of tools, techniques and resources but there is a dearth of data sets that can support quantitative prediction of performance resulting from the way in which capability interacts with the demand made by product features. | Langdon, P., Johnson, D., Huppert, F., & Clarkson, P. J. (2015). A framework for collecting inclusive design data for the UK population. Applied ergonomics, 46, 318-324. |
| 127 | Design/ Engineering | Inclusive Design | "Inclusive Design as neither a new genre of design, nor a separate specialism, but as a general approach to designing in which designers ensure that their products and services address the needs of the widest possible audience, irrespective of age or ability. Inclusive Design (also known [in Europe] as Design for All and as Universal Design in the USA) is in essence the inverse of earlier approaches to designing for disabled and elderly people as a sub-set of the population, and an integral part of a more recent international trend towards the integration of older and disabled people in the mainstream of society. " (Design Council, 2008). | Clarkson, P. J., & Coleman, R. (2015). History of inclusive design in the UK. Applied ergonomics, 46, 235-247. |
| 128 | Design/ Engineering | Inclusive Design | Inclusive Design is in essence the inverse of earlier approaches to designing for disabled and elderly people as a sub-set of the population, and an integral part of a more recent international trend towards the integration of older and disabled people in the mainstream of society. | Clarkson, P. J., & Coleman, R. (2015). History of inclusive design in the UK. Applied ergonomics, 46, 235-247. |
| 129 | Design/ Engineering | Inclusive Design | By using the Inclusive Design Cube model and the user-aware design that it promotes, designers can better understand users' capabilities and create intuitive interfaces, easy-open packaging, well structured, logical and clear signage, power assisted steering and braking, and many other products that are regularly taken for granted. Modular and customisable designs can greatly enhance usability. | Clarkson, P. J., & Coleman, R. (2015). History of inclusive design in the UK. Applied ergonomics, 46, 235-247. |
| 130 | Design/ Engineering | Inclusive Design | Inclusively designed products, services and environments will likely benefit the whole population and can lead to an increase of business for companies. | Clarkson, P. J., & Coleman, R. (2015). History of inclusive design in the UK. Applied ergonomics, 46, 235-247. |
| 131 | Design/ Engineering | Inclusive Design | The economic case for Inclusive Design is built on two key factors. First, the Potential Support Ratio (PSR) e the numberof people aged 15-64 who could support one person over 65 e is declining rapidly, particularly in the developed world, while care costs as aproportion of GDP are escalating. Second, without an effective consumeroffer addressing people's lifestyle, needs and aspirations, older people in particular will have little incentive to spend what disposable income they have, removing what could be a significant economic driver. Inclusive Design, especially in the workplace, offers the possibility for older and disabled people to enter or continue work in gainful employment and, therefore, extend in dependent living, which in turn can contribute to lowering care costs and help stimulate the economy. | Clarkson, P. J., & Coleman, R. (2015). History of inclusive design in the UK. Applied ergonomics, 46, 235-247. |
| 132 | Design/ Engineering | Inclusive Design | In addition, there is a strong social case for Inclusive Design based on: (1) the desirability of social cohesion and inclusivity and (2) the accessibility of public buildings, spaces and services, which can promote social inclusion. | Clarkson, P. J., & Coleman, R. (2015). History of inclusive design in the UK. Applied ergonomics, 46, 235-247. |
| 133 | Design/ Engineering | Inclusive Design | "fresh approaches of the kind referred to here are needed to bridge the present gulf between main stream design and design for the elderly, especially with regard to the scale of demographic change. The concept of Inclusive Design coupled with story telling and scenario building techniques can turn what is often considered as a branch of design for disability into an exciting gateway to product innovation and a more user-friendly future for all. As attitudes towards ageing and disability change an important role is emerging for ergonomics in the design and assessment of everyday products and environments, to ensure that they allow for the broadest possible range of abilities in their user profiles" (Coleman, 1994). | Clarkson, P. J., & Coleman, R. (2015). History of inclusive design in the UK. Applied ergonomics, 46, 235-247. |
| 134 | Design/ Engineering | Inclusive Design | 'a simple concept to help them [manufacturers and retailers] see potential commercial benefits for their businesses', along with 'examples of how that concept could be applied in practice'. | Clarkson, P. J., & Coleman, R. (2015). History of inclusive design in the UK. Applied ergonomics, 46, 235-247. |
| 135 | Design/ Engineering | Inclusive Design | to measure design exclusion, the project realised the multi-faceted nature of exclusion: the combination of capability levels demanded by products and services across a variety of environmental contexts and the diversity of capability across the whole population and particularly across the lifespan. This led to a summative approach tothe measureof design exclusion, to account for the multiple, minor capacity deficiencies typical of ageing populations, and revealed important differences between the 'disabled' population and older adults. | Clarkson, P. J., & Coleman, R. (2015). History of inclusive design in the UK. Applied ergonomics, 46, 235-247. |
| 136 | Design/ Engineering | Inclusive Design | The concept of design exclusion as a quantifiable aspect of products and services is unique to Inclusive Design and differentiates it from the more aspirational Universal Design and Design for All approaches by assuming as a start point that no design will work perfectly for everyone. | Clarkson, P. J., & Coleman, R. (2015). History of inclusive design in the UK. Applied ergonomics, 46, 235-247. |
| 137 | Design/ Engineering | Inclusive Design | Inclusive Design based on countering design exclusion. By identifying users vulnerable to design exclusion and designing to accommodate their capability levels, products can often be improved for all users and thus compete more effectively and appeal to a wider range of consumers. | Clarkson, P. J., & Coleman, R. (2015). History of inclusive design in the UK. Applied ergonomics, 46, 235-247. |
| 138 | Design/ Engineering | Inclusive Design | Inclusive Design can therefore be seen as an iterative process of knowledge acquisition leading tocontinuous designimprovement, increased customer satisfaction and hence brand loyalty. | Clarkson, P. J., & Coleman, R. (2015). History of inclusive design in the UK. Applied ergonomics, 46, 235-247. |
| 139 | Design/ Engineering | Inclusive Design | Inclusive Design approaches for new product and service development, incorporating better data for designers that will result in products and services that facilitate independence at home, at work, and in other environments and contexts of use (Persadetal.,2007;EltonandNicolle, 2010). To achieve this there has been a need to: (1) understand the capability demand made by a product within its operating environment; (2) define a specification for and collect new population based capability data; (3) calculate levels of product exclusion and difficulty; and (4) present such data in anaccessible anduseful way. | Clarkson, P. J., & Coleman, R. (2015). History of inclusive design in the UK. Applied ergonomics, 46, 235-247. |
| 140 | Design/ Engineering | Inclusive Design | The conceptof Inclusive Design, as described by the matching of product demandstousercapabilities in a given environment is also transferable to other domains. | Clarkson, P. J., & Coleman, R. (2015). History of inclusive design in the UK. Applied ergonomics, 46, 235-247. |
| 141 | Design/ Engineering | Inclusive Design | Good design is about making conscious and well-informed decisions throughout the designprocess. A great product or service is typically built on a foundation of understanding the real needs of the userandother stakeholders. In short, good design and Inclusive Design should be seen as inseparable and essential. | Clarkson, P. J., & Coleman, R. (2015). History of inclusive design in the UK. Applied ergonomics, 46, 235-247. |
| 142 | Design/ Engineering | Inclusive Design | inclusive design applies an understanding of customer diversity to inform decisions throughout the development process, in order to better satisfy the needs of more people. Products that are more inclusive can reach a wider market, improve customer satisfaction and drive business success. | Waller, S., Bradley, M., Hosking, I., & Clarkson, P. J. (2015). Making the case for inclusive design. Applied ergonomics, 46, 297-303. |
| 143 | Design/ Engineering | Inclusive Design | Every design decision has the potential to include or exclude customers. Inclusive design emphasises the contribution that understanding user diversity makes to informing these decisions. | Waller, S., Bradley, M., Hosking, I., & Clarkson, P. J. (2015). Making the case for inclusive design. Applied ergonomics, 46, 297-303. |
| 144 | Design/ Engineering | Inclusive Design | Understanding diversity within the population. Responding to this diversity with informed design decisions | Waller, S., Bradley, M., Hosking, I., & Clarkson, P. J. (2015). Making the case for inclusive design. Applied ergonomics, 46, 297-303. |
| 145 | Design/ Engineering | Inclusive Design | Inclusive design does not suggest that it is always possible (or appropriate) to design one product to address the needs of the entire population. Instead, inclusive design guides an appropriate design response to diversity in the population through:<br>Developing a portfolio of products and derivatives to provide the best possible coverage of diversity within the 'Population Pyramid'.<br>Ensuring that each individual product has clear and distinct target users.<br>Making informed decisions to improve the success criteria for each product. | Waller, S., Bradley, M., Hosking, I., & Clarkson, P. J. (2015). Making the case for inclusive design. Applied ergonomics, 46, 297-303. |
| 146 | Design/ Engineering | Inclusive Design | inclusive design applies an understanding of customer diversity to the design of mainstream products, in order to better satisfy the needs of more people and deliver commercial success. | Waller, S., Bradley, M., Hosking, I., & Clarkson, P. J. (2015). Making the case for inclusive design. Applied ergonomics, 46, 297-303. |
| 147 | Design/ Engineering | Inclusive Design | Inclusive design is about choosing an appropriate target population for a particular design, and making informed decisions to maximise the success criteria for the target market. | Waller, S., Bradley, M., Hosking, I., & Clarkson, P. J. (2015). Making the case for inclusive design. Applied ergonomics, 46, 297-303. |
| 148 | Design/ Engineering | Inclusive Design | Why design inclusively? Having introduced and defined inclusive design, this section contains the business case. The key messages are:- Ageingpopulations lead to growingopportunities for inclusive products.- Inclusive design mitigates business risk.- Simplicity can offer competitive advantage | Waller, S., Bradley, M., Hosking, I., & Clarkson, P. J. (2015). Making the case for inclusive design. Applied ergonomics, 46, 297-303. |
| 149 | Design/ Engineering | Inclusive Design | products that are more inclusive can reach a wider market, improve customer satisfaction and drive business success, especially given the ageing population. | Waller, S., Bradley, M., Hosking, I., & Clarkson, P. J. (2015). Making the case for inclusive design. Applied ergonomics, 46, 297-303. |
| 150 | Design/ Engineering | Inclusive Design | successful in clusive design requires informed decision-making at the concept stage, because it can become prohibitively expensive to make changes later on. The process of concept generation is summarised here according to the core activities of exploration, creativity, evaluation and project management.<br>- Whatare the needs?- Howcan the needs be met?- Howwell are the needs met?- Whatshould we do next? | Waller, S., Bradley, M., Hosking, I., & Clarkson, P. J. (2015). Making the case for inclusive design. Applied ergonomics, 46, 297-303. |
| 151 | Design/ Engineering | Inclusive Design | the specific emphases for inclusive design are:<br>Understanding the true diversity of user and business needs.<br>Applying this knowledge to better inform the design decisions taken throughout the development process.<br>Evaluating rough prototypes with real users, before all the important decisions are finalised. | Waller, S., Bradley, M., Hosking, I., & Clarkson, P. J. (2015). Making the case for inclusive design. Applied ergonomics, 46, 297-303. |
| 152 | Design/ Engineering | Inclusive Design | The 'Design Wheel' shows the specific activities that help to answer the four fundamental questions of concept design. - Whatare the needs?- Howcan the needs be met?- Howwell are the needs met?- Whatshould we do next? the particularly critical activities, namely 'observe users', 'generate personas', 'test with experts', 'test with users' and 'estimate exclusion'. | Waller, S., Bradley, M., Hosking, I., & Clarkson, P. J. (2015). Making the case for inclusive design. Applied ergonomics, 46, 297-303. |
| 153 | Design/ Engineering | Inclusive Design | Estimating exclusion identifies the task steps where a product or prototype places the highest demands on the following user capabilities:<br>Vision<br>Hearing<br>Thinking<br>Reach and Dexterity<br>Mobility.<br>The process of estimating exclusion highlights the causes of frustration and exclusion, and prioritises these on a population basis. An exclusion calculator is freely available on the Inclusive Design Toolkit website (Clarkson et al., 2011). | Waller, S., Bradley, M., Hosking, I., & Clarkson, P. J. (2015). Making the case for inclusive design. Applied ergonomics, 46, 297-303. |
| 154 | Design/ Engineering | Inclusive Design | At the heart of inclusive design is a better understanding of diversity in the population and its relevance for design. | Hosking, I., Waller, S., & Clarkson, P. J. (2010). It is normal to be different: Applying inclusive design in industry. Interacting with computers, 22(6), 496-501. |
| 155 | Design/ Engineering | Inclusive Design | The aim of inclusive design is to extend the reach of mainstream products as far up the pyramid as possible, while maintaining commercial viability, and without compromising the design for those at the bottom of the pyramid | Hosking, I., Waller, S., & Clarkson, P. J. (2010). It is normal to be different: Applying inclusive design in industry. Interacting with computers, 22(6), 496-501. |
| 156 | Design/ Engineering | Inclusive Design | The aim of inclusive product design is to successfully integrate human factors in the product development process with the intention of making products accessible for the largest possible group of users. | Kirisci, P. T., Thoben, K. D., Klein, P., & Modzelewski, M. (2011). Supporting inclusive product design with virtual user models at the early stages of product development. In DS 68-9: Proceedings of the 18th International Conference on Engineering Design (ICED 11), Impacting Society through Engineering Design, Vol. 9: Design Methods and Tools pt. 1, Lyngby/Copenhagen, Denmark, 15.-19.08. 2011 (pp. 80-90). |
| 157 | Design/ Engineering | Inclusive Design | Inclusive design is a process that results in inclusive products or environments which can be used by everyone regardless of age, gender or disability [1]. The main barriers for adopting inclusive product design include technical complexity, lack of time, lack of knowledge and techniques, and lack of guidelines [2]. | Kirisci, P. T., Thoben, K. D., Klein, P., & Modzelewski, M. (2011). Supporting inclusive product design with virtual user models at the early stages of product development. In DS 68-9: Proceedings of the 18th International Conference on Engineering Design (ICED 11), Impacting Society through Engineering Design, Vol. 9: Design Methods and Tools pt. 1, Lyngby/Copenhagen, Denmark, 15.-19.08. 2011 (pp. 80-90). |
| 158 | Design/ Engineering | Inclusive Design | According to Clarkson and Coleman (2015), inclusive design emerged "(...) not as a new approach to design, but rather as a synthesis of initiatives, experiments and insights dating back to the 1960s and beyond". | Carli Lorenzini, G., & Olsson, A. (2015). Design towards better life experience: closing the gap between pharmaceutical packaging design and elderly people. In DS 80-9 Proceedings of the 20th International Conference on Engineering Design (ICED 15) Vol 9: User-Centred Design, Design of Socio-Technical systems, Milan, Italy, 27-30.07. 15 (pp. 065-076). |
| 159 | Design/ Engineering | Inclusive Design | As stated by the European Design for All e-Accessibility Network (EDeAN, 2007), inclusive design is a "process-driven approach whereby designers and industry ensure that products and services address the needs of the widest possible consumer base, regardless of age or ability. Emphasis is placed on working with critical users' to stretch design brief". | Carli Lorenzini, G., & Olsson, A. (2015). Design towards better life experience: closing the gap between pharmaceutical packaging design and elderly people. In DS 80-9 Proceedings of the 20th International Conference on Engineering Design (ICED 15) Vol 9: User-Centred Design, Design of Socio-Technical systems, Milan, Italy, 27-30.07. 15 (pp. 065-076). |
| 160 | Design/ Engineering | Inclusive Design | Inclusive design works in respect to the full range of human diversity in terms of ability, language, culture, gender, age and other forms of human difference (Inclusive Design Institute, 2014). | Carli Lorenzini, G., & Olsson, A. (2015). Design towards better life experience: closing the gap between pharmaceutical packaging design and elderly people. In DS 80-9 Proceedings of the 20th International Conference on Engineering Design (ICED 15) Vol 9: User-Centred Design, Design of Socio-Technical systems, Milan, Italy, 27-30.07. 15 (pp. 065-076). |
| 161 | Design/ Engineering | Inclusive Design | What universal design, inclusive design and design for all have in common is the principle of integration of elderly and disabled people into mainstream society, instead of viewing them as sub-sets of the population (Clarkson and Coleman, 2015). These three design approaches share similar issues and interests, and have clarified that there are possibilities to achieve market goals through a more social orientation | Carli Lorenzini, G., & Olsson, A. (2015). Design towards better life experience: closing the gap between pharmaceutical packaging design and elderly people. In DS 80-9 Proceedings of the 20th International Conference on Engineering Design (ICED 15) Vol 9: User-Centred Design, Design of Socio-Technical systems, Milan, Italy, 27-30.07. 15 (pp. 065-076). |
| 162 | Design/ Engineering | Inclusive Design | the implementation of inclusivity struggles with lack of time, budget limitations, lack of knowledge and tools for practicing, and lack of perception of inclusive design as a need for the end user (Goodman et al., 2006). | Carli Lorenzini, G., & Olsson, A. (2015). Design towards better life experience: closing the gap between pharmaceutical packaging design and elderly people. In DS 80-9 Proceedings of the 20th International Conference on Engineering Design (ICED 15) Vol 9: User-Centred Design, Design of Socio-Technical systems, Milan, Italy, 27-30.07. 15 (pp. 065-076). |
| 163 | Design/ Engineering | Inclusive Design | The capability-demand approach provides a more pragmatic way of understanding the interplay between users' cognition and product features, because it directly relates the two within a product interaction context. According to this approach, products place demands on their users' capabilities. A user whose capability does not meet a demand will not be able to use the product effectively and can be considered to be 'excluded' from its use (Clarkson et al., 2015). | Ning, W., Goodman-Deane, J., & Clarkson, P. J. (2019, July). Addressing cognitive challenges in design–a review on existing approaches. In Proceedings of the design society: International conference on engineering design (Vol. 1, No. 1, pp. 2775-2784). Cambridge University Press. |
| 164 | Design/ Engineering | Inclusive Design | This approach (The capability-demand approach ) originates from the field of inclusive design. Inclusive design is 'the design of mainstream products and/or services that are accessible to, and usable by, people with the widest range of abilities within the widest range of situations without the need for special adaptation or design' (BSI, 2005). | Ning, W., Goodman-Deane, J., & Clarkson, P. J. (2019, July). Addressing cognitive challenges in design–a review on existing approaches. In Proceedings of the design society: International conference on engineering design (Vol. 1, No. 1, pp. 2775-2784). Cambridge University Press. |
| 165 | Design/ Engineering | Inclusive Design | Inclusive design employs commonly-used HCD methods, but has also developed additional design supports particularly for tackling the incompatibility between products' demands and users' capabilities. The capability-demand approach is one example of this. In this approach, population-based capability data is used to assess design exclusion. | Ning, W., Goodman-Deane, J., & Clarkson, P. J. (2019, July). Addressing cognitive challenges in design–a review on existing approaches. In Proceedings of the design society: International conference on engineering design (Vol. 1, No. 1, pp. 2775-2784). Cambridge University Press. |
| 166 | Design/ Engineering | Inclusive Design | 'The design of mainstream products and/or services that are accessible to, and usable by, as many people as reasonably possible ... without the need for special adaptation or specialised design (BSI, 2005)'. | Li, F., & Dong, H. (2019, July). The economic explanation of inclusive design in different stages of product life time. In Proceedings of the Design Society: International Conference on Engineering Design (Vol. 1, No. 1, pp. 2377-2386). Cambridge University Press. |
| 167 | Design/ Engineering | Inclusive Design | Design exclusion occurs when the capability required for using a product (in a broad sense) exceeds the user's actual capability (Clarkson and Keates, 2002). The purpose of inclusive design is often to eliminate the barriers and exclusion, so that products, services and built environment could meet as diverse needs as possible. | Li, F., & Dong, H. (2019, July). The economic explanation of inclusive design in different stages of product life time. In Proceedings of the Design Society: International Conference on Engineering Design (Vol. 1, No. 1, pp. 2377-2386). Cambridge University Press. |
| 168 | Design/ Engineering | Inclusive Design | barriers, such as disability; and older people often have one or more types of physical capability loss or decline, which leads to difficulties when using products designed only for 'ordinary people' or 'mainstream people' (Hitchcock et al., 2001). | Li, F., & Dong, H. (2019, July). The economic explanation of inclusive design in different stages of product life time. In Proceedings of the Design Society: International Conference on Engineering Design (Vol. 1, No. 1, pp. 2377-2386). Cambridge University Press. |
| 169 | Design/ Engineering | Inclusive Design | The BSI definition of inclusive design is given at the beginning of this paper, from which we can derive design exclusion as: some people's needs are not considered from the existing mainstream products, services and environment, and they are not satisfied. | Li, F., & Dong, H. (2019, July). The economic explanation of inclusive design in different stages of product life time. In Proceedings of the Design Society: International Conference on Engineering Design (Vol. 1, No. 1, pp. 2377-2386). Cambridge University Press. |
| 170 | Design/ Engineering | Inclusive Design | inclusive design hopes to bridge the inequality gap caused by diversity and enable as many users as possible to gain the value and experience provided by the design; on the other hand, the universal design makes diversified users reduce diversity to universality and reduce the richness of choice through a materialized uniqueness (Winance, 2014). | Li, F., & Dong, H. (2019, July). The economic explanation of inclusive design in different stages of product life time. In Proceedings of the Design Society: International Conference on Engineering Design (Vol. 1, No. 1, pp. 2377-2386). Cambridge University Press. |
| 171 | Design/ Engineering | Inclusive Design | study the inclusion and exclusion from the law of human behaviour, and focus on analysing the mechanism that affects inclusion/exclusion and comparing different situations. | Li, F., & Dong, H. (2019, July). The economic explanation of inclusive design in different stages of product life time. In Proceedings of the Design Society: International Conference on Engineering Design (Vol. 1, No. 1, pp. 2377-2386). Cambridge University Press. |
| 172 | Design/ Engineering | Inclusive Design | There is a need for responsible engineering design to accommodate the diverse user requirements that come with the global phenomenon of population ageing. Inclusive design can address these diverse requirements through the consideration of a wide diversity of user needs within the design process. | Wilson, N., Thomson, A., Thomson, A., & Holliman, A. F. (2019, July). Understanding inclusive design education. In Proceedings of the Design Society: International Conference on Engineering Design (Vol. 1, No. 1, pp. 619-628). Cambridge University Press. |
| 173 | Design/ Engineering | Inclusive Design | The Design Council defines inclusive design (ID) as “a general approach to designing in which designers ensure their products and services address the needs of the widest population possible, irrespective of age and ability” (Design Council, 2008). This philosophy includes all users within the engineering design process to ensure products and services are usable and accessible to as wide a range of the population as possible. With the diverse capabilities resulting from disability and population ageing, there is an immediate opportunity ID can address. | Wilson, N., Thomson, A., Thomson, A., & Holliman, A. F. (2019, July). Understanding inclusive design education. In Proceedings of the Design Society: International Conference on Engineering Design (Vol. 1, No. 1, pp. 619-628). Cambridge University Press. |
| 174 | Design/ Engineering | Inclusive Design | Inclusive design (ID) looks to provide designers with a more accurate understanding of the requirements of different user groups, with the aim of driving informed decisions throughout the engineering design process (Waller et al., 2015). The primary objective of ID is to create equality within society through the eradication of social exclusion (Clarkson & Coleman, 2015). | Wilson, N., Thomson, A., Thomson, A., & Holliman, A. F. (2019, July). Understanding inclusive design education. In Proceedings of the Design Society: International Conference on Engineering Design (Vol. 1, No. 1, pp. 619-628). Cambridge University Press. |
| 175 | Design/ Engineering | Inclusive Design | ID is an approach that can tackle the diverse population needs that come with disability and ageing, such as declines in physical and cognitive capabilities (Bouma, 2013) that make it difficult for older users to carry out simultaneous tasks. Heterogeneity also increases with age - as the population gets older, the needs and capabilities of users becomes more diverse (Johnson et al., 2010). ID aims to make products more usable by the wider population (Dong et al., 2004a), helping individuals become less reliant on healthcare and social services (Cremers et al., 2014) and maintaining independence for longer. | Wilson, N., Thomson, A., Thomson, A., & Holliman, A. F. (2019, July). Understanding inclusive design education. In Proceedings of the Design Society: International Conference on Engineering Design (Vol. 1, No. 1, pp. 619-628). Cambridge University Press. |
| 176 | Design/ Engineering | Inclusive Design | ID relies heavily on the collection of user information to justify design decisions (Keates et al., 2000). Involvement of the end user throughout the design process drives informed decision making (Vavik & Keitsch, 2010) and ensures their needs are incorporated within the design process, making the designer less reliant on their own perception of the problem. | Wilson, N., Thomson, A., Thomson, A., & Holliman, A. F. (2019, July). Understanding inclusive design education. In Proceedings of the Design Society: International Conference on Engineering Design (Vol. 1, No. 1, pp. 619-628). Cambridge University Press. |
| 177 | Design/ Engineering | Inclusive Design | An ID approach benefits the design process as a whole, with additional time and resources in the early stages of the process ensuring user requirements are satisfied and minimising the need for costly design alterations later in the process (Waller et al., 2015). | Wilson, N., Thomson, A., Thomson, A., & Holliman, A. F. (2019, July). Understanding inclusive design education. In Proceedings of the Design Society: International Conference on Engineering Design (Vol. 1, No. 1, pp. 619-628). Cambridge University Press. |
| 178 | Design/ Engineering | Inclusive Design | An ID approach also ensures user needs are satisfied early in the design process, minimising the need for costly alterations and rework later in the process (Waller et al., 2015). | Wilson, N., Thomson, A., Thomson, A., & Holliman, A. F. (2019, July). Understanding inclusive design education. In Proceedings of the Design Society: International Conference on Engineering Design (Vol. 1, No. 1, pp. 619-628). Cambridge University Press. |
| 179 | Design/ Engineering | Inclusive Design | Inclusive design is for those who want to make great products for the greatest number of people. | Microsoft. Inclusive Design. Available at: https://inclusive.microsoft.design/ |
| 180 | Design/ Engineering | Inclusive Design | Inclusive design: A design methodology that enables and draws on the full range of human diversity. | Microsoft. Inclusive Design. Available at: https://inclusive.microsoft.design/ |
| 181 | Design/ Engineering | Inclusive Design | Designing inclusively doesn't mean you're making one thing for all people.<br>You're designing a diversity of ways for everyone to participate in an experience with a sense of belonging. | Microsoft. Inclusive Design. Available at: https://inclusive.microsoft.design/ |
| 182 | Design/ Engineering | Inclusive Design | Recognize exclusion<br>Learn from diversity<br>Solve for one, extend to many | Microsoft. Inclusive Design. Available at: https://inclusive.microsoft.design/ |

**Supplementary_4**

| No | Domain | Definition | References |
|---|---|---|---|
| 1 | Management | Defined as the individual experience of increased self-determination and efficacy, empowerment generally leads to increased trust in the empowering the person or organization and an enhanced tendency to repeat the empowered behavior [12,33,39] | Füller, J., Mühlbacher, H., Matzler, K., & Jawecki, G. (2009). Consumer empowerment through internet-based co-creation. Journal of management information systems, 26(3), 71-102. |
| 2 | Management | empowerment is often equated with the sharing of power with subordinates and with participative management. In this sense, empowerment describes the perceived power or control that an individual actor or organizational subunit has over others [12]. | Füller, J., Mühlbacher, H., Matzler, K., & Jawecki, G. (2009). Consumer empowerment through internet-based co-creation. Journal of management information systems, 26(3), 71-102. |
| 3 | Management | In this context, empowerment refers to how the new technologies enable people to interact with the world on different levels (personal, dyad, group, or community) and to do or to achieve things that they found difficult to do or to achieve before. | Füller, J., Mühlbacher, H., Matzler, K., & Jawecki, G. (2009). Consumer empowerment through internet-based co-creation. Journal of management information systems, 26(3), 71-102. |
| 4 | Management | Perceived empowerment is conceptualized as consumers' perceived influence on the product design and decision-making [112]. When consumers feel enabled and competent to solve the product development task assigned to them, when they feel they have some impact on the NPD decisions, they may feel empowered [33]. | Füller, J., Mühlbacher, H., Matzler, K., & Jawecki, G. (2009). Consumer empowerment through internet-based co-creation. Journal of management information systems, 26(3), 71-102. |
| 5 | Management | Empowerment becomes the process by which a leader or manager shares his or her power with subordinates. | Conger, J. A., & Kanungo, R. N. (1988). The empowerment process: Integrating theory and practice. Academy of management review, 13(3), 471-482. |
| 6 | Management | Power, in this context, is interpreted as the possession of formal authority or control over organizational resources. | Conger, J. A., & Kanungo, R. N. (1988). The empowerment process: Integrating theory and practice. Academy of management review, 13(3), 471-482. |
| 7 | Management | The emphasis is primarily on the notion of sharing authority. | Conger, J. A., & Kanungo, R. N. (1988). The empowerment process: Integrating theory and practice. Academy of management review, 13(3), 471-482. |
| 8 | Management | To empower, implies the granting of power-delegation of authority | Burke, W. (1986). Leadership as empowering others. Executive power/Jossey-Bass. |
| 9 | Management | empower as "to enable." (the Oxford English dictionary) | Conger, J. A., & Kanungo, R. N. (1988). The empowerment process: Integrating theory and practice. Academy of management review, 13(3), 471-482. |
| 10 | Management | a process of enhancing feelings of self-efficacy among organizational members through the identification of conditions that foster powerlessness and through their removal by both formal organizational practices and informal techniques of providing efficacy information. | Conger, J. A., & Kanungo, R. N. (1988). The empowerment process: Integrating theory and practice. Academy of management review, 13(3), 471-482. |
| 11 | Management | Fuchs and Schreier (2011) introduced the concept 'consumer empowerment' that signifies customers' involvement into a firm's internal NPD processes, as well as the impact of customer perception of a particular firm in the market. | Sarmah, B., & Rahman, Z. (2017). Transforming jewellery designing: Empowering customers through crowdsourcing in India. Global Business Review, 18(5), 1325-1344.<br><br>Fuchs, C., & Schreier, M. (2011). Customer empowerment in new product development. Journal of Product Innovation Management, 28(1), 17–32. |
| 12 | Management | Consumers are turned designers, and through positive word of mouth can also play the role of a marketer. Thus, it is an empowering process of the customers and elevates their positions from a mere purchaser to an internal employee and a marketer. | Sarmah, B., & Rahman, Z. (2017). Transforming jewellery designing: Empowering customers through crowdsourcing in India. Global Business Review, 18(5), 1325-1344. |
| 13 | Psychology | An iterative process in which a person who lacks power sets a personally meaningful goal-oriented toward increasing power, takes action toward that goal, and observes and reflects on the impact of this action, drawing on his or her evolving self-efficacy, knowledge, and competence related to the goal. | Cattaneo, L. B., & Chapman, A. R. (2010). The process of empowerment: a model for use in research and practice. American psychologist, 65(7), 646. |
| 14 | Psychology | empowerment; the motivational concept of self-efficacy (Conger and Kanungo (1988)) | Spreitzer, G. M. (1995). Psychological empowerment in the workplace: Dimensions, measurement, and validation. Academy of management Journal, 38(5), 1442-1465. |
| 15 | Psychology | Thomas and Velthouse (1990) defined empowerment as increased intrinsic task motivation manifested in a set of four cognitions reflecting an individual's orientation to his or her work role: meaning, competence (which is synonymous with Conger and Kanungo's self-efficacy), self-determination, and impact. | Spreitzer, G. M. (1995). Psychological empowerment in the workplace: Dimensions, measurement, and validation. Academy of management Journal, 38(5), 1442-1465. |
| 16 | Marketing | "customer empowerment" reflects consumers' enhanced ability to access, understand and share information. | Pires, G. D., Stanton, J., & Rita, P. (2006). The internet, consumer empowerment and marketing strategies. European journal of marketing, 40(9/10), 936-949. |
| 17 | Marketing | As a process, empowerment requires mechanisms for individuals to gain control over issues that concern them, including opportunities to develop and practice skills necessary to exert control over their decision-making. | Pires, G. D., Stanton, J., & Rita, P. (2006). The internet, consumer empowerment and marketing strategies. European journal of marketing, 40(9/10), 936-949. |
| 18 | Marketing | Empowerment as an outcome is subjective. Empowered individuals "would be expected to feel a sense of control, understand their sociopolitical environment, and become active in efforts to exert control" (Zimmerman and Warschausky, 1998, p. 6). | Pires, G. D., Stanton, J., & Rita, P. (2006). The internet, consumer empowerment and marketing strategies. European journal of marketing, 40(9/10), 936-949. |
| 19 | Design & Psychology | Empowerment is a process by which people, organizations, and communities gain mastery over issues they perceive as concerning them (Rappaport 1987). | Hussain, S. (2010). Empowering marginalised children in developing countries through participatory design processes. *CoDesign* , *6* (2), 99-117. |
| 20 | Design & Psychology | Empowerment is a multilevel construct with all levels of analysis being interdependent with each other. | Hussain, S. (2010). Empowering marginalised children in developing countries through participatory design processes. *CoDesign* , *6* (2), 99-117. |
| 21 | Design & Psychology | As pointed out by Zimmerman (1995), psychological empowerment and power are not the same: 'Power suggests authority, whereas psychological empowerment is a feeling of control, a critical awareness of one's environment, and an active engagement in it. [. . .] Actual power or control is not necessary for empowerment because in some context and for some populations real control or power may not be the desired goal. Rather, goals such as being more informed, more skilled, more healthy [sic.], or more involved in decision making may be the desired outcome' (pp. 592–593). | Hussain, S. (2010). Empowering marginalised children in developing countries through participatory design processes. *CoDesign* , *6* (2), 99-117. |
| 22 | Design & Psychology | Physiological empowerment is context-dependent. Therefore, the development of a universal and global measure of it is not an appropriate goal; empowerment does not mean the same thing for all people, everywhere, at any time (Zimmerman 1995). This means that in each project, designers have to decide how changes in psychological empowerment should be evaluated. | Hussain, S. (2010). Empowering marginalised children in developing countries through participatory design processes. *CoDesign* , *6* (2), 99-117. |
| 23 | Design & Psychology | 'Psychological empowerment [. . .] includes beliefs that goals can be achieved, awareness about recourses and factors that hinder or enhance one's efforts to achieve those goals, and efforts to fulfil the goals' (Zimmerman 1995, p. 582). | Hussain, S. (2010). Empowering marginalised children in developing countries through participatory design processes. *CoDesign* , *6* (2), 99-117. |
| 24 | Design & Psychology | Psychological empowerment is not a static trait, but an open-ended dynamic construct. Every individual has the potential of experiencing both empowering and disempowering processes. The perceptions, skills, or actions for increasing someone's sense of empowerment depend on a person's age, socioeconomic status, gender, etc. | Hussain, S. (2010). Empowering marginalised children in developing countries through participatory design processes. *CoDesign* , *6* (2), 99-117. |
| 25 | Design | "the process of increasing the capacity of individuals or groups to make choices and to transform those choices into desired actions and outcomes." *the World Bank* | Fladvad Nielsen, B. (2012). Participate! A critical investigation into the relationship between participation and empowerment in design for development. In DS 71: Proceedings of NordDesign 2012, the 9th NordDesign conference, Aarlborg University, Denmark. 22-24.08. 2012. |
| 26 | Design | Zimmermann [25] for example describes psychological empowerment as a potential within the individual, while Ife[26] focuses also on top-down conditions for empowerment. | Fladvad Nielsen, B. (2012). Participate! A critical investigation into the relationship between participation and empowerment in design for development. In DS 71: Proceedings of NordDesign 2012, the 9th NordDesign conference, Aarlborg University, Denmark. 22-24.08. 2012. |
| 27 | Design | "exists only as exercised by some on others, only when it is put into action" [27] | Fladvad Nielsen, B. (2012). Participate! A critical investigation into the relationship between participation and empowerment in design for development. In DS 71: Proceedings of NordDesign 2012, the 9th NordDesign conference, Aarlborg University, Denmark. 22-24.08. 2012. |
| 28 | Design | The purpose of empowerment should also include an understanding of how the individual empowerment can increase decision power and in which circumstances it can represent a real and not only potential change that they did not have before the outsider intervened; if not, one may even argue that our efforts are unethical and indeed neocolonialist in nature. | Fladvad Nielsen, B. (2012). Participate! A critical investigation into the relationship between participation and empowerment in design for development. In DS 71: Proceedings of NordDesign 2012, the 9th NordDesign conference, Aarlborg University, Denmark. 22-24.08. 2012. |
| 29 | Design | Real empowerment means that we increase someone's power, which depends on both top-down and bottom-up approaches[26] to change while design for capabilities is focused on the individual and perhaps already assuming that the individual has the opportunity to do what they are capable of. | Fladvad Nielsen, B. (2012). Participate! A critical investigation into the relationship between participation and empowerment in design for development. In DS 71: Proceedings of NordDesign 2012, the 9th NordDesign conference, Aarlborg University, Denmark. 22-24.08. 2012. |
| 30 | Marketing | Consumer empowerment has been defined (Wathieu et al., 2002) as letting consumers take control of variables that are conventionally pre-determined by marketers; one of these variables is brand meaning. | Cova, B., & Pace, S. (2006). Brand community of convenience products: new forms of customer empowerment–the case "my Nutella The Community". European journal of marketing, 40(9/10), 1087-1105. |
| 31 | Marketing | Ferrero enables Nutella fans to live their passion. It is in this sense that consumer empowerment has occurred. Conversely, Ferrero does not encourage exchange and community-oriented creations, remaining relatively distant from any form of communal empowerment. | Cova, B., & Pace, S. (2006). Brand community of convenience products: new forms of customer empowerment–the case "my Nutella The Community". European journal of marketing, 40(9/10), 1087-1105. |
| 32 | Management | customer empowerment – engagement strategies that give customers a  sense of control over a brand's general offerings | Acar, O. A., & Puntoni, S. (2016). Customer empowerment in the digital age. Journal of Advertising Research, 56(1), 4-8. |
| 33 | Management | Customer empowerment therefore means a deeper connection with the brand. Customers who are empowered develop positive attitudes towards a brand. | Acar, O. A., & Puntoni, S. (2016). Customer empowerment in the digital age. Journal of Advertising Research, 56(1), 4-8. |
| 34 | Marketing | consumer's ability to specify and adjust the choice context. According to this view, the experience of empowerment derives not from more choices, but from one's flexibility in defining one's choices. | Wathieu, L., Brenner, L., Carmon, Z., Chattopadhyay, A., Wertenbroch, K., Drolet, A., ... & Wu, G. (2002). Consumer control and empowerment: a primer. Marketing Letters, 13, 297-305. |
| 35 | Engineering & Education | "multi-dimensional social process that helps people gain control over their own lives" by fostering power in people to operate the changes they may want "in their own lives, their communities, and in their society" (Page and Czuba, 1999) | Kleba, J.B., Cruz, C.C. (2021). From empowerment to emancipation - A framework for empowering sociotechnical interventions. International Journal of Engineering, Social Justice and Peace, 8(2), pages 28-49. |
| 36 | Engineering & Education | "while we cannot give people power [or] make them "empowered," we can provide the opportunities, resources and support that they need to become involved themselves" (Page and Czuba, 1999, p.4) | Kleba, J.B., Cruz, C.C. (2021). From empowerment to emancipation - A framework for empowering sociotechnical interventions. International Journal of Engineering, Social Justice and Peace, 8(2), pages 28-49. |
| 37 | Political Science | Community empowerment implies that people will have the necessary information, as well as power and influence to exercise some control over the future of their area. | Perrons, D., & Skyers, S. (2003). Empowerment through participation? Conceptual explorations and a case study. International Journal of Urban and Regional Research, 27(2), 265-285. |
| 38 | Psychology | Empowerment - the process by which people gain some control over valued events, outcomes, and resources - is an important construct for understanding and improving the lives of people of marginal status. | Fawcett, S. B., White, G. W., Balcazar, F. E., Suarez-Balcazar, Y., Mathews, R. M., Paine-Andrews, A., ... & Smith, J. F. (1994). A contextual-behavioral model of empowerment: Case studies involving people with physical disabilities. *American Journal of Community Psychology* , *22* (4), 471-496. |
| 39 | Psychology | "...to enhance the possibilities for people to control their own lives..." (Rappaport, 1981, p. 15), whether as individuals or as groups of people sharing common experiences, turf, or concerns. | Fawcett, S. B., White, G. W., Balcazar, F. E., Suarez-Balcazar, Y., Mathews, R. M., Paine-Andrews, A., ... & Smith, J. F. (1994). A contextual-behavioral model of empowerment: Case studies involving people with physical disabilities. *American Journal of Community Psychology* , *22* (4), 471-496. |
| 40 | Psychology | Empowerment is viewed as both an individual or psychological process (Zimmerman, 1990; Zimmerman & Rappaport, 1988) and a group experience (Chavis & Wandersman, 1990). | Fawcett, S. B., White, G. W., Balcazar, F. E., Suarez-Balcazar, Y., Mathews, R. M., Paine-Andrews, A., ... & Smith, J. F. (1994). A contextual-behavioral model of empowerment: Case studies involving people with physical disabilities. *American Journal of Community Psychology* , *22* (4), 471-496. |
| 41 | Engineering/ Industrial management and organisation | Empowerment does not mean that the management has no role to play or no responsibility. In fact, the management has more responsibilities. They have to monitor the skills continuously required for carrying out the ever-changing complexity of jobs of the teams. The management must be willing to help the teams when they are unable to solve issues. The responsibilities of the management are to control the processes and not the individual team members. | Thamizhmanii, S., & Hasan, S. (2010). A review on an employee empowerment in TQM practice. Journal of Achievements in Materials and Manufacturing Engineering, 39(2), 204-210. |
| 42 | Engineering/ Industrial management and organisation | Empowerment is a concept that links individual strengths and competencies, natural helping systems and proactive behaviour to social policy and social change. In the other words, empowerment links individual and his or her well-being to wider social and political environment in which he or she functions [11]. | Thamizhmanii, S., & Hasan, S. (2010). A review on an employee empowerment in TQM practice. Journal of Achievements in Materials and Manufacturing Engineering, 39(2), 204-210. |
| 43 | Engineering/ Industrial management and organisation | Empowerment is, an organizational state, where people are obliged to direct business and understand their performance boundaries, thus it enables them to take responsibility and ownership while seeking improvements, identifying the best course of action and imitative steps to meet customer requirements [12]. | Thamizhmanii, S., & Hasan, S. (2010). A review on an employee empowerment in TQM practice. Journal of Achievements in Materials and Manufacturing Engineering, 39(2), 204-210. |
| 44 | Engineering/ Industrial management and organisation | Empowerment means engaging employees in the thinking processes of an organization. Involvement means having input. Empowerment means having input that is heard and seriously considered. | Thamizhmanii, S., & Hasan, S. (2010). A review on an employee empowerment in TQM practice. Journal of Achievements in Materials and Manufacturing Engineering, 39(2), 204-210. |
| 45 | Engineering/ Industrial management and organisation | Empowerment requires a change in an organization culture, but does not mean that top management  abdicate their responsibility or authority. An employee empowerment is necessary for the effective functioning of the skill of employee. | Thamizhmanii, S., & Hasan, S. (2010). A review on an employee empowerment in TQM practice. Journal of Achievements in Materials and Manufacturing Engineering, 39(2), 204-210. |
| 46 | Social and Behavioral Sciences | Community empowerment is one of the main purpose of development in developing countries. While development only result in a ratio between the “before” and “after” condition, empowerment can keep the continuity of the development through the good relationship between related parties, in terms of knowledge production and field application. | Sianipar, C. P. M., & Widaretna, K. (2012). NGO as Triple-Helix axis: Some lessons from Nias community empowerment on cocoa production. Procedia-Social and Behavioral Sciences, 52, 197-206. |
| 47 | Environmental Sciences | Outsiders give hoe to local people, teach them how to use and maintain it with better method, and also teach them how to make it by themself, adjusting the function based on required conditions. These efforts will sustain the sustainable development, empower local people. When the outsiders leave them, local people will maintain the sustainable development by themself. | Sianipar, C. P. M., Yudoko, G., Adhiutama, A., & Dowaki, K. (2013). Community empowerment through appropriate technology: Sustaining the sustainable development. Procedia Environmental Sciences, 17, 1007-1016. |
| 48 | Environmental Sciences | appropriate technology can fill in the gap to transform community development into community empowerment. The appropriateness of appropriate technology can help community members in maintaining the ongoing development (sustaining the development) as well as in making adjustment of technology to overcome future conditions in changing environment. | Sianipar, C. P. M., Yudoko, G., Adhiutama, A., & Dowaki, K. (2013). Community empowerment through appropriate technology: Sustaining the sustainable development. Procedia Environmental Sciences, 17, 1007-1016. |
| 49 | Environmental Sciences | Ferguson [2] stated that empowerment is categorized as transformation process of a community. In order to fasten the process, technology is always needed to bring multiplier effect for the transformation process [3]. | Sianipar, C. P. M., Yudoko, G., Adhiutama, A., & Dowaki, K. (2013). Community empowerment through appropriate technology: Sustaining the sustainable development. Procedia Environmental Sciences, 17, 1007-1016. |
| 50 | Environmental Sciences | While development often interpreted as the flow of resources from outside into community, empowerment push-and-pull full participation of all community members to change their world by themself, from inside to outside [2]. | Sianipar, C. P. M., Yudoko, G., Adhiutama, A., & Dowaki, K. (2013). Community empowerment through appropriate technology: Sustaining the sustainable development. Procedia Environmental Sciences, 17, 1007-1016. |
| 51 | Environmental Sciences | Community with lack of development must be developed through community development. Then, the increasing of knowledges will sustain the development. After that, the transformation process should be continued to reach beyond sustainable development, that is empowerment. | Sianipar, C. P. M., Yudoko, G., Adhiutama, A., & Dowaki, K. (2013). Community empowerment through appropriate technology: Sustaining the sustainable development. Procedia Environmental Sciences, 17, 1007-1016. |
| 52 | Environmental Sciences | Empowerment as transformation process of a community should be understood completely through scientific thoughts as well as real field problems. | Sianipar, C. P. M., Yudoko, G., Adhiutama, A., & Dowaki, K. (2013). Community empowerment through appropriate technology: Sustaining the sustainable development. Procedia Environmental Sciences, 17, 1007-1016. |
| 53 | Environmental Sciences | As a transformation process [2], empowerment will ensure the sustainability of given sustainable development. | Sianipar, C. P. M., Yudoko, G., Adhiutama, A., & Dowaki, K. (2013). Community empowerment through appropriate technology: Sustaining the sustainable development. Procedia Environmental Sciences, 17, 1007-1016. |
| 54 | Environmental Sciences | The development maybe still maintained someway, but community members will face many problems because outsiders do not teach them any capabilities to solve their future problems, how to adjust the ongoing development in the changes environment. Through this gap, empowerment approach provides the solution. In that approach, outsiders teach community members to 'make' the development by themselves. It is the next keyword after the 'give' and 'maintain.' By teaching such making activities related to the ongoing sustainable development, community members can always adapt their efforts in sustaining the development by themself, even when otsiders have already leave their areas and/or the project has ended. That is "empowerment" level. | Sianipar, C. P. M., Yudoko, G., Adhiutama, A., & Dowaki, K. (2013). Community empowerment through appropriate technology: Sustaining the sustainable development. Procedia Environmental Sciences, 17, 1007-1016. |
| 55 | Sustainability | empowering consumers by giving them more possibilities of repairing their products instead of discarding them. - Right to repair | Hernandez, R. J., Miranda, C., & Goñi, J. (2020). Empowering sustainable consumption by giving back to consumers the ‘right to repair’. Sustainability, 12(3), 850. |
| 56 | Sustainability | (we believe are underexplored, particularly about the impact directives like “right to repair” can have on empowering sustainable consumption from a consumer point of view and not only as a production mechanism.) - Right to repair | Hernandez, R. J., Miranda, C., & Goñi, J. (2020). Empowering sustainable consumption by giving back to consumers the ‘right to repair’. Sustainability, 12(3), 850. |
| 57 | Sustainability | “empower consumers to make informed choices and play an active role in the ecological transition” [39] | Hernandez, R. J., Miranda, C., & Goñi, J. (2020). Empowering sustainable consumption by giving back to consumers the ‘right to repair’. Sustainability, 12(3), 850. |
| 58 | Sustainability | Empowerment, we believe, will be based on fostering more community action and autonomy. | Hernandez, R. J., Miranda, C., & Goñi, J. (2020). Empowering sustainable consumption by giving back to consumers the ‘right to repair’. Sustainability, 12(3), 850. |
| 59 | Manufacturing | Empowerment is one of the important step adopted by top management and has achieved desired results in terms of productivity and waste minimization. Further, empowerment has improved information sharing and coordination among workers assigned to different work stations or within the same work station (e.g. [68, 70]). Empowerment includes different activities and training to create adaptability, foster innovation, improve collaboration and enhance speed [65]. | Dubey, R., & Gunasekaran, A. (2015). Agile manufacturing: framework and its empirical validation. The International Journal of Advanced Manufacturing Technology, 76, 2147-2157. |
| 60 | Manufacturing | Empowerment of workforce may help to achieve AM (agile manufacturing) environment, and thus, it can be said based on preceding discussions that, empowerment is an important construct of the AM framework. | Dubey, R., & Gunasekaran, A. (2015). Agile manufacturing: framework and its empirical validation. The International Journal of Advanced Manufacturing Technology, 76, 2147-2157. |
| 61 | Innovation Management | User empowerment refers to the types of new product decisions that are outsourced to users. In various ways are users invited to take more active part in corporate NPD. | Jespersen, K. R. (2011). Online channels and innovation: Are users being empowered and involved?. International Journal of Innovation Management, 15(06), 1141-1159. |
| 62 | Sustainable Transition (Environmental Policy & Planning) | In the broadest sense of the word, empowerment refers to the process of gaining power. There is a diversity of understandings of empowerment in transition studies, ranging from empowerment as a systemic transition pattern (De Haan & Rotmans, 2011) and functional property of niches (Smith & Raven, 2012) to empowerment at the level of individual or group capacities (Avelino, 2009,2011). | Avelino, F., & Wittmayer, J. M. (2016). Shifting power relations in sustainability transitions: a multi-actor perspective. Journal of environmental policy & planning, 18(5), 628-649. |
| 63 | Sustainable Transition (Environmental Policy & Planning) | Thomas and Velthouse (1990) operationalize empowerment in terms of ‘intrinsic motivation’ and argue that the extent to which individuals are intrinsically motivated to engage in an activity depends on the extent to which they have a sense of impact, competence, meaning and choice regarding that activity. | Avelino, F., & Wittmayer, J. M. (2016). Shifting power relations in sustainability transitions: a multi-actor perspective. Journal of environmental policy & planning, 18(5), 628-649. |
| 64 | Business & Industrial Marketing | Team empowerment, which refers to the authority and power the team has in order to manage and lead itself (Manz and Sims, 1991), is also imperative for team reflexivity. | Dayan, M., & Basarir, A. (2009). Antecedents and consequences of team reflexivity in new product development projects. Journal of Business & Industrial Marketing, 25(1), 18-29. |
| 65 | Design Education | teaching relevant design thinking methods and processes to community members, empowering them to design for and by themselves. | Hasselknippe, K. S., Flygenring, T., & Kirah, A. (2017). Empowering refugee and host-community youth with design thinking skills for community development. In DS 88: Proceedings of the 19th International Conference on Engineering and Product Design Education (E&PDE17), Building Community: Design Education for a Sustainable Future, Oslo, Norway, 7 & 8 September 2017 (pp. 092-097). |
| 66 | Purchasing and Materials Management | Empowerment is a philosophy congruent with the goals of Total Quality Management (TQM) and continuous improvement. | Giunipero, L. C., & Vogt, J. F. (1997). Empowering the purchasing function: moving to team decisions. International Journal of Purchasing and Materials Management, 33(4), 8-15. |
| 67 | Purchasing and Materials Management | Part of empowerment is maintaining one’s positive self-esteem and connectedness to others and to the organization. | Giunipero, L. C., & Vogt, J. F. (1997). Empowering the purchasing function: moving to team decisions. International Journal of Purchasing and Materials Management, 33(4), 8-15. |
| 68 | Purchasing and Materials Management | Empowerment means to enable, to allow, or to permit, and can be either self-initiated or initiated by others.3<br>It is a method for creating and redistributing power.4<br>In the traditional sense, power is one’s ability to control or change another’s behavior. Empowerment occurs when the power goes to the employees, who then experience a sense of ownership and control over their jobs.5 | Giunipero, L. C., & Vogt, J. F. (1997). Empowering the purchasing function: moving to team decisions. International Journal of Purchasing and Materials Management, 33(4), 8-15. |
| 69 | Design / Education | Empowering patients through their grip on surfaces -  designing aesthetic products by using 3D printer | Lyche, W., & Berg, A. (2017). Empowerment through Product Design: Digital textile pattern design for grip development in healthcare. In DS 88: Proceedings of the 19th International Conference on Engineering and Product Design Education (E&PDE17), Building Community: Design Education for a Sustainable Future, Oslo, Norway, 7 & 8 September 2017 (pp. 568-573). |
| 70 | Manufacturing | Employee empowerment, therefore, requires a complete realignment of power relations within which managers emphasize interdisciplinary collaboration and leadership, shared values, and motivation for knowledge diversity (Fawcett and Myers 2001; Forsythe 1997). | Gunasekaran, A., Yusuf, Y. Y., Adeleye, E. O., Papadopoulos, T., Kovvuri, D., & Geyi, D. A. G. (2019). Agile manufacturing: an evolutionary review of practices. International Journal of Production Research, 57(15-16), 5154-5174. |
| 71 | Social Science | Empowerment is defined as a group's or individual's capacity to make effective choices, that is, to make choices and then to transform those choices into desired actions and outcomes. | Alsop, R., Bertelsen, M. F., & Holland, J. (2006). Empowerment in practice: From analysis to implementation. World Bank Publications. |
| 72 | Finance | women's empowerment is defined as improving the ability of women to access the constituents of development - in particular health, education, earning opportunities, rights, and political participation.<br>Empowerment can, in other words, accelerate development. | Duflo, E. (2012). Women empowerment and economic development. Journal of Economic literature, 50(4), 1051-1079. |
| 73 | Management | The best way to define empowerment is to consider it as part of a process or an evolution – an evolution that goes on whenever you have two or more people in a relationship, personally or professionally. | Pastor, J. (1996). Empowerment: what it is and what it is not. Empowerment in organizations, 4(2), 5-7. |
| 74 | Management | empowerment is the management style where managers share with the rest of the organisational members their influence in the decision making process – that is to say, the collaboration in the decision making process is not limited to those positions with formal power –, with certain characteristics as far as information systems, training, rewarding, power sharing, leadership style and organisational culture concerns. | del Val, M. P., & Lloyd, B. (2003). Measuring empowerment. Leadership & organization development journal, 24(2), 102-108. |
| 75 | Sustainable Development | Empowerment can be defined as a "multi-dimensional social process that helps people gain control over their own lives. It is a process that fosters power (that is, the capacity to implement) in people, for use in their own lives, their communities, and in their society, by acting on issues that they define as important". | Lohani, M., & Aburaida, L. (2017). Women empowerment: A key to sustainable development. The Social ION, 6(2), 26-29. |
| 76 | Sustainable Development | Women's empowerment means women gaining more power and control over their own lives. This entails the idea of women's continued disadvantage compared to men which is apparent in different economic, socio-cultural and<br>political spheres. Therefore, women's empowerment can also be seen as an important process in reaching gender equality, which is understood to mean that the "rights, responsibilities and opportunities of individuals will not depend on whether they are born male or female". According to the UN Population Fund, an empowered woman has a sense of self-worth. She can determine her own choices, and has access to opportunities and resources providing her with an array of options she can pursue. She has control over her own life, both within<br>and outside the home and she has the ability to influence the direction of social change to create a more just social and economic order, both nationally and internationally. | Lohani, M., & Aburaida, L. (2017). Women empowerment: A key to sustainable development. The Social ION, 6(2), 26-29. |
| 77 | Social Science and Technology | Community empowerment is the process of increasing control by groups over consequences that are important to their members and to others in the broader community. For example, by forming a tenants' rights organization individual residents of a public housing project may increase their ability to redress grievances and obtain improvements in housing conditions.<br><br>Community empowerment requires; a critical balancing act between extending opportunities for control over community decisions and maximizing - efficiency in community functioning. | Fawcett, S. B., Seekins, T., Whang, P. L., Muiu, C., & Balcazar, Y. S. D. (1984). Creating and using social technologies for community empowerment. Prevention in Human Services, 3(2-3), 145-171. |
| 78 | Social Science and Technology | From a behavioral perspective, the ability to produce empowerment may be a function of such critical variables as knowledge of problems and solution alternatives and skill in presenting issues, leading groups, and implementing other empowerment tactics. Further, it may be related to the capacity to control consequences for critical actors in the system, including positive consequences (e.g., votes for elected officials) and negative ones (e.g., unfavorable newspaper reports). Finally, empowerment may be enhanced by environmental or structural variables, such as the opportunity to gain access to agendas of meetings of elected officials. An analysis of those with and without power might suggest the importance of similar variables in community empowerment. | Fawcett, S. B., Seekins, T., Whang, P. L., Muiu, C., & Balcazar, Y. S. D. (1984). Creating and using social technologies for community empowerment. Prevention in Human Services, 3(2-3), 145-171. |
| 79 | Social Science and Technology | the empowerment strategy as it relates to the choice of goals, the selection of the targets of change, the choice of means, and the analysis of consequences for the targets, change agents, and society in general. | Fawcett, S. B., Seekins, T., Whang, P. L., Muiu, C., & Balcazar, Y. S. D. (1984). Creating and using social technologies for community empowerment. Prevention in Human Services, 3(2-3), 145-171. |
| 80 | Social Science and Technology | the empowerment effort should increase the power of the target group relative to other groups in society | Fawcett, S. B., Seekins, T., Whang, P. L., Muiu, C., & Balcazar, Y. S. D. (1984). Creating and using social technologies for community empowerment. Prevention in Human Services, 3(2-3), 145-171. |
| 81 | Marketing | To ensure effective outcomes, marketers need to understand the factors that affect consumers’ mental statement of being empowered and constitute customers’ perception of appropriate empowerment. We label this customer psychological empowerment (CPE) in the current study. | Prentice, C., Han, X. Y., & Li, Y. Q. (2016). Customer empowerment to co-create service designs and delivery: Scale development and validation. Services Marketing Quarterly, 37(1), 36-51. |
| 82 | Management | For one thing, customers are now empowered with greater access to information — thanks, in part, to the Internet — so that many want to have a greater say about the products they purchase. | Ogawa, S., & Piller, F. T. (2006). Reducing the risks of new product development. MIT Sloan management review. |
| 83 | Health | The etymological denvation of 'empowerment' suggests 'to be able' or 'to enable' as meanings of the word Chandler (1991) defines 'to empower' as 'to enable to act' | Rodwell, C. M. (1996). An analysis of the concept of empowerment. Journal of advanced nursing, 23(2), 305-313. |
| 84 | Psychology | One’s degree of empowerment depends jointly on the affordances of one’s ecology and one’s capacities in relation to one’s goals. Hence, agency – the ability to act –is not the same as power. One can act (or have agency) without being able to fulfil one’s goals – to be empowered – because of deficiencies in one’s context, capacities, or their relation. One’s degree of empowerment is variable across agents, time, and location. One’s degree of agency is universal, singular, and constant. | Pratto, F. (2016). On power and empowerment. British Journal of Social Psychology, 55(1), 1-20. |
| 85 | Psychology | The empowerment approach provides a clear general way to gauge people’s life conditions: Their level of well-being and the quality of people’s choice sets. | Pratto, F. (2016). On power and empowerment. British Journal of Social Psychology, 55(1), 1-20. |
| 86 | Management | The way to truly empower a woman is to make her less poor, financially independent, and better educated; we need social and cultural changes that eliminate the prejudices that are the cause of her deprivations. | Karnani, A. G. (2006). Fortune at the Bottom of the Pyramid: A Mirage. California Management Review, Forthcoming. |

**Supplementary_5**

| No | Reference - Case Studies |
|---|---|
| 1 | Singh, D. P., & Gosain, A. K. (2019). Development of a Low-Cost Groundwater-Level Measuring Device. In Rural Technology Development and Delivery: RuTAG and Its Synergy with Other Initiatives (pp. 225-235). Springer Singapore. |
| 2 | Tak, P. P., Nuli, D. T., Ghosh, S., & Rao, A. B. (2019). Evolution of “Floating Fish Cages for Inland Waters” Developed by RuTAG IIT Bombay. In Rural Technology Development and Delivery: RuTAG and Its Synergy with Other Initiatives (pp. 237-247). Springer Singapore. |
| 3 | Khanolkar, R. S., Rao, A. B., & Ghosh, S. (2018). RuTAG IIT Bombay floating fish cages for livelihood opportunities for tribals in dimbhe area. In Techno-Societal 2016: Proceedings of the International Conference on Advanced Technologies for Societal Applications (pp. 27-35). Springer International Publishing. |
| 4 | Kotwal, V., Satya, S., Naik, S. N., Dahiya, A., & Kumar, J. (2019). Street Food Cart Design: A Critical Component of Food Safety. In Rural Technology Development and Delivery: RuTAG and Its Synergy with Other Initiatives (pp. 263-278). Springer Singapore. |
| 5 | Bhat, S., Doshi, N., Bharadwaj, C. D., Singh, S. N., Patel, Y., & Saha, S. K. (2019). Design of a Low-Cost Full-Face Mask for Stone Carvers. In Rural Technology Development and Delivery: RuTAG and Its Synergy with Other Initiatives (pp. 279-285). Springer Singapore. |
| 6 | Tak, P. P., Bhandakkar, T. K., & Khanolkar, R. S. (2019). Designing a Cow Lift for Downer Cow: Experience of Working on a Rural Technology. In Rural Technology Development and Delivery: RuTAG and Its Synergy with Other Initiatives (pp. 323-333). Springer Singapore. |
| 7 | Saini, R. P., Singal, S. K., Ali, I., & Joshi, R. C. (2019). Development of Modified Bageshwari Wool Charkha. In Rural Technology Development and Delivery: RuTAG and Its Synergy with Other Initiatives (pp. 347-357). Springer Singapore. |
| 8 | Tak, P. P., Haque, T., Guha, A., Rao, A. B., Shah, N., & Khanolkar, R. S. (2019). Study of supply chain, production potential of hirda and design of hirda decortication machine for livelihood generation for tribal people. In Rural Technology Development and Delivery: RuTAG and Its Synergy with Other Initiatives (pp. 249-261). Springer Singapore. |
| 9 | Haque, T., Tak, P., & Rao, A. B. (2016, December). Better livelihood opportunities for tribals through supply chain interventions of Hirda. In 2016 IEEE Region 10 Humanitarian Technology Conference (R10-HTC) (pp. 1-5). IEEE. |
| 10 | Ogawa, S., and F. T. Piller. 2006. Reducing the risks of new product development. Sloan Management Review 47 (2): 65–71. |
| 11 | Fuchs, C., & Schreier, M. (2011). Customer empowerment in new product development. Journal of product innovation management, 28(1), 17-32. |
| 12 | Nishikawa, H., Schreier, M., & Ogawa, S. (2013). User-generated versus designer-generated products: A performance assessment at Muji. International Journal of Research in Marketing, 30(2), 160-167. |
| 13 | Mukherji, S. (2022). Inclusive business models: transforming lives and creating livelihoods (Chapter 6). Cambridge University Press. |

**Supplementary_6**

The steps are as follows:

1. Function Identification
- Identification of functions with involved stakeholders and their roles
2. Needs and Resources
- Identification of needs and resources of stakeholders for each function
3. Matching function explanations with the aspects of sustainability
- Measure sustainability value

---

**1. Function Analysis**

Analyse functions from the content based on the generalisation models of function; see Appendix 1 – 2 – 3.

- Function analysis starts with the function of "Request Task" or "Request information". The following function should be "Receive the task" or "Receive information, "Appendix 1. The following functions should consider the evaluation of information or decision on attempting another task to detail functions. A maximum of 2 times detailing is required in function analysis. Content may involve required information or detailing that can be completed based on the tasks to be completed in related phases of a product lifecycle process.
- The roles of stakeholders should be changed by adding the function "Change in the role Receiver to Provider" or "Change in the role Provider to Receiver" based on the role of the stakeholder group. For example, If stakeholder group A requests a task from stakeholder group B. Stakeholder A is the provider; Stakeholder B is the receiver. When B is providing the outcome, the role of B should be changed from receiver to provider. This change is shown in diagrams in Appendix 1 – 2 – 3.
- In function pairs, the role of the stakeholders should be considered for each function. Providers ensure resources or an opportunity for receivers to benefit from them. Receivers access this opportunity based on their resources and needs, and according to their capabilities and conditions, they can be involved in the function pairs.
- There can be functions which involve only one stakeholder group (individual functions). A function of change in the role is not required for these functions.

It is required to detail functions a max of 2 times if detailed information is not available in the text. Example:

Stakeholder A (provider): Provide information

Stakeholder B (receiver): Receive information

Stakeholder A (provider): Request Task

Stakeholder B (receiver): Receive Task

Stakeholder B(individual function, no need to mention the role): Evaluate information (Detailing 1)

Stakeholder B(individual function, no need to mention the role): Preparation for providing outcome (Detailing 2)

Stakeholder B: Change in the role RP

Stakeholder A: Change in the role PR

Stakeholder B (Provider): Provide outcome

Stakeholder A (receiver): Receive outcome

Stakeholder A (individual function): Evaluate the outcome (1)

- **Identification of functions**

In order to identify functions, it is required to read statements or text. The generalised model of function analysis has three versions.

Example: Based on **the feedback from the users after using the first prototype** and observations, the second prototype was made. "... **Trials** were conducted on the **second prototype after fabrication**. The results of trials show that **the machine performance has increased** and it is more user-friendly than the other two. " (Tak et al., 2019a).

| | | | | |
|---|---|---|---|---|
| F286&F287 | F286 | Provide training about the 2nd modified machine | Stakeholder A | Provider |
| | F287 | Receive information | Stakeholder B | Receiver |
| F288&F289 | F288 | Request to use the modified machine | Stakeholder A | Provider |
| | F289 | Receive the request | Stakeholder B | Receiver |
| X | F290 | Use the machine | Stakeholder B | - |
| X | F291 | Observe the users | Stakeholder A | - |
| F292&F293 | F292 | Request to provide feedback about the modified machine | Stakeholder A | Provider |
| | F293 | Receive the task | Stakeholder B | Receiver |
| X | F294 | Change in the role PR | Stakeholder A | Receiver |
| X | F295 | Change in the role RP | Stakeholder B | Provider |

- **Stakeholders**

All stakeholders are considered as a group.

Example: "The "floating fish cages for inland waters" developed by RuTAG IIT Bombay has impacted local fisherman cooperative society in terms of livelihood generation, increasing women participation in cage fishing due to increase in safety factor [4]..." (Tak et al, 2019b).

Based on the text, the stakeholder groups are "RuTAG" and the "local fishermen cooperative society". We do not focus on the internal hierarchy of an organisation. RuTAG must have different departments to attempt various functions. However, we accept it as one stakeholder group without considering internal relations. Our focus is on function detailing as a first goal.

## 2. Needs and Resources

What is the reason for being involved in a function?

Use the Matrix of Needs and Satisfiers (Max Neef, 1992) to identify the group and number of needs.

- What is the need of Stakeholder A to attempt the function (286) - Provide training about the 2nd modified machine?
  - Responsibility, duties, work - check the matrix; Participation and Having
- What are the resources of Stakeholder A to attempt the function (286)- Provide training about the 2nd modified machine.
  - Position

| F. Pair | F. No | F. Explanation | Stakeholder | Role | Groups of the needs | # need groups | Groups of the sources | # resource groups |
|---|---|---|---|---|---|---|---|---|
| F286& F287 | F286 | Provide training about the 2nd modified machine | Stakeholder A | Provider | Participation and Having | 2 | Position | 1 |
| | F287 | Receive information | Stakeholder B | Receiver | Participation and Having | 2 | Experience | 1 |

We keep "resources and needs" limited to the function explanation. It is required to consider only the explanation of the particular function to identify needs and resources.

Resources can be position, experience and capability. It is important to consider which resources have priority in completing the function. For example, Stakeholder A has the experience and position to attempt Function 286. However, the position has priority in this situation to attempt the function. It can be considered as a hierarchy of the resources – which should come first. Due to the position of Stakeholder A, Stakeholder B was involved in the training about the modified machine.

## 3. Matching function explanations with the aspects of sustainability

For each function:

Read the content/case study (from where you identified the function explanation), and focus on the detailed information in the content. – What is it about: economic, social or environmental situations?

Read definitions of social, environmental and economic sustainability.

Which information is used in content: what is required to be sustained with the function?

Identify the potential contribution of function on the aspects of sustainability.

Example:

The information mentioned in Function 1 follows:

"Shashwat, a non-government organisation, has been working with tribal people in this block for many years. Livelihood generation is one of the major focuses of this NGO. An ayurvedic medicine manufacturing company from Pune asked Shashwat for decorticated hirda for making triphala churna. The NGO initiated the practice of manual decortication of hirda in Gadewadi village. Manual decortication meant breaking the shell of Hirda by striking it with a stone. After decortication, the self-help group (SHG) got a rate of Rs. 35 per kg for decorticated shells of hirda. Manual decortication of hirda is a slow process. A person can decorticate only 5–7 kg of hirda in 8 h." (Tak et al., 2019a). The information which is provided in this content matches the definitions of economic and social sustainability as clarified with the keywords and reference numbers in the following table:

| Reference (Supplementary_1) of Keywords | Keywords (from definitions) & Functions | SUSTAINABILITY | | | | **Function Groups** | **Function No** | **Function Explanation** | **Stakeholder** | **Role** |
|---|---|---|---|---|---|---|---|---|---|---|
| | | ENV | SOC | ECO | SV | | | | | |
| [2] -[35]- [59] -[29] | resources and services needs, network, source of livelihood (income) | 1 | 1 | 1 | 1 | F1&F12 | F1 | Provide information about the requirement of upgrading the machine | Stakeholder A | Provider |
| [2] -[35]- [59] -[29] | resources and services needs, network, source of livelihood (income) | 1 | 1 | 1 | 1 | | F2 | Receive information | Stakeholder B | Receiver |

**Appendix 1:**

Generalisation of Function Analysis

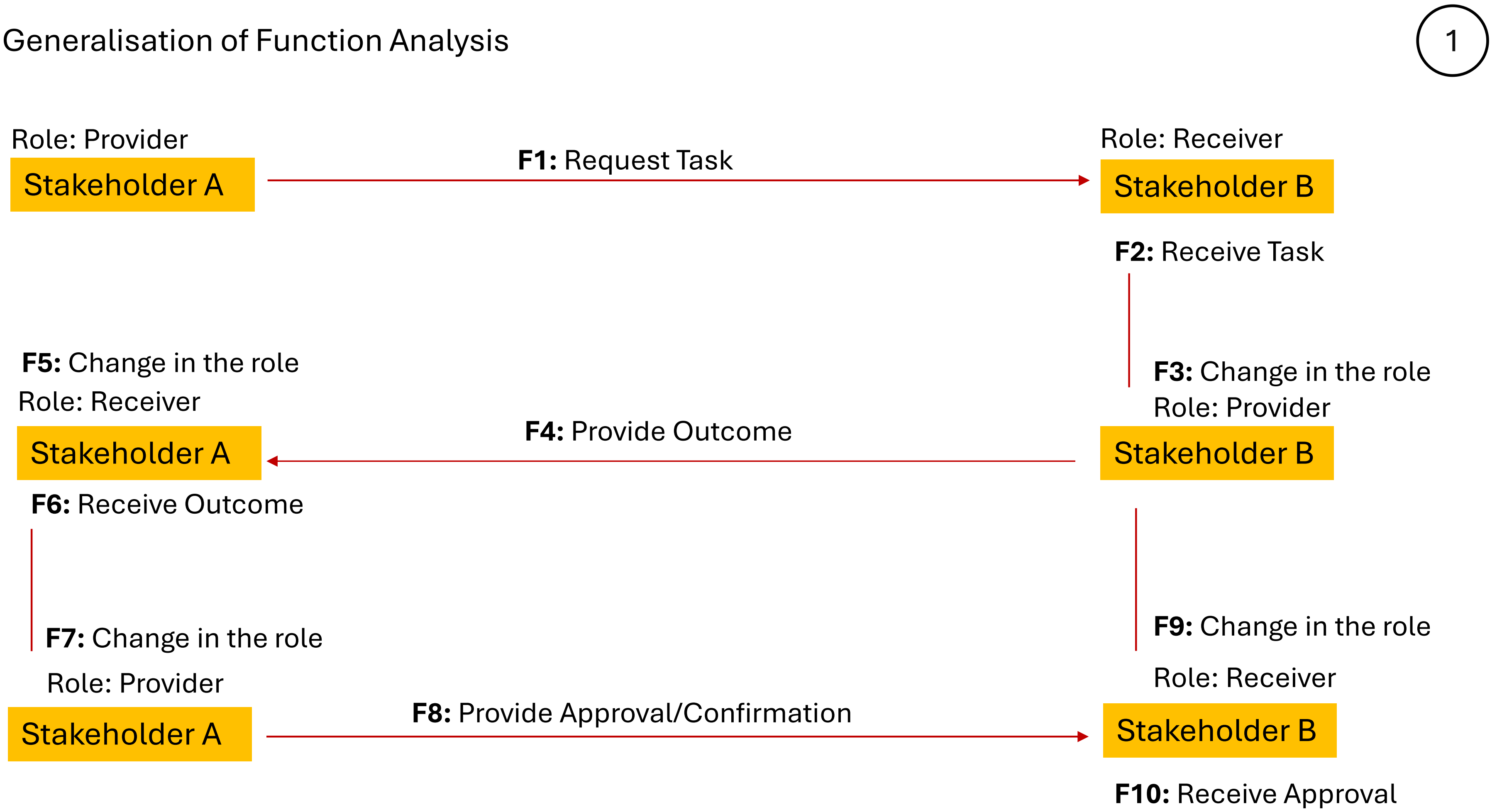

**Appendix 2**

Generalisation of Function Analysis

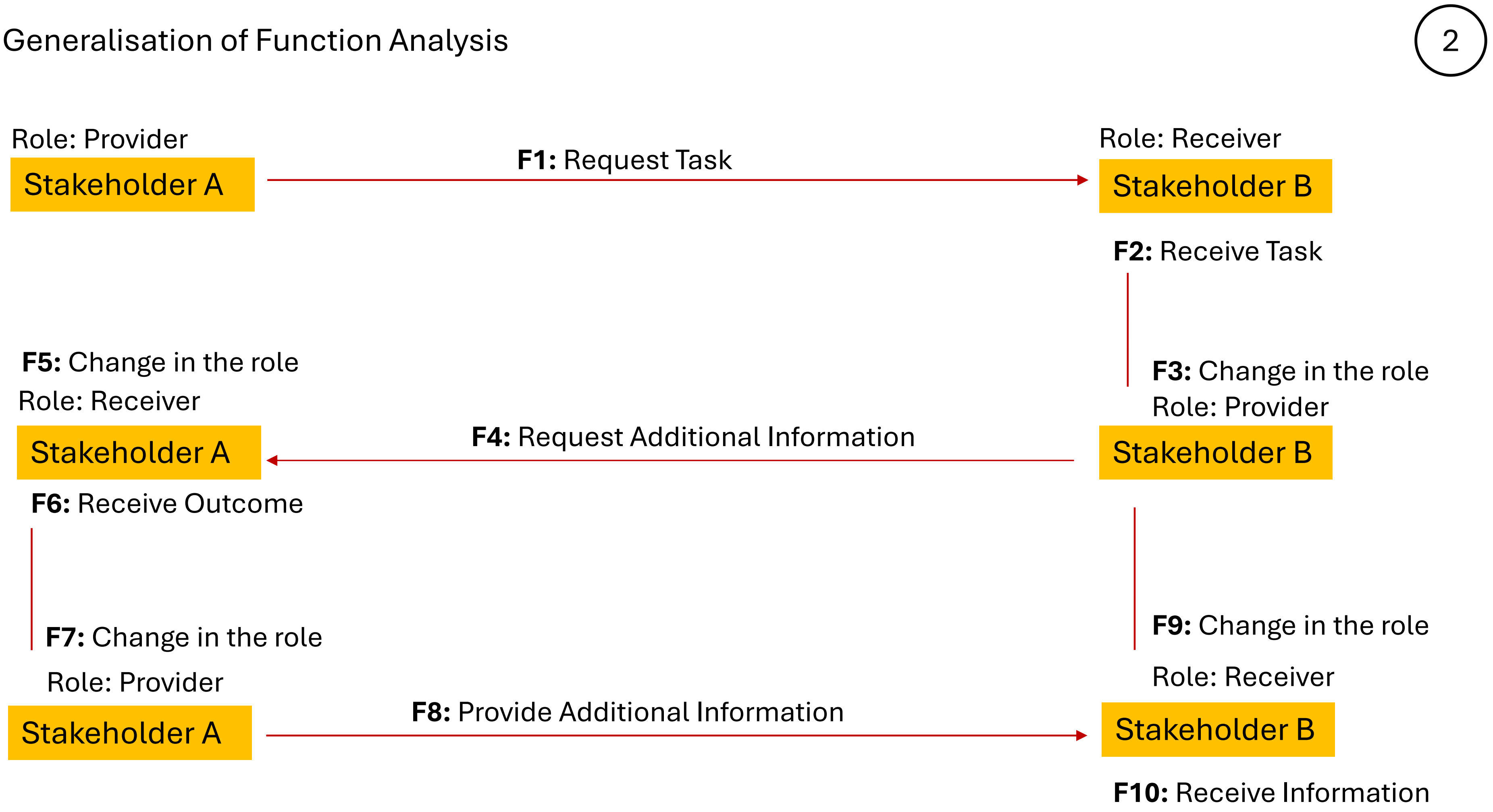

**Appendix 3**

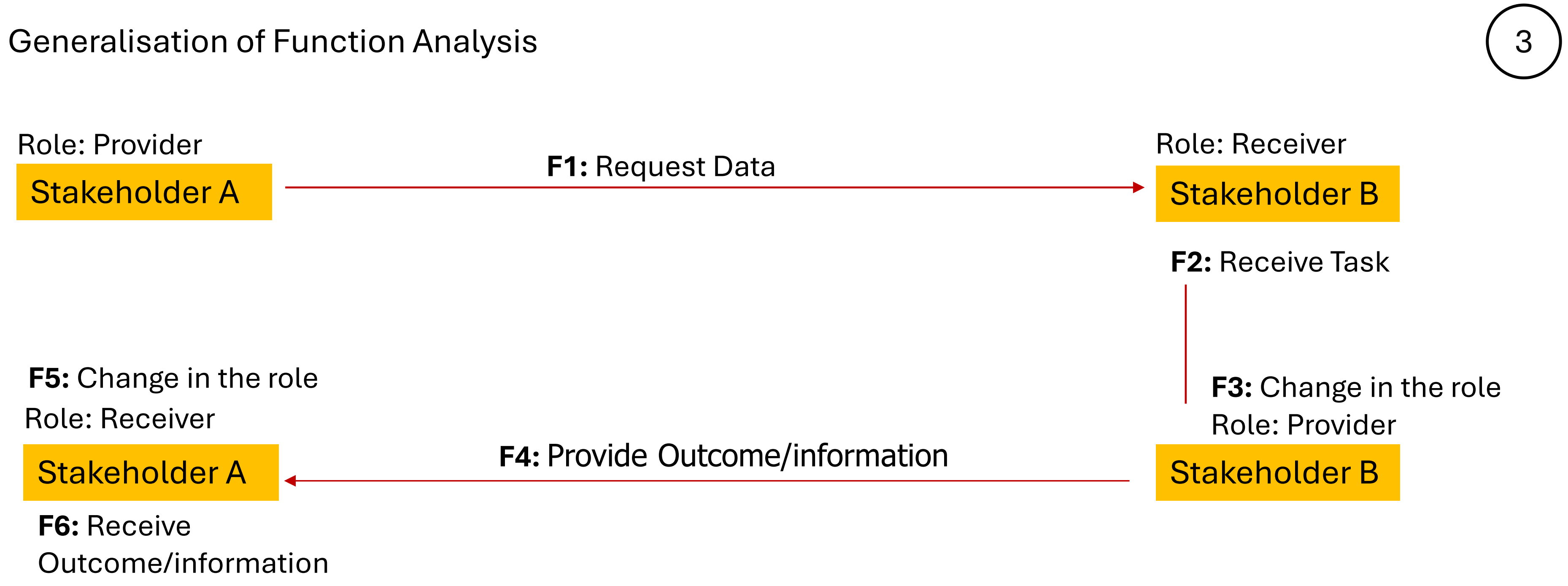

**Supplementary_7**

The steps are as follows:

1. Function Identification
- Identification of functions with involved stakeholders and their roles
2. Needs and Resources
- Identification of needs and resources of stakeholders for each function
3. Matching function explanations with the aspects of sustainability
- Measure sustainability value

---

**1. Function Analysis**

Analyse functions from the content based on the generalisation models of function; see Appendix 1 – 2 – 3.

- Function analysis starts with the function of "Request Task" or "Request information". The following function should be "Receive the task" or "Receive information, "Appendix 1. The following functions should consider the evaluation of information or decision on attempting another task to detail functions. A maximum of 2 times detailing is required in function analysis. Content may involve required information or detailing that can be completed based on the tasks to be completed in related phases of a product lifecycle process.
- The roles of stakeholders should be changed by adding the function "Change in the role Receiver to Provider" or "Change in the role Provider to Receiver" based on the role of the stakeholder group. For example, If stakeholder group A requests a task from stakeholder group B. Stakeholder A is the provider; Stakeholder B is the receiver. When B is providing the outcome, the role of B should be changed from receiver to provider. This change is shown in diagrams in Appendix 1 – 2 – 3.
- In function pairs, the role of the stakeholders should be considered for each function. Providers ensure resources or an opportunity for receivers to benefit from them. Receivers access this opportunity based on their resources and needs, and according to their capabilities and conditions, they can be involved in the function pairs.
- There can be functions which involve only one stakeholder group (individual functions). A function of change in the role is not required for these functions.

It is required to detail functions a max of 2 times if detailed information is not available in the text. Example:

Stakeholder A (provider): Provide information

Stakeholder B (receiver): Receive information

Stakeholder A (provider): Request Task

Stakeholder B (receiver): Receive Task

Stakeholder B(individual function, no need to mention the role): Evaluate information (Detailing 1)

Stakeholder B(individual function, no need to mention the role): Preparation for providing outcome (Detailing 2)

Stakeholder B: Change in the role RP

Stakeholder A: Change in the role PR

Stakeholder B (Provider): Provide outcome

Stakeholder A (receiver): Receive outcome

Stakeholder A (individual function): Evaluate the outcome (1)

- **Identification of functions**

In order to identify functions, it is required to read statements or text. The generalised model of function analysis has three versions.

Example: Based on **the feedback from the users after using the first prototype** and observations, the second prototype was made. "... **Trials** were conducted on the **second prototype after fabrication**. The results of trials show that **the machine performance has increased** and it is more user-friendly than the other two. " (Tak et al., 2019a).

Table 1: Function identification

| | | | | |
|---|---|---|---|---|
| F286&F287 | F286 | Provide training about the 2nd modified machine | Stakeholder A | Provider |
| | F287 | Receive information | Stakeholder B | Receiver |
| F288&F289 | F288 | Request to use the modified machine | Stakeholder A | Provider |
| | F289 | Receive the request | Stakeholder B | Receiver |
| X | F290 | Use the machine | Stakeholder B | - |
| X | F291 | Observe the users | Stakeholder A | - |
| F292&F293 | F292 | Request to provide feedback about the modified machine | Stakeholder A | Provider |
| | F293 | Receive the task | Stakeholder B | Receiver |
| X | F294 | Change in the role PR | Stakeholder A | Receiver |
| X | F295 | Change in the role RP | Stakeholder B | Provider |

- **Stakeholders**

All stakeholders are considered as a group.

Example: "The "floating fish cages for inland waters" developed by RuTAG IIT Bombay has impacted local fisherman cooperative society in terms of livelihood generation, increasing women participation in cage fishing due to increase in safety factor [4]..." (Tak et al, 2019b).

Based on the text, the stakeholder groups are "RuTAG" and the "local fishermen cooperative society". We do not focus on the internal hierarchy of an organisation. RuTAG must have different departments to attempt various functions. However, we accept it as one stakeholder group without considering internal relations. Our focus is on function detailing as a first goal.

**Summarise: Follow these steps:**

1. Read the text
2. Highlight the activities (observe, survey, field stay)
3. Identify stakeholders involved in highlighted activities
4. Detail each activity (a max of 2 times)
5. Write function explanations based on the generalisation models (follow Appendix 1-2-3)
6. Identify stakeholders for each function.

Example: Based on **the feedback from the users after using the first prototype** and observations, the second prototype was made. "... **Trials** were conducted on the **second prototype after fabrication**. The results of trials show that **the machine performance has increased** and it is more user-friendly than the other two. " (Tak et al., 2019a).

**the feedback from the users after using the first prototype –** the feedback can be asked after the use of the second prototype (Stakeholder A requests feedback from Stakeholder B).

**trials –** sharing information about how to use the machine (Stakeholder A (can be the design team); Stakeholder B (can be the user group))

**the machine performance has increased –** reply to the feedback request (Stakeholder B (provider); Stakeholder A (receiver))

The interaction Diagram can be as follows:

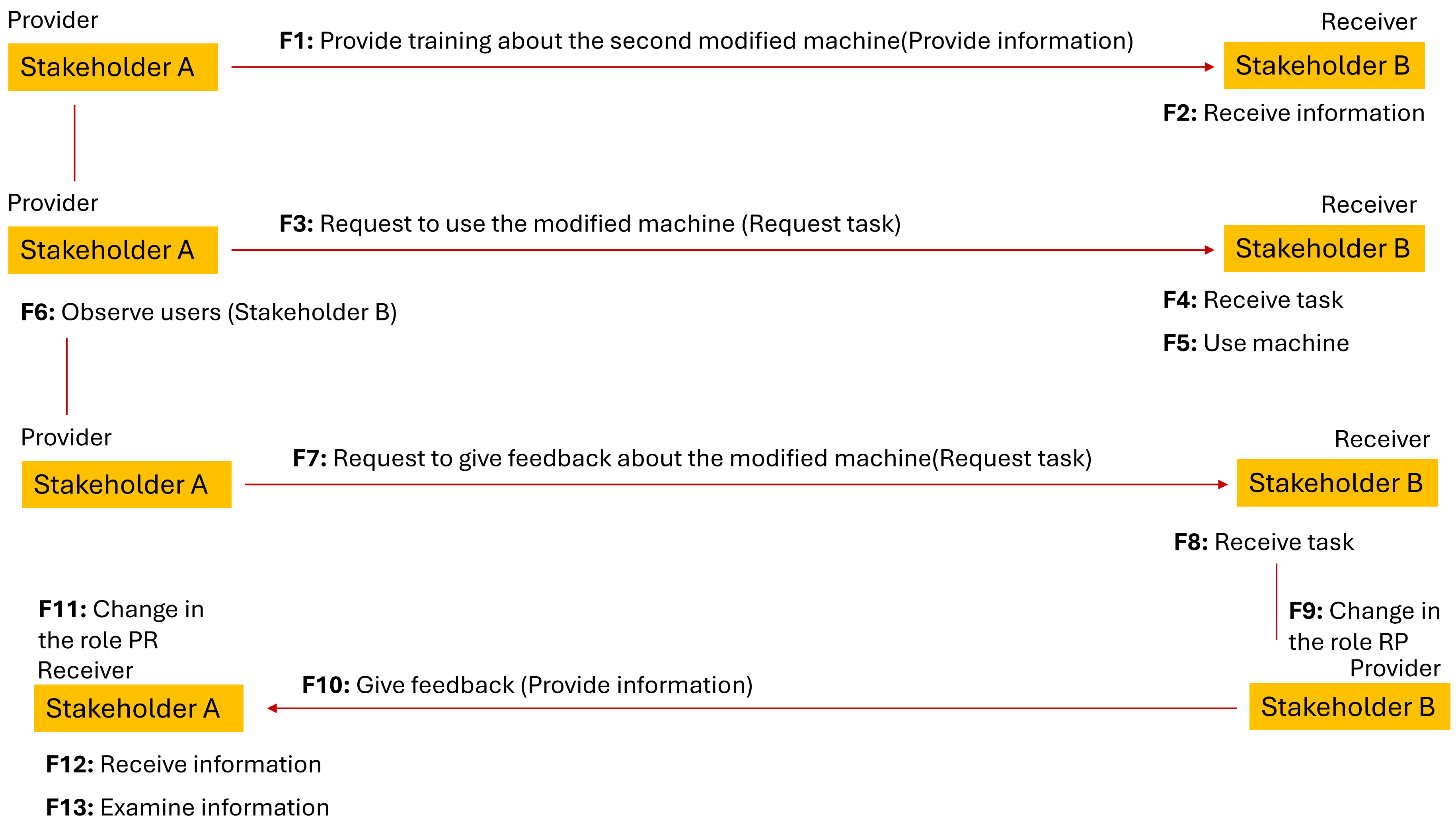

Figure 1: Interaction diagram for function analysis

## 2. Needs and Resources

What is the reason for being involved in a function?

Use the Matrix of Needs and Satisfiers (Max-Neef, 1992) to identify the group and number of needs.

- What is the need of Stakeholder A to attempt the function (286) - Provide training about the 2nd modified machine?
  - Responsibility, duties, work - check the matrix; Participation and Having
- What are the resources of Stakeholder A to attempt the function (286)- Provide training about the 2nd modified machine.
  - Position

Table 2: Needs and resources of stakeholders

| F. Pair | F. No | F. Explanation | Stakeholder | Role | Groups of the needs | # need groups | Groups of the (re)sources | # resource groups |
|---|---|---|---|---|---|---|---|---|
| F286 & F287 | F286 | Provide training about the 2nd modified machine | Stakeholder A | Provider | Participation and Having | 2 | Position | 1 |
| | F287 | Receive information | Stakeholder B | Receiver | Participation and Having | 2 | Experience | 1 |

We keep "resources and needs" limited to the function explanation. It is required to consider only the explanation of the particular function to identify needs and resources.

Resources can be position, experience and capability. It is important to consider which resources have priority in completing the function. For example, Stakeholder A has the experience and position to attempt Function 286. However, the position has priority in this situation to attempt the function. It can be considered as a hierarchy of the resources – which should come first. Due to the position of Stakeholder A, Stakeholder B was involved in the training about the modified machine.

In Figure 1, the need for Stakeholder A to be involved in function F7 is related to the duty of Stakeholder A; therefore, the need is for achievements based on the responsibilities. As proposed in the Matrix of Needs and Satisfiers, this can be grouped under 'Participation and Having' (Max-Neef, 1992). The resource of Stakeholder A to attempt function F7 is the position. It might be considered as experience or capability. However, due to the position (job), Stakeholder A can be involved in F7. The impact of F7 on Stakeholder B is F8. There is a need to attempt F8 to fulfil the requirement, a modified version of the existing machine. As proposed in the Matrix of Needs and Satisfiers, this need can be connected with group Participation and Having' (Max-Neef, 1992). Because the machine can help the users (Stakeholder B) generate their income. In this context, it can be related to responsibilities under the need group of Participation and Having. The resource of Stakeholder B for F8 is the experience of using the previous and modified version of the machine.

## 3. Matching function explanations with the aspects of sustainability

The mobilised resources must be highlighted based on the function explanation to measure sustainability value. Then, read the text (related parts are highlighted on the paper) and read the definitions of the aspects of sustainability (Supplementary_1). Then, you can find the contribution of function and which aspects of sustainability match based on the content and context. Quntify the functions based on the contribution to the aspects of sustainability. If functions contribute to environmental sustainability, it can be numbered as 1. Otherwise, it can be numbered as 0. If function contributes to all aspects of sustainability, the sustainability value is 1. Otherwise, it is 0.

For each function:

Read the content/case study (from where you identified the function explanation) and focus on the detailed information in the content. – What is it about: economic, social or environmental situations?

Read definitions of social, environmental and economic sustainability.

Which information is used in content: what is required to be sustained with the function?

Identify the potential contribution of function on the aspects of sustainability.

Example:

The information mentioned in Function 1 follows:

"Shashwat, a non-government organisation, has been working with tribal people in this block for many years. Livelihood generation is one of the major focuses of this NGO. An ayurvedic medicine manufacturing company from Pune asked Shashwat for decorticated hirda for making triphala churna. The NGO initiated the practice of manual decortication of hirda in Gadewadi village. Manual decortication meant breaking the shell of Hirda by striking it with a stone. After decortication, the self-help group (SHG) got a rate of Rs. 35 per kg for decorticated shells of hirda. Manual decortication of hirda is a slow process. A person can decorticate only 5–7 kg of hirda in 8 h." (Tak et al., 2019a). The information which is provided in this content matches the definitions of economic and social sustainability as clarified with the keywords and reference numbers in the following table:

Table 3: Sustainability value of functions

| Reference (Supplementary_1) of Keywords | Keywords (from definitions) & Functions | SUSTAINABILITY | | | | **Function Groups** | **Function No** | **Function Explanation** | **Stakeholder** | **Role** |
|---|---|---|---|---|---|---|---|---|---|---|
| | | ENV | SOC | ECO | SV | | | | | |
| [2] -[35]- [59] -[29] | resources and services needs, network, source of livelihood (income) | 1 | 1 | 1 | 1 | F1&F12 | F1 | Provide information about the requirement of upgrading the machine | Stakeholder A | Provider |
| [2] -[35]- [59] -[29] | resources and services needs, network, source of livelihood (income) | 1 | 1 | 1 | 1 | | F2 | Receive information | Stakeholder B | Receiver |

**Appendix 1:**

Generalisation of Function Analysis

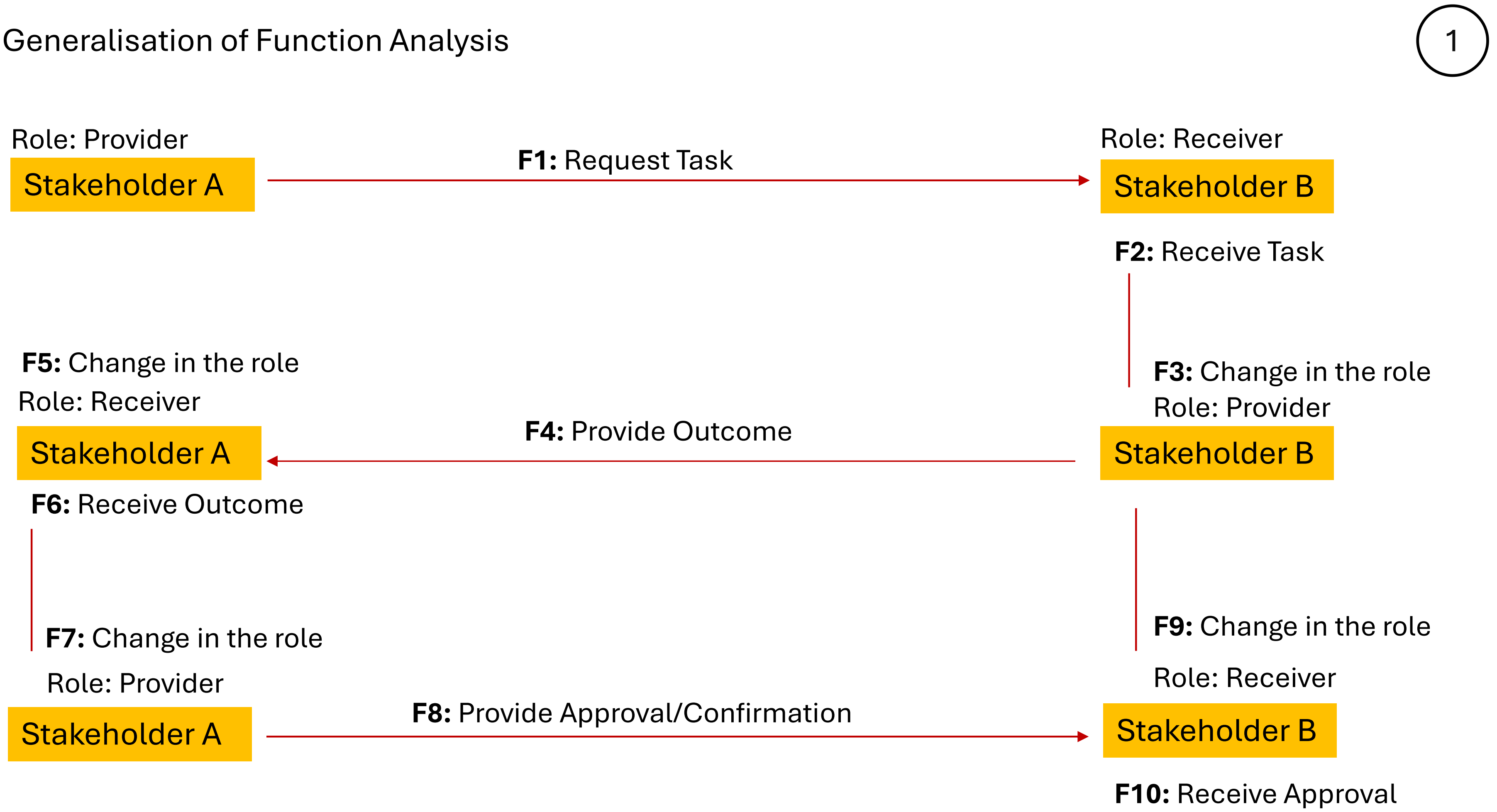

**Appendix 2**

Generalisation of Function Analysis

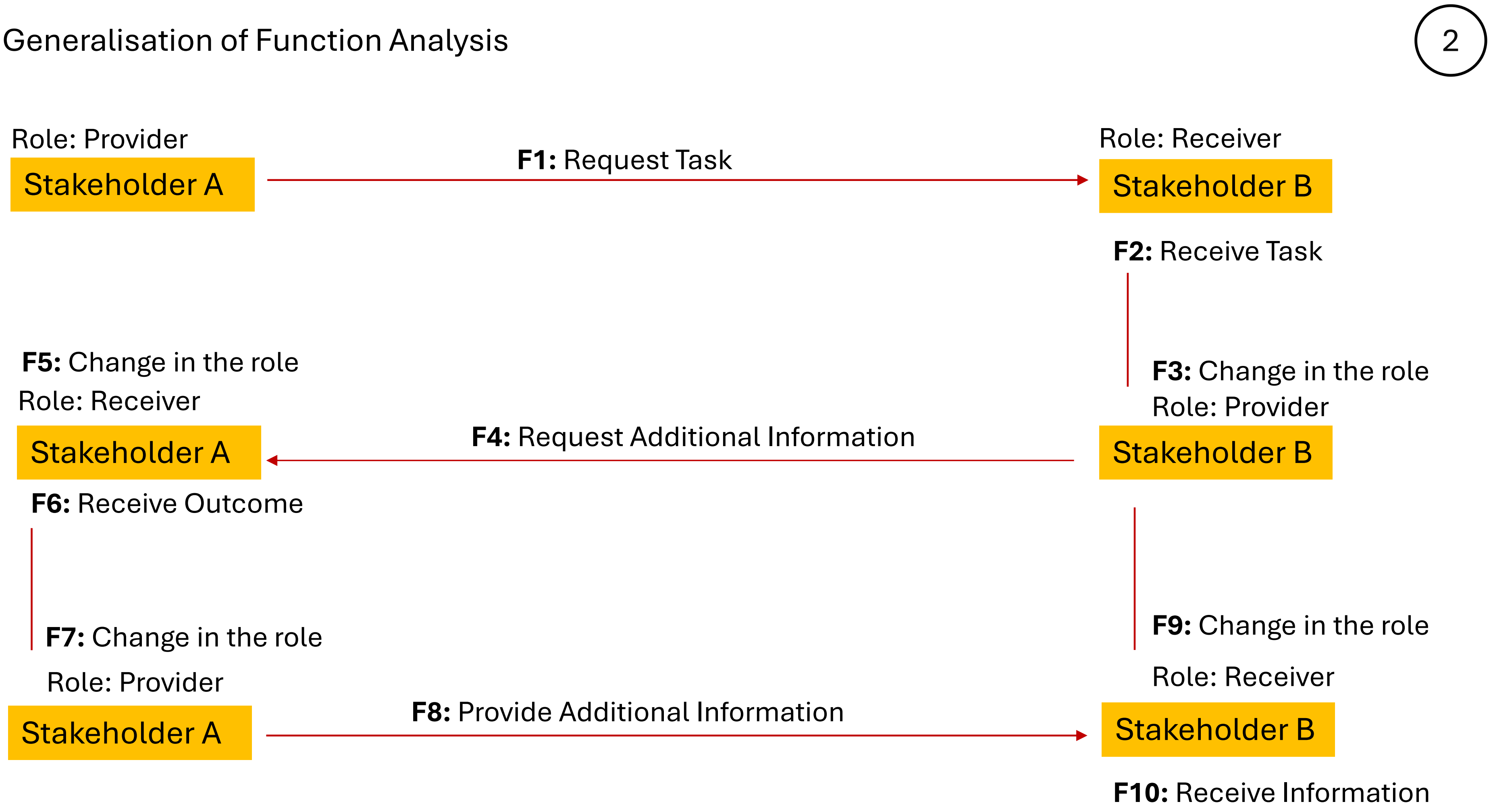

**Appendix 3**

Generalisation of Function Analysis

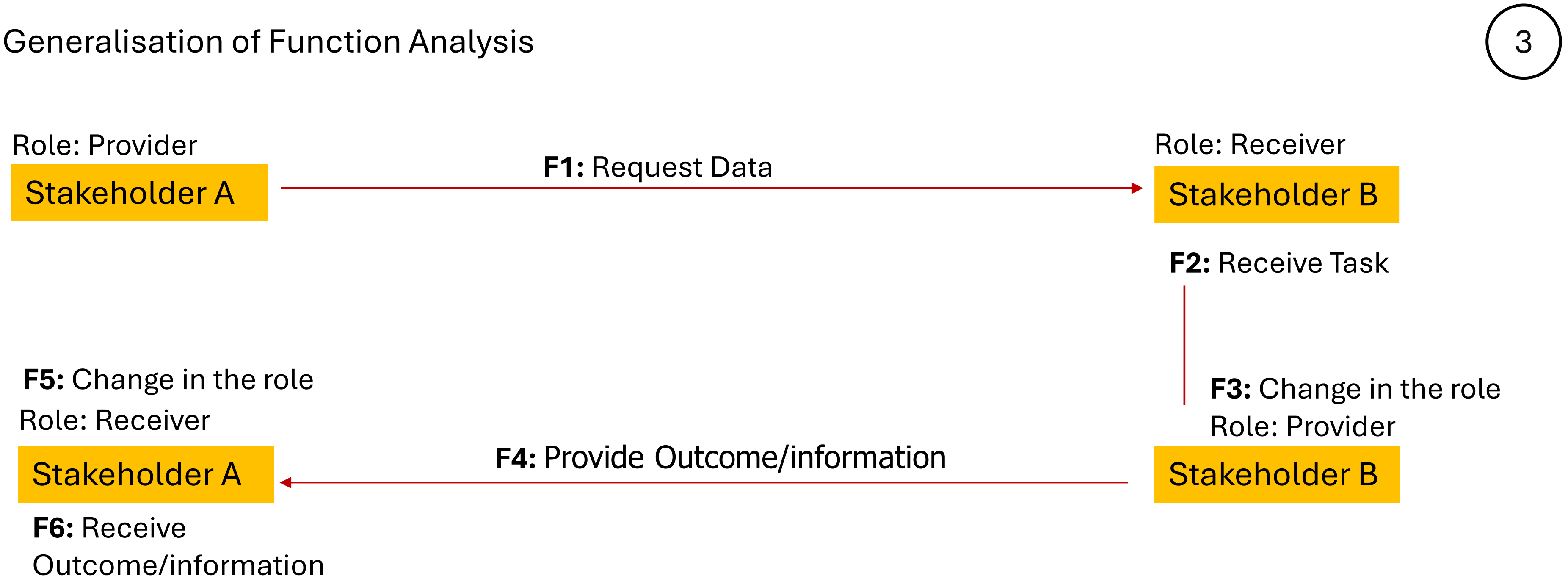